\documentclass[%
 reprint,
superscriptaddress,
 amsmath,amssymb,
 aps,
pra,
longbibliography
]{revtex4-2}

\usepackage{graphicx}
\graphicspath{{./pdf_eps/}}
\DeclareGraphicsExtensions{.eps,.png}

\usepackage{dcolumn}
\usepackage{bm}
\usepackage{natbib}
\usepackage{textcase}
\usepackage{url}
\usepackage{amsmath}
\usepackage{bbold}
\usepackage{calc}
\usepackage{dirtytalk}
\usepackage{mathtools}
\usepackage{array}
\usepackage{siunitx}
\usepackage{hhline}
\usepackage{xcolor}

\DeclarePairedDelimiter{\abs}{\lvert}{\rvert}

\newcommand{\sX}{\mathcal{X}}

\begin{document}

\preprint{APS/123-QED}

\title{The effect of loops on the mean square displacement of Rouse-model chromatin}

\author{Tianyu Yuan}
\affiliation{%
    Integrated Graduate Program in Physical and Engineering Biology, Yale University, New Haven, Connecticut 06520, USA
}
\affiliation{%
    Department of Physics, Yale University, New Haven, Connecticut 06520, USA
}%

\author{Hao Yan}
\affiliation{%
    Integrated Graduate Program in Physical and Engineering Biology, Yale University, New Haven, Connecticut 06520, USA
}
\affiliation{%
    Department of Physics, Yale University, New Haven, Connecticut 06520, USA
}

\author{Mary Lou P. Bailey}
\affiliation{%
    Integrated Graduate Program in Physical and Engineering Biology, Yale University, New Haven, Connecticut 06520, USA
}
\affiliation{%
    Department of Applied Physics, Yale University, New Haven, Connecticut 06520, USA
}

\author{Jessica F. Williams}
\affiliation{
    Department of Cell Biology, Yale School of Medicine, New Haven, Connecticut 06520, USA
}

\author{Ivan Surovtsev}
\affiliation{%
    Department of Physics, Yale University, New Haven, Connecticut 06520, USA
}
\affiliation{
    Department of Cell Biology, Yale School of Medicine, New Haven, Connecticut 06520, USA
}

\author{Megan C. King}
\affiliation{%
    Integrated Graduate Program in Physical and Engineering Biology, Yale University, New Haven, Connecticut 06520, USA
}
\affiliation{
    Department of Cell Biology, Yale School of Medicine, New Haven, Connecticut 06520, USA
}
\affiliation{
    Department of Molecular, Cell and Developmental Biology, Yale University, New Haven, Connecticut 06511, USA
}

\author{Simon G. J. Mochrie}
\affiliation{%
    Integrated Graduate Program in Physical and Engineering Biology, Yale University, New Haven, Connecticut 06520, USA
}
\affiliation{%
    Department of Physics, Yale University, New Haven, Connecticut 06520, USA
}
\affiliation{%
    Department of Applied Physics, Yale University, New Haven, Connecticut 06520, USA
}
\email{simon.mochrie@yale.edu}


\date{April 21, 2023}

\begin{abstract}
Chromatin exhibits polymer properties and its dynamics are now commonly described using the classical Rouse model. 
The subsequent discovery, however, of intermediate scale chromatin organization known as topologically associating domains (TADs) in experimental Hi-C contact maps for chromosomes across the tree of life, together with the proposed loop extrusion factor (LEF) model that aims to explain TAD formation, motivates efforts to understand the effect of loops and loop extrusion on chromatin dynamics.
This paper seeks to fulfill this need by  combining  LEF-model  simulations with extended Rouse-model polymer simulations to investigate the dynamics of chromatin with loops and dynamic loop extrusion.
Specifically, we extend the classical Rouse model by modifying the polymer's dynamical matrix to incorporate extra springs that represent loop bases.
We also theoretically generalize the friction coefficient matrix so that the Rouse beads with non-uniform friction coefficients are compatible with our Rouse model simulation method. 
We show that loops significantly suppress the averaged mean square displacement (MSD) of a gene locus, consistent with recent experiments that track fluorescently-labelled chromatin loci. We also find that loops reduce the MSD's stretching exponent from the classical Rouse-model value of $1/2$  to a loop-density-dependent value in
the 0.45-0.40 range. Remarkably, stretching exponent values in this range have also been observed  in recent experiments
[S. C. Weber, A. J. Spakowitz, and J. A. Theriot, Phys. Rev. Lett. 104, 238102 (2010)
and M. L. P. Bailey, I. Surovtsev, J. F. Williams, H. Yan, T. Yuan, S. G. Mochrie, and M. C. King, Mol. Biol. Cell (in press)].
We also show that for a wide range of plausible loop-extrusion parameters, the dynamics of loop extrusion itself negligibly affects chromatin mobility.
By studying a static ``rosette'' loop configuration, we also demonstrate that chromatin MSDs and stretching exponents depend on the location of the locus in question
relative to the position of the loops.
Finally, we show that  non-uniformity in friction coefficients for different parts of a chromatin polymer does not significantly modify its averaged dynamics.
However, unsurprisingly, individual loci with distinctly high friction coefficients exhibit correspondingly reduced MSDs.
\end{abstract}

\maketitle

\section{Introduction}
\label{intro}
The classical Rouse model for the dynamics of a polymer in a viscous fluid
depicts the polymer as an array of overdamped beads,
connected together by nearest-neighbor springs~\cite{rouse1953}.
Each bead actually represents a sub-polymer, whose end-to-end distance follows a Gaussian distribution,
which determines the Rouse model spring stiffness.
Despite its simplicity,
in cases where interaction between different polymers restores $N^\frac{1}{2}$-scaling of the polymer end-to-end distance~\cite{deGennes1979}
($N$ is the number of monomers in the polymer in question),
and where hydrodynamic interactions between different submolecules are screened by other polymers~\cite{ahlrichs2001},
and where the polymers are shorter than their entanglement length~\cite{putz2000},
the Rouse model's predictions
match experimental measurements on many polymer melts~\cite{richter1993,pearson1994,wischnewski2003} and solutions~\cite{weiss2004,di2018,tamm2015}.

Beyond synthetic polymers, the possibility that the Rouse model might also provide an appropriate description of the dynamics of chromatin
in living cells has emerged from experiments which quantitatively
characterize the motion of a fluorescently-labelled gene locus via its mean-square displacement (MSD)~\cite{hediger2003,cabal2006,weber2010a,weber2012a,weber2012b,hajjoul2013,verdaasdonk2013,backlund2014,backlund2015,wang2015,rolls2017,shukron2017,osmanovic2017,socol2019}.
In many of these experiments, the experimentally measured mean-square displacement  of a labelled gene locus
behaves similarly to the Rouse-model prediction that the MSD initially increases as $t^\frac{1}{2}$ with increasing time, $t$.
However, chromatin, which is comprised of DNA and its myriad associated proteins, is now understood to possess a more elaborate spatial organization than the simple random-walk polymer envisioned in the classical Rouse model.
It is unclear whether the classical Rouse model predictions should be expected to
apply to such a polymer.

Until recently, chromatin organization was well established only at the two extremes of the genome scale - with DNA wrapped around histones in a nucleosome at the molecular scale (hundreds of DNA base pairs)~\cite{kaplan2009,zhang2011,garcia2017}, and with each chromosome largely occupying its own space in the nucleus in chromosome territories (millions of base pairs)~\cite{zink1998,fritz2019,ghosh2021}.
However, the advent of chromatin configuration capture (Hi-C) methods has now unveiled an inhomogeneous, hierarchical, domains-within-domains, organization at intermediate scales ($10^4-10^6$~bps)~\cite{aiden2009,berkum2010,dixon2012,dixon2016,sexton2012,mizuguchi2014,dekker2014,dekker2015,pollard2016,jerkovic2021}.
Gene loci within the same domain (termed topologically associated domains or TADs) have a much higher probability to come into contact, even if they are genomically distant, than do loci from different TADs~\cite{dekker2016}.
High contact probability between two distant loci supports the idea that the two loci in question are likely to be at the base of a chromatin loop, consistent with the long-standing hypothesis that loops are a fundamental organizing principle for chromatin~\cite{schleif1992,yokota1995,dekker2008}.

The loop extrusion factor (LEF) model has emerged as the preferred candidate mechanism for TAD formation~\cite{alipour2012,sanborn2015,fudenberg2016,goloborodko2016.1,goloborodko2016.2,nuebler2018,banigan2020,davidson2021,oudelaar2021}.
In this model, LEFs bind to chromatin and then initiate dynamic loop extrusion, until they either stall when they encounter another LEF or at specific boundary elements (generally identified as CCCTC-binding factor or CTCF binding sites), which mark TAD boundaries, or until they unbind.
Thus, a population of LEFs establishes a dynamic steady-state, which largely recapitulates contact probabilities determined by experimental Hi-C maps~\cite{bonev2017}.
On the experimental side, the structural maintenance of chromosomes (SMC) complexes, cohesin and condensin, have been
identified as possible LEFs, with TADs disappearing from Hi-C maps in the absence of cohesin~\cite{rao2017,wutz2017}.
At larger, sub-chromosomal scales, phase separation of different regions has been proposed to further organize chromatin into chromatin ``compartments''~\cite{ghosh2021,zhang2019,misteli2020,ahn2021,erdel2018}.

Although a number of modifications to the classical Rouse model have been proposed to better describe chromatin dynamics~\cite{weber2010b,osmanovic2017,socol2019}, to-date how polymer loops might affect the predictions of the Rouse model have not been described, to our knowledge.
In this paper, motivated by the LEF model, we consider a modified version of the Rouse model that includes loops in order to investigate how loops may affect chromatin MSDs.
Specifically, we augment the classical nearest-neighbor Rouse model by adding an additional spring (of the same spring constant) between the pair of monomers at the base of each loop.
The number, sizes, and locations of these additional springs evolve according to LEF model simulations. 
Because the additional springs lead to far-from-diagonal terms in the Rouse-model dynamical matrix, our modified Rouse model is no longer analytically tractable.
Nevertheless, it is straightforward to simulate in an exact manner as follows.
First, we diagonalize the modified dynamical matrix numerically, based on the current loop configuration, and find its eigenvalues and eigenvectors, which define the coordinate transformation to normal coordinates.
The time evolution of each independent normal coordinate is then simulated, using a version of the method described in Ref.~\cite{gillespie1996},
assuming equipartition with an effective temperature.
Next, the bead positions versus time are recovered by the inverse coordinate transformation.
The above steps are repeated for each different loop configuration to obtain the time series of bead positions.
Finally, the MSD of each bead, representing the MSD of the corresponding gene locus, is determined from the time series of bead positions.

This method applied to the classical Rouse model reproduces the well-known analytical results for the behavior of the MSD:
At early times, the 1-D MSD varies as $2D t^\alpha$ with the stretching exponent $\alpha$, taking on the predicted value of $\alpha=1/2$ and the amplitude, $D$, taking on the predicted value of $D=k_BT/\sqrt{\pi\kappa\zeta}$, where $\kappa$ is the effective spring constant between subpolymers and $\zeta$ is the friction coefficient for each subpolymer; at long times the MSD achieves a (boundary-condition dependent) limiting value.
By contrast, for the Rouse model with loops, the MSD is significantly reduced.
In addition, it shows a noticeably reduced and subtly time-dependent stretching exponent with values that clearly fall below one-half.
Interestingly, these results are reminiscent of recent experimental results, that examine the dynamics of gene loci in living cells~\cite{weber2010a,weber2012a,weber2012b,bailey2023}.

Beyond the LEF model, our approach is applicable to any polymer configuration involving loops, by appropriately picking the locations of the additional springs in the Rouse-model dynamical matrix.
Another loop configuration that has attracted attention is a ``rosette'', in which similarly-sized loops emanate from an organizing center.
For example, rosettes are believed to be relevant to  the {\em E. coli} nucleoid~\cite{hinnebusch1997,macvanin2012}.
Examining a fixed-loop-configuration rosette facilitates investigation of the MSDs for several distinct genomic locations, relative to the loops.
Thus, we find that the tip of a loop exhibits the largest mobility, at early times appearing unconstrained by loops, while the mobility at the base of a loop is constrained the most. 
The mobility of loci on the polymer backbone, between loops, resembles that of the tips at early times but approaches the mobility of the loop bases at longer times.

Our approach also allows us to assign different friction coefficients to individual beads, granting us the ability to investigate polymer dynamics in environments with inhomogeneous viscosity. 
Extending the Rouse model to embrace non-uniform friction environments is motivated  by the observation
that chromatin itself is locally heterogeneous. 
We assign each bead a fixed friction coefficient drawn from a log-normal distribution with mean $\mu_0$ and standard deviation of $\sigma_0=\mu_0$ and calculate the MSDs and stretching exponents for each bead. 
The averaged MSD and stretching exponent over 60 uniformly-spaced beads are calculated as well. 
We find that although the MSD and stretching exponent for each individual bead vary in a wide interval from bead to bead, the 60-bead-averaged MSD and stretching exponent shows insignificant deviation from the averaged MSD and stretching exponent for the polymer with uniform friction coefficient of $\mu_0$.
So-called replication factories may provide another example, where inhomogeneous friction is important.
In a replication factory, chromatin is hypothesized to be folded into a rosette-like structure, at whose center  transcription factors and SMCs form a phase-separated droplet and assist the simultaneous replication of the rosette loops~\cite{newport1996,ma1998,jackson1998,frouin2003,guillou2010,saner2013,mangiameli2018,geiger2021}.
Plausibly, the phase separation gives rise to a locally-high viscosity.
For this scenario, unsurprisingly, we find that the mobility of  chromatin loci within the high-friction central cluster  is significantly reduced compared to the case when the friction is the same for all beads, suggesting a potential role for locus dynamics in elucidating phase separation within the nucleus~\cite{weber2019,strom2019,razin2020}.

This paper is organized as follows.
In Sec.~\ref{nbead}, we examine theoretically the dynamics of $N$-coupled beads, subject to
a random force, for general, symmetric, dynamical and friction coefficient matrices
in terms of the appropriate normal coordinates.
In Sec.~\ref{equipThm}, we use the equipartition theorem of statistical mechanics to express
the  mean-square amplitude of each normal coordinate in terms of an effective temperature
and an eigenvalue of the dynamical matrix.
Sec.~\ref{analysoln} applies the results of Sec.~\ref{nbead} and \ref{equipThm} to the classical Rouse model with free ends (free boundary conditions),
reproducing the well-known analytic results for the Rouse-model MSD.
Sec.~\ref{scalinglaw} reviews the scaling of $\kappa$,  $\zeta$, and $D$ with the
size of the subpolymers and explains how  experimental measurements of $D$ and literature estimates of the chromatin persistance length
and of the number of DNA base pairs per unit chromatin contour length allow us to estimate appropriate values of $\kappa$ and $\zeta$ for our polymer simulations.
Sec.~\ref{loopcombrouse} describes two slightly different versions of the LEF model, which we call the random loop model and the CTCF model, respectively,
and which are the basis of our loop extrusion factor (LEF) simulations.
Sec.~\ref{loopcombrouse2} explains how we carry out Rouse model simulations,
incorporating loops and loop extrusion.
In Sec.~\ref{R&D}, we present our results.
Sec.~\ref{loopvsnoloop} examines the simulated dynamics of the chromatin polymer with dynamic loops, driven by the LEF model.
Comparison between polymers with and without loops reveals that loops significantly reduce chromatin mobility, as measured by the mean-square displacement (MSD). Loops also reduce the stretching exponent $\alpha$ from a plateau value of 0.5 without loops to a plateau value lying between about $0.45$ and $0.4$, depending on the density of loops.
By examining the behavior of polymers with loops but without loop extrusion, {\em i.e.} polymers with static loops,
Sec.~\ref{loopvsnoloop} also reveals that the dynamics of loop extrusion itself
has a minimal effect on the dynamics of the chromatin polymer, as measured by both the MSD and stretching exponent, suggesting that although LEFs actively extrude loops, their dynamics do not  dictate how chromatin polymers move.
Sec.~\ref{rosette} examines the simulated dynamics of static loops in a rosette, revealing that mobility is reduced near the base of a loop
in comparison to the mobility at the tip of a loop.
In Sec.~\ref{LogNormal}, we examine Rouse-model polymers without loops, when the beads' friction coefficients are randomly distributed according to a log-normal distribution.
We calculate the MSDs and stretching exponents for individual beads as well as the averaged MSD and stretching exponent for 60 of these beads, uniformly distributed along the polymer.
Comparing these averaged results to those for a polymer with uniform friction, we find that there are insignificant differences for both MSD and stretching exponent between the non-uniform-friction and the uniform-friction case.
Sec.~\ref{NUFEnv} shows that polymers with rosette structures subject to locally high friction at the central cluster, mimicing  replication factories, exhibit significantly reduced MSDs in the high-friction region (rosette bases) but exhibit similar MSDs at the tips of loops, where the friction environment is unchanged.
Finally, we conclude in Sec.~\ref{Concl}.

\section{Theoretical background}
\label{teocsv}
\subsection{Dynamics of $N$-coupled beads}
\label{nbead}
The Rouse model describes a polymer as $N$ subpolymers connected together into a chain.
Each subpolymer is conceived to be sufficiently long that its end-to-end distance is a Gaussian random variable.
Consequently, each subpolymer acts as a Hooke's-law spring with a spring constant equal to \(\kappa = d k_BT/\langle R_{EE}^2\rangle\),
where $k_B$ is the Boltzmann constant, $T$ is the effective temperature, $d$ is the number of spatial dimensions, and \(\langle R_{EE}^2\rangle\) is the mean square end-to-end distance of a subpolymer \cite{strobl1997}.
It is often convenient to conceive the center of mass of each subpolymer as a bead.
This picture leads to a chain comprised of \(N\) beads, each with friction coefficient, \(\zeta\), and connected to its neighbors via springs of spring constant, \(\kappa\).

The equations of motion for $N$ beads, coupled together by
springs, in the low-Reynolds-number, overdamped regime, may be expressed as the matrix equation
\begin{equation}
    Z \dot{X}=-KX+F,
    \label{newtoncoupled}
\end{equation}
where \(X\) is the vector of bead coordinates, \(Z\) is the friction matrix, \(K\) is the dynamical matrix,
and \(F\) is a vector of stochastic forces.
To ensure that our discussion is applicable to a Rouse polymer with loops, we envision $K$ and $Z$ to be as general as possible.
Nevertheless,
Newton's Third Law requires that \(K\) and \(Z\) are both symmetric;
in addition, $Z$ must be positive
definite in order to ensure dissipation,
and $K$ must be positive semi-definite to ensure that the minimum potential energy corresponds to zero displacements relative to the center of mass.
It is convenient to introduce dimensionless versions of $Z$ and $K$, namely $J$ and $A$, respectively,  via \(Z = \zeta J\) and \(K = \kappa A\). In terms of  $J$ and $A$,
Eq.~\ref{newtoncoupled} reads
\begin{equation}
    \label{stochLangevin}
    \frac{dX}{dt}=-\frac{\kappa}{\zeta}J^{-1}A X+\frac{1}{\zeta}J^{-1}F.
\end{equation}
To proceed, we
diagonalize Eq.~\ref{stochLangevin}
(assuming that \(J^{-1}A\) is diagonalizable).
Supposing  that \(S\) is the matrix that diagonalizes \(J^{-1}A\), {\em i.e.}   \(S^{-1}J^{-1}AS = D_1\),
where $D_1$ is a diagonal matrix,
then Eq.~(\ref{stochLangevin}) becomes
\begin{equation}
    \frac{dX}{dt} = -\frac{\kappa}{\zeta}S D_1 S^{-1}X + \frac{1}{\zeta}J^{-1}F.
\end{equation}
Introducing the normal coordinates,  \(\sX = S^{-1}X\), we obtain
\begin{equation}
    \frac{d\sX}{dt} = -\frac{\kappa}{\zeta} D_1 \sX + \frac{1}{\zeta} \widetilde{F},
\end{equation}
where $\widetilde{F} = S^{-1}J^{-1}F$.
Thus, $N$ coupled equations of motion  decouple into \(N\) independent equations of normal coordinates, $\sX_m$, which satisfies, for each $m$,
%
\begin{equation}
    \frac{d\sX_m}{dt} = -\frac{\kappa}{\zeta} \Lambda_m\sX_m + \frac{1}{\zeta}\widetilde{F}_m,
    \label{EQ5}
\end{equation}
where \(\Lambda_m\) is the \(m\)-th eigenvalue of \(J^{-1}A\).
Eq.~\ref{EQ5} leads to the following results:
\begin{equation}
    \left < \sX_m(t \to \infty) \right > = \left< \sX_m \right> = 0,
    \label{EQ6}
\end{equation}
\begin{equation}
    \left< \sX^2_m(t \to \infty) \right> = \sigma^2_{ \sX_m},
    \label{EQ7}
\end{equation}
and
\begin{equation}
    \left< \sX_m(t + s) \sX_m(s) \right>_{s \to \infty} = \sigma^2_{ \sX_m} e^{-\frac{\kappa}{\zeta}\Lambda_m t},
    \label{EQ8}
\end{equation}
where $\left < .... \right >$ denotes an ensemble average, and $\sigma^2_{ \sX_m}$ and $\langle \sX_m \rangle$ are the variance and mean of eigenmode $m$ at steady state
\cite{gillespie1996}.

\subsection{Equipartition and theoretical MSD for $N$-coupled beads}
\label{equipThm}
To calculate $\sigma^2_{ \sX_m}$, we turn to statistical mechanics,
assuming an effective temperature, $T$.
To this end, first, we express the $N$-bead potential energy, $U$, in terms of the normal coordinates:
\begin{equation}
    \label{qForm}
    \begin{aligned}
        U&=\frac{1}{2}X^TKX \\
        &=\frac{\kappa}{2}\sX^TS^TAS\sX \\
        &=\frac{\kappa}{2}\sX^T(S^TAS)\sX.
    \end{aligned}
\end{equation}
Evidently, the $N$-bead potential energy is a quadratic form of the normal coordinates.
Because $S$ was chosen to diagonalize $J^{-1} A$, not $A$, it is not obvious from Eq.~\ref{qForm} that the potential energy is a function of ${ \sX_m}^2$-terms only.
However, if we can prove \(S^TAS\) is diagonal, then the potential energy is guaranteed to decouple in the normal coordinates \(\sX\), and we can straightforwardly calculate $\sigma^2_{ \sX_m}$, using the equipartition theorem.
Because $A$ is positive semi-definite and symmetric, we may write
\begin{equation}
    S^T A S = (A^{\frac{1}{2}} S)^T A^{\frac{1}{2}} S.
    \label{EQsquarerootA}
\end{equation}
Eq.~\ref{EQsquarerootA} informs us that
for $S^T A S$ to be diagonal  requires that
$\{ A^{\frac{1}{2}} v_m \}$ are orthogonal vectors, where $\{ v_m \}$  are the eigenvectors of $J^{-1} A$
and constitute the columns of $S$.
To show that $\{ A^\frac{1}{2} v_m \}$ are orthogonal,
we start with the eigenvalue equation:
\begin{equation}
    J^{-1}A v_m = \Lambda_m v_m.
    \label{eigenvalueequation}
\end{equation}
Multiplying both sides of Eq.~\ref{eigenvalueequation} by $A^\frac{1}{2}$, we find
\begin{equation}
    A^\frac{1}{2}J^{-1}A^\frac{1}{2} (A^\frac{1}{2}v_m)
    = \Lambda_m (A^\frac{1}{2} v_m).
\end{equation}
Thus, we see that $A^\frac{1}{2} v_m$ is an eigenvector of the matrix, $A^\frac{1}{2}J^{-1}A^\frac{1}{2}$, with eigenvalue, $\Lambda_m$.
The matrix, $A^\frac{1}{2}J^{-1}A^\frac{1}{2}$, is symmetric because $A$ and $J$ are symmetric and raising a matrix to a power commutes with transposing the matrix.
Because a symmetric matrix is guaranteed to possess orthogonal eigenvectors, it follows
that
$\{ A^\frac{1}{2} v_m \}$ are indeed orthogonal. (In the case of two equal eigenvalues,
it is always possible to pick orthogonal eigenvectors in that eigen-subspace.)
Defining the diagonal matrix $D_2 = S^T A S$, we thus have

\begin{equation}
   U=\frac{\kappa}{2}\sX^T D_2 \sX = \frac{\kappa}{2}\sum_m {\lambda}_m \sX_m^2.
   \label{PE_decouple}
\end{equation}
where $\lambda_m$ is the $m$-th eigenvalue (diagonal entry) of $D_2=S^T A S$.
Thus, for non-zero $\lambda_m$'s, the equipartition theorem gives 
\begin{equation}
    \langle\sX_m^2\rangle = \sigma^2_{\sX_m}= \frac{k_BT}{\kappa \lambda_m},
    \label{EQ14}
\end{equation}
and Eq.~\ref{EQ7} and~\ref{EQ8} become
\begin{equation}
\left < \sX_m^2(t)    \right > =\frac{k_BT}{\kappa \lambda_m},
\label{EQ15}
\end{equation}
and
\begin{equation}
\left < \sX_m(t) \sX_m(0)    \right > =\frac{k_BT}{\kappa \lambda_m} e^{-\frac{\kappa}{\zeta} \Lambda_m t},
\label{EQ16}
\end{equation}
respectively.
For $\lambda_m = 0$,
we set the corresponding normal coordinate, $\sX_m$, which is proportional to the polymer's center of mass, to zero, which eliminates the overall diffusion,
mimicking chromatin's confinement within the nucleus. 

$X(t)$ may be expressed in terms of $\sX(t)$:
\begin{equation}
X_n(t) = \sum_{m=1}^N S_{nm} \sX_m(t).
\end{equation}
Using the fact that different normal coordinates are uncorrelated,  it
immediately follows that
\begin{align}
    \left <X_n^2(t) \right > &= \sum_{m=1}^N \sum_{p=1}^N S_{nm} S_{np} \left < \sX_m(t) \sX_p(t) \right > \nonumber \\
    &= \sum_{m=1}^N \ S^2_{nm} \frac{k_BT}{\kappa \lambda_m},
\label{EQ18}
\end{align}
and
\begin{align}
    \left <X_n(t) X_m(0) \right > &= \sum_{m=1}^N \sum_{p=1}^N S_{nm} S_{np} \left < \sX_m(t) \sX_p(0) \right > \nonumber \\
    &= \sum_{m=1}^N  S^2_{nm} \left < \sX_m(t) \sX_m(0)\right > \nonumber \\
    &= \sum_{m=1}^N  S^2_{nm}\frac{k_BT}{\kappa \lambda_m} e^{-\frac{\kappa}{\zeta} \Lambda_m t}.
\end{align}
The MSD of bead $n$ is, therefore,
\begin{align}
    \left < [ X_n(t)-X_n(0) ]^2 \right > &= 2 \left < X^2_n(0) \right > -2 \left < X_n(t) X_n(0) \right > \nonumber \\
    &=\frac{2 k_BT}{\kappa } \sum_{m=1}^N \ \frac{S^2_{nm}}{\lambda_m}  (1-e^{-\frac{\kappa}{\zeta} \Lambda_m t}).
\label{EQ20}
\end{align}
These results are applicable in general when \(K\) and $Z$ (or \(A\) and $J$) are both symmetric, \(K\) (or \(A\)) is positive semi-definite and \(Z\) (or \(J\)) is positive-definite. 
We also require that \(J^{-1}A\) is  diagonalizable.
An important special case is when $J$ is the identity matrix.
In this instance, since $D_2=S^TAS=S^T S S^{-1}J^{-1}A S = S^T S D_1 = D_1$,
$\lambda_m$ reduces to \(\Lambda_m\). 

\subsection{Classical Rouse model}
\label{analysoln}
For the classical nearest-neighbor Rouse model of $N$-beads with free ends, the dynamical matrix, $K$, is given by,
\begin{equation}
    K = \kappa A = \kappa
    \begin{bmatrix}
        1 & -1 & 0 & \cdots & \cdots & 0 \\
        -1 & 2 & -1 & \cdots & \cdots & 0 \\
        0 & -1 & 2 & \cdots & \cdots & 0 \\
        \vdots & \vdots & \vdots & \ddots & \ddots & \vdots \\
        \vdots & \vdots & \vdots & \ddots & 2 & -1 \\
        0 & 0 & 0 & \cdots & -1 & 1
    \end{bmatrix}
    ,
\end{equation}
and the friction matrix, \(Z\), is diagonal,
\begin{equation}
    Z = \zeta J =\zeta
    \begin{bmatrix}
        1 & 0 & 0 & \cdots & \cdots & 0 \\
        0 & 1 & 0 & \cdots & \cdots & 0 \\
        0 & 0 & 1 & \cdots & \cdots & 0 \\
        \vdots & \vdots & \vdots & \ddots & \ddots & \vdots \\
        \vdots & \vdots & \vdots & \ddots & 1 & 0 \\
        0 & 0 & 0 & \cdots & 0 & 1 
    \end{bmatrix}
   .
\end{equation}
In this case, $K$, \(A\), and $J^{-1}A$ are diagonalized by the orthogonal matrix, $S$, given by:
\begin{equation}
 S=\sqrt{\frac{2}{N}} 
    \begin{bmatrix}
        1/\sqrt{2} & \cos\frac{1\pi/2}{N} & \cos\frac{1\pi}{N} & \cdots \\
        1/\sqrt{2} & \cos\frac{3\pi/2}{N} & \cos\frac{3\pi}{N} & \cdots \\
        1/\sqrt{2} & \cos\frac{5\pi/2}{N} & \cos\frac{5\pi}{N} & \cdots \\
        \vdots & \vdots & \vdots & \vdots \\
        \vdots & \vdots & \vdots & \vdots \\
    \end{bmatrix}
    ,
\end{equation}
i.e. 
 \begin{equation}
    S_{nm}=
    \begin{cases}
        \sqrt{\frac{2}{N}}\cos\left[\frac{\pi\left(n+\frac{1}{2}\right)m}{N}\right]\quad & m\neq 0 \\
        \sqrt{\frac{1}{N}}\quad & m=0,
    \end{cases}
\end{equation}
given the matrix index range as $\{n,m\}\in\{ 0,1,2,\dots,N-1 \}$, and $N$ as the number of beads~\cite{keesman2013,sato2021}.
Matrix $S$ diagonalizes \(A \), {\em i.e.} \(SAS=D_1\).
The resultant diagonal matrix \(D_1\) contains the eigenvalues of $A$ on its diagonal: 
\begin{equation}
D_1 =    \begin{bmatrix}
        0 & 0 & 0 & \cdots \\
        0 & 2-2\cos\frac{\pi}{N} & 0 & \cdots \\
        0 & 0 & 2-2\cos\frac{2\pi}{N} & \cdots \\
        \vdots & \vdots & \vdots & \ddots \\
    \end{bmatrix}
    ,
\end{equation}
i.e.
\begin{equation}
    \label{eigenVfree}
    \Lambda_m=2-2\cos\frac{m\pi}{N}.
\end{equation}

Using these results (and setting the normal coordinate with zero eigenvalue to zero), Eq.~\ref{EQ18} and Eq.~\ref{EQ20} become
\begin{align}
    \langle X_n^2(0) \rangle &= \langle X_n^2(t) \rangle \nonumber \\
    &= \frac{k_BT}{\kappa N} \sum_{m=1}^{N-1}
    \cos^2 \left[\frac{\pi (n+\frac{1}{2})m}{N}\right]
    \frac{1}{\left[1-\cos\left(\frac{m\pi}{N}\right)\right]} \nonumber \\
    &= \frac{k_BT}{\kappa N} 
    \left[ \left(n-\frac{1}{2}(N-1)\right)^2 + \frac{1}{12}(N^2-1) \right],
    \label{MSD0}
\end{align}
and
\begin{align}
    \langle &\left [ X_n(t)-X_n(0)\right ] ^2 \rangle =\text{MSD}_n(t) =  \nonumber \\
    &\frac{2k_BT}{\kappa N} 
    \sum_{m=1}^{N-1} \cos^2 \left[\frac{\pi (n+\frac{1}{2})m}{N}\right]
    \frac{1-e^{-\frac{2\kappa}{\zeta}\left(1-\cos(\frac{m\pi}{N})\right)\abs*{t}}}
    {1-\cos \left(\frac{m\pi}{N}\right)},
    \label{MSD1}
\end{align}
respectively.
The index of the first bead starts from $n = 0$.
At early times, Eq.~\ref{MSD1} may be shown to be well-described by
\begin{equation}
    \text{MSD}_n(t) = \left < [X_n(t)-X_n(0)]^2 \right > = 2 D t^\frac{1}{2},
    \label{MSDdiffusionform}
\end{equation}
with
\begin{equation}
    D
    =\frac{k_BT}{\sqrt{\pi \zeta \kappa}},
    \label{RouseDiffusionCoefficient}
\end{equation}
where $D$ plays a role similar to a diffusion coefficient~\cite{doi1988}. 
Conventionally, one defines the polymer time-scale to be the quantity, $\tau_p = \zeta/(4 \kappa)$, which is proportional to the inverse of the largest eigenvalue and proportional to the shortest relaxation time. The one-dimensional MSD thus can also be expressed as
\begin{equation}
    \text{MSD}_n(t) = \left < [X_n(t)-X_n(0)]^2 \right > = \frac{k_BT}{\sqrt{\pi}\kappa} \sqrt{\frac{t}{\tau_p}}.
    \label{MSD1000}
\end{equation}
The form of Eq.~\ref{MSD1} depends on the boundary conditions (see Eq.~\ref{MSD1periodic} and Eq.~\ref{MSD1fixed} for periodic and fixed boundary conditions).
By contrast, Eq.~\ref{MSD1000} turns out to be independent of boundary conditions.
\subsection{Rouse model with loops}
\label{rousewithloops}
\begin{figure}[htp]
    \includegraphics[scale=0.5]{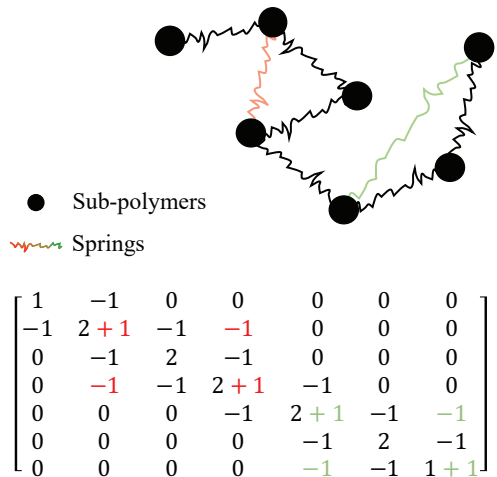}
    \caption{\label{loop_matrix_illus}
    Simplified illustration of a (free-boundary-condition) Rouse polymer with LEFs and its corresponding dynamical matrix $A$. The two LEFs are modeled as springs and color-coded as red and green, and their effects on the dynamical matrix are color-coded accordingly. The subpolymers are numbered from left to right.
    }
\end{figure}

To incorporate chromatin loops into our Rouse simulation, we model each loop base as an additional spring that connects the two loci to which it is bound. 
For simplicity, we choose the additional springs to have spring constants equal to those of the usual nearest-neighbor springs.
We then modify the dynamical matrix, \(K\), accordingly, as illustrated in Fig.~\ref{loop_matrix_illus}. 
Because $K$ remains symmetric upon adding loops, the theory in Sec.~\ref{nbead} remains applicable.
\subsection{Parameter scaling}
\label{scalinglaw}
There is considerable laxity in how to pick the subpolymers (and, therefore, $N$).
However, the Rouse model parameters are constrained by the following scaling: If the size of the subpolymer is chosen a factor $f$ larger, then the number of subpolymers becomes $\frac{N}{f}$, 
the spring constant of the new subpolymer becomes $\frac{\kappa}{f}$,
the friction coefficient of the new subpolymer becomes $f \zeta$,
and the characteristic polymer time of the Rouse polymer becomes $f^2 \tau_p$ \cite{doi1988}.
 Under this scaling $D$ is invariant (see Eq.~\ref{RouseDiffusionCoefficient}), as must be the case for a measurable quantity.

We chose to model  6~Mb regions of the mouse genome using $N=600$ beads, corresponding to subpolymers of size $N_1=10$~kb.
The spring constant between these beads may be expressed as
\begin{equation}
    \kappa = d\frac{k_BT}{(N_1/C) l_k},
    \label{10k_kappa}
\end{equation}
where $C$ is the number of base pairs per unit chromatin contour length (so that $\frac{N_1}{C} $ is the contour length of a subpolymer) and \(l_k\) is the chromatin Kuhn length.
The chromatin is considered as a freely-jointed chain of Kuhn segments of length, $l_k$.
Combining Eq.~\ref{RouseDiffusionCoefficient} and Eq.~\ref{10k_kappa} yields
\begin{equation}
    \zeta=\frac{\frac{N_1}{C} l_k k_BT}{d \pi D^2}.
    \label{10k_zeta}
\end{equation}
Ref.~\cite{arbona2017} estimates \(C=50\)~bp nm$^{-1}$ and \( l_k=138 \)~nm, while $D$ is measured to be roughly $D \simeq 4.1 \times 10^{-3}~\mu$m$^2$ s$^{-\frac{1}{2}}$ for $d=2$~\cite{bailey2023}.
These results lead to the following numerical estimates for a chromatin Rouse polymer built from subpolymers comprising 10~kb of DNA:
$\frac{\kappa}{k_BT} \simeq
7.25 \times10^{-5}$~nm$^{-2}$,
$\frac{\zeta}{k_B T} \simeq 2.61\times 10^{-4}$~nm$^{-2}$ s,
and $\tau_p=\zeta/(4\kappa) = 0.9$~s. 
We used these numerical values in the simulations shown below in Sec.~\ref{R&D}.

\section{Methods}
\label{methods}
\subsection{LEF model simulations}
\label{loopcombrouse}
To generate dynamic loop configurations, we carried out simulations of the loop extrusion factor (LEF) model, closely following Refs.~\cite{alipour2012,sanborn2015,fudenberg2016,goloborodko2016.1,goloborodko2016.2}, where the LEF model is introduced and described in detail.
In our implementation, the chromatin polymer is conceived as a linear array of LEF binding sites.
Each LEF possesses two legs, and binds at random, initially occupying two empty, neighboring binding sites.
Subsequently, the LEF starts to extrude a loop. Each loop extrusion event moves a site that was previously a backbone site into a loop. 
Only outward steps, that grow the loop, are permitted, with growth to the left or to the right occurring randomly and independently with equal probability. 
A LEF cannot occupy a site that is already occupied by another LEF. 
Therefore, LEF binding and loop extrusion are blocked by other LEFs.
A LEF can also unbind, dissipating its loop.
In our simulations, a LEF immediately rebinds after dissociation, so that the overall number of bound LEFs remains fixed throughout our simulations.

We implemented two versions of the LEF model. One is just as described in the previous paragraph, which we call the random loop model. 
The second, we call the CTCF model.
In the CTCF model,
loop extrusion is partially blocked at specific locations along the genome, where
CCCTC-binding factors (CTCFs) bind to specific DNA sequences.
As shown in the simulations of Ref.~\cite{fudenberg2016},
this model is able to reproduce several of the key features seen in chromosome conformation capture (Hi-C) experiments,
including, in particular, TAD boundaries, separating different genomic regions of high self contact probability.
By construction, the random loop model does not exhibit TAD boundaries.

To incorporate CTCF-related TAD boundaries in simulations, following
Ref.~\cite{fudenberg2016}, we binned the experimental mouse CTCF ChIP-Seq coverage data from Ref.~\cite{bonev2017}, in 10-kb bins, and summed over the coverage data in each bin.
In our simulations, each such bin is located between neighboring LEF binding sites, so that CTCF abundance scales down the loop extrusion rate from one LEF binding sites to the next by multiplying a factor
\begin{equation}
    p(x) = \frac{1}{1+e^{x/x_0-\mu}},
\end{equation}
where $x$ is the experimental CTCF ChIP-Seq coverage of the bin, $x_0=20$, and $\mu=3$, based on values given in Ref.~\cite{fudenberg2016}.
In this paper, we chose to focus on three
regions of the mouse genome, namely 32-38 Mb on Chromosome 12, 4.8-10.8 Mb on Chromosome 13, and 52-58 Mb on Chromosome 18. 
These three genomic regions have 58, 41, and 62 CTCF peaks with varying strengths, respectively, based on the peak-calling protocol described in Ref.~\cite{bonev2017}.

We simulated these LEF models via a Gillespie-type algorithm~\cite{gillespie1977},
implemented in MATLAB.
For each possible next event,
say event \(k\), a random variable is generated representing the time \(t_k\),
at which event $k$ would occur.
In each case,  $t_k$ is drawn from an exponential distribution
characterized by the appropriate rate for the event in question, given the current loop configuration.
Which of these possible next events is
actually realized is the one with the smallest value of $t_k$, say, $t_{\tilde{k}}$, advancing the simulation by a time $t_{\tilde{k}}$.
This process is repeated for the duration of the simulation.
The parameters of our simulations are presented in Table~\ref{LEFparam}.
We picked the loop extrusion velocity in either left or right direction to be 60 bp/s, giving a total loop extrusion velocity of 120 bp/s,
which is equal or comparable to the velocities of 100 bp/s estimated in Ref.~\cite{banigan2020} and 125 bp/s in Ref.~\cite{gabriele2022}, respectively.
We picked the number of LEFs and the loop dissociation rate to make corresponding processivity and loop density to lie near the center of the ranges that are given in Ref.~\cite{fudenberg2016}.

\begin{center}
\begin{table}
\begin{tabular}{wl{4.28cm}wr{4.1cm}}
    \hline
    \hline
    LEF simulation parameter & Typical value \\
    \hline
    number of LEFs &  24, 48, 72 \\
    simulated chromatin length & 6 Mb (600 sites) \\
    loop extrusion rate & 120 bps/s (0.012 sites/s) \\
    loop dissociation rate & 0.0005~s$^{-1}$ \\
    LEF processivity & 240 kb\\
    mean LEF separation & 250kb, 125kb, 83kb \\
    \hline
    \hline
\end{tabular}
\caption{\label{LEFparam}LEF simulation parameters. 
Each lattice site represents 10 kb. 
The loop extrusion rate given here is for both ends of a LEF combined; for each end of a LEF, the extrusion rate is 60 bps/s.
The LEF density values given here is defined by the ratio of total simulated genomic length over total number of LEFs.
}
\end{table}
\end{center}

\begin{figure}[t]
      \includegraphics[width=0.45\textwidth]{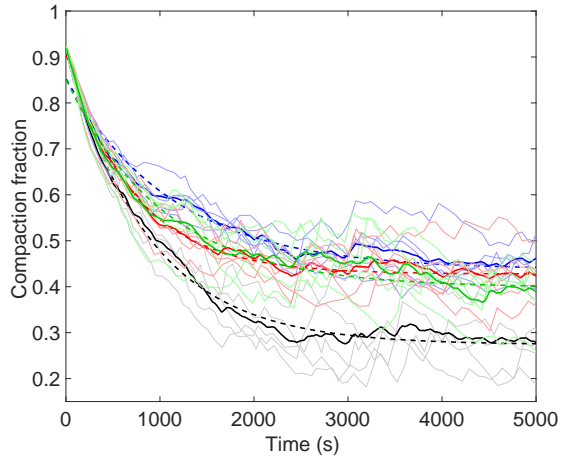}
\caption{
\label{four_RoG} 
Time evolution of polymer compaction by loops. 
Compaction is defined as the ratio of mean squared radius of gyration $\langle R_G^2\rangle$ (Eq.~\ref{rG}) of a looped polymer to that of a Gaussian polymer without loop.
The colored lines represent CTCF loop extrusion simulations run on three different genomic regions using: red, 52-58 Mb of Chr 18; green, 4.8-10.8 Mb of Chr 13; blue, 32-38 Mb of Chr 12. 
The parameters used in the CTCF LEF simulations are given in Table~\ref{LEFparam}, and each polymer has 48 LEFs bound to it.
The black line represents the loop extrusion simulation using the random loop model.
The light thin lines represent independent simulation runs.
The dark thick lines are the averages of 5 independent simulation runs.
The dashed lines are exponential fit to the mean curves, which give steady-state $\langle R_G^2\rangle$ values of 0.43, 0.40, 0.44, and 0.28 for red, green, blue and black, respectively.
The fitted decay times are: red, 718 sec; green, 1030 sec; blue, 1126 sec; black, 876 sec.
}
\end{figure}

A key question for any simulation is the time needed to reach a steady-state.
To assess how long it takes our LEF simulations to reach steady-state, we exploited the
observation that
treating chromatin as a Gaussian polymer permits us to determine
the polymer's mean squared radius of gyration, $\langle R_{G}^2\rangle$,
directly from the time-dependent loop configuration
(assuming that the polymer time is sufficiently small that each loop configuration is
sufficiently explored by the polymeric degrees of freedom).
Under these conditions, 
the mean squared radius of gyration between sites A and B along the genome is given by
\begin{equation}
    \langle R_{G}^2\rangle = \frac{l_k^2}{2N(N+1)} \sum_{i,j=A}^{B} (N_{eff})_{ij},
    \label{rG}
\end{equation}
where 
$(N_{eff})_{ij}$ is the effective genomic distance between location $i$ and $j$, which can be determined directly from LEF simulations. Appendix~\ref{RoGformula} presents a derivation of Eq.~\ref{rG} and a detailed explanation of $(N_{eff})_{ij}$.
\begin{figure*}[htp]
    \includegraphics[width=\textwidth]{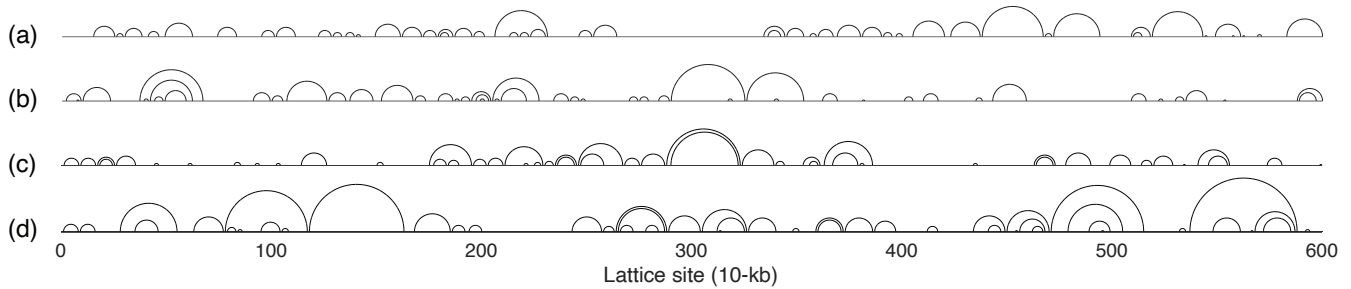}
    \caption{\label{loop_config} Abstract representations, following Ref.~\cite{bundschuh2002}, of three randomly-chosen, steady-state loop configurations (a, b, and c) taken from three independent CTCF LEF simulations of a 6~Mb region of mouse Chr 12 (32-38~Mb), Chr 13 (4.8-10.8~Mb) and Chr 18 (52-58~Mb), respectively, and (d) taken from the LEF simulation using random loop model.
    The loop extrusion simulations are done on 600 lattice sites, each representing 10-kb of genomic length.
    The black, horizontal line represents stretched-out chromatin; the semicircles represent loops extruded on the chromatin polymer. 
    There are 48 loops in each loop configuration.
    }
\end{figure*}

Fig.~\ref{four_RoG} shows the polymer compaction as a function of time as measured by the relaxation of $\langle R_G^2\rangle$, normalized to the $\langle R_G^2\rangle$ for a Gaussian polymer without loops, for four different LEF simulations.
Each simulation starts with 48 LEFs bound to neighboring lattice cites and ready to undergo loop extrusion.
The solid, black lines show the relaxation of the normalized $\langle R_G^2\rangle$ for a chromatin polymer with loops extruding according to the random loop model. 
The thin black lines show the relaxation of the normalized $\langle R_G^2\rangle$ for five independent LEF simulations, while the thicker black line shows their average.
The  red, green, and blue lines show the relaxation behaviors of the normalized $\langle R_G^2\rangle$, according to the CTCF model, for three different genomic regions, namely 52-58 Mb of Chr 18, 4.8-10.8 Mb of Chr 13, and 32-38 Mb of Chr 12, respectively.

By fitting each mean relaxation curve to a single exponential function, shown as the dashed lines in the figure, we find that the best-fit relaxation times for the three genomic regions, 52-58 Mb of Chr 18, 4.8-10.8 Mb of Chr 13, and 32-38 Mb of Chr 12, are 718, 1030, and 1126 seconds, respectively, while the relaxation time in the case of the random loop model is 876 seconds.
The corresponding steady-state compaction factors for the three different genomic regions are 0.43, 0.40, and 0.44, respectively, while the steady-state compaction factor in the case of the random loop model is 0.28.
Evidently, in every case, the simulations achieve a steady-state within a few thousand seconds.
Accordingly, because the loop extrusion simulations are computationally inexpensive, to ensure a loop extrusion steady state prior to starting data collection, we run all of our LEF simulations until the total elapsed time exceeds $2\cdot\num{e4}$ seconds,
prior to data collection.

Fig.~\ref{loop_config} shows abstract representations of three randomly-chosen, steady-state loop configurations from CTCF-model simulations of mouse chromosome 12, 13, and 18, and one from the LEF simulation using random loop model, all using parameters in Table~\ref{LEFparam}.
The loop configuration originated from the random loop model shows a higher level of compaction by possessing denser and bigger loops along the backbone.
We ascribe the reduced steady-state radius of gyration in the random loop model, in comparison to the CTCF model, to the absence of limitations on loop extrusion, provided by CTCF boundary elements.

\subsection{Rouse model simulations}
\label{loopcombrouse2}
For a random variable obeying Eq.~\ref{EQ5},
Gillespie \cite{gillespie1996} showed that
\begin{align}
    \label{updateFormula}
    \sX_m(t&+\Delta t) \nonumber \\
    =~& \text{\bf{N}} \left( \sX_m(t) e^{-\frac{\kappa}{\zeta} \Lambda_m\Delta t},\sigma^2_{ \sX_m}(1-e^{-2 \frac{\kappa}{\zeta} \Lambda_m \Delta t}) \right) \nonumber \\
    =~& \text{\bf{N}} \left( \sX_m(t) e^{-\frac{\kappa}{\zeta} \Lambda_m\Delta t},\frac{k_BT}{\kappa \lambda_m}(1-e^{-2 \frac{\kappa}{\zeta} \Lambda_m \Delta t}) \right) \nonumber \\
    =~& \sX_m(t)e^{-\frac{\kappa}{\zeta}\Lambda_m\Delta t} + \text{\bf{N}}(0,1)\sqrt{\frac{k_BT}{\kappa {\lambda}_m}(1-e^{-\frac{2\kappa}{\zeta}\Lambda_m\Delta t})}.
\end{align}
where $\text{\bf{N}}(\mu,\sigma^2)$ is the Gaussian distribution with mean $\mu$ and variance $\sigma^2$, and where we used Eq.~\ref{EQ14} after the second equality.
Eq.~\ref{updateFormula} provides a prescription for how to simulate a Rouse-model polymer, either with or without loops.
We chose to simulate Rouse-model polymers containing 600 beads, matching the number of beads to the number of LEF binding sites in the LEF simulations. 
The parameters used in our Rouse-model simulations are given in Table~\ref{Rouseparam}. 

To progress a simulation of a Rouse-model polymer from time $t_1$, to time $t_2$,
we first determine the transformation matrix from bead positions to normal coordinates, using the loop configuration at $t_1$.
Using the bead positions at $t_1$, $X(t_1)$,
we then use this $t_1$-transformation matrix to determine the normal coordinates at
$t_1$, $\sX(t_1)$.
As long as there is no change in the loop configuration between $t_1$ and $t_2$,
we evolve the normal coordinates according to Eq.~\ref{updateFormula} to get normal coordinates at $t_2$, $\sX(t_2)$.
To then determine the bead positions at $t_2$, $X(t_2$), we calculate the inverse of the $t_1$-transformation matrix, and apply it to  $\sX(t_2)$.
Alternatively, if there is a change in the loop configuration at time $t_3$, intermediate between $t_1$ and $t_2$, we progress the simulation as
just described from $t_1$ to $t_3$. We then recalculate the $t_3$-transformation matrix, based on the new loop configuration. Using this new $t_3$-transformation matrix, we then progress the simulation from $t_3$ to $t_2$.

During each simulation, the dynamical matrix is repeatedly updated to reflect the current loop configuration,
and a new corresponding set of eigenvalues and normal coordinates are repeatedly calculated. 
Thus, the entire Rouse simulation is separated by loop extrusion events, into sub-simulations, each with its own stochastic update formula specified by the eigensystem of the current loop configuration. 
For any two consecutive sub-simulations, the former's final conditions serve as the latter's initial conditions, ensuring continuity of the entire Rouse simulation. 
As noted in Ref.~\cite{gillespie1996}, this procedure represents an ``exact'' simulation, with no small-time approximation in the sense that there is no constraint on the size of time steps used ($t_2-t_1$ or $t_3-t_1$ or $t_2-t_3$), provided each different loop configuration is properly included and accounted for.

The Rouse simulation is run for loop configurations that have already relaxed to steady state (Fig.~\ref{four_RoG}).
Specifically, as discussed above, we used loop configurations from beyond $2\cdot\num{e4}$ seconds into each loop extrusion simulation.
The starting bead positions in the Rouse simulation are initialized by transforming the initial normal coordinates, which are randomly drawn from the normal distribution given in Eq.~\ref{updateFormula}, with $\sX_m(t=0) = 0$ and $\Delta t = \infty$.
The initial bead positions calculated in this way are guaranteed to follow the statistics at the equilibrium, subject to a mean position of zero.
We chose the update time step in our Rouse simulations to be 1~s, comparable to the Rouse polymer time, and we ran each Rouse-model simulation for $\num{e4}$~s.

\begin{center}
    \begin{table}
    \begin{tabular}{wl{5.0cm}wr{3.38cm}}
        \hline
        \hline
        Rouse simulation parameter & Typical value \\
        \hline
        friction coefficient, $\zeta$ & $1.08\times 10^{-6}$ Ns/m \\
        spring constant, $\kappa$ & $3\times 10^{-7}$ N/m \\
        temperature, T & 300 K\\
        polymer time, $\tau_p$ & 0.9 sec\\
        update time step &  1 sec \\
        simulation length & $\sim \num{e4}$ sec \\
        \hline
        \hline
    \end{tabular}
    \caption{\label{Rouseparam}Rouse simulation parameters. }
    \end{table}
\end{center}

\begin{figure}[htp]
    \includegraphics[width=0.45\textwidth]{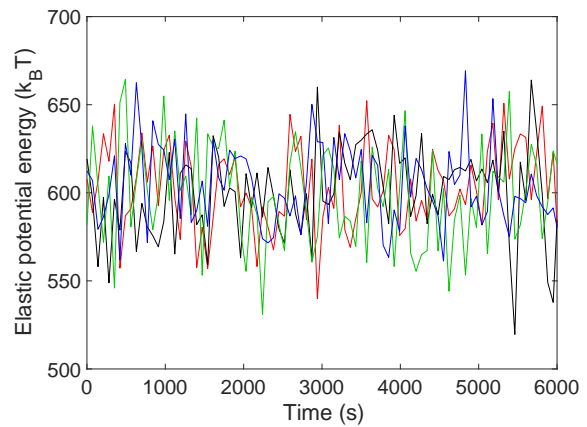}
    \caption{\label{energy_series} 
        Time series of Rouse polymers' potential energy for different loop extrusion rates.
        The red line shows the simulated spring potential energy versus time for a free Rouse polymer without loops, using the parameters given in Table ~\ref{Rouseparam}.
        The black line shows the simulated spring potential energy versus time for a random-loop-model Rouse polymer, for 48 LEFs and the parameters given in Table~\ref{LEFparam} and~\ref{Rouseparam}.
        The blue and green lines show the potential energy versus time for random-loop-model polymers, populated by 48 LEFs, when all of the loop extrusion rates given in Table~\ref{LEFparam} are increased by a factor of 10 and 100, respectively, while the polymer parameters give in
        Table~\ref{Rouseparam} are held fixed.
        The resolution of the line points in the plot is set to 70 seconds to ensure a clear view.
    }
\end{figure}

One possible caveat to our approach is that
it is possible to envisage the development of out-of-equilibrium values
of the Rouse model spring potential energy, if the increase in potential energy
as a result of loop extrusion outpaces Rouse-model energy relaxation and dissipation.
To investigate this possibility, in Fig.~\ref{energy_series}, we plot examples of the potential energy versus time for several different situations.
The red line represents the potential energy from simulations of a classical, two-dimensional Rouse-model polymer without loops,
where we find that
the potential energy fluctuates about a mean value of 603 $k_BT$.
In comparison, the theoretically-expected value, from the equipartition theorem, is (600 beads)$\times$(2 dimensions)$\times (\frac{1}{2}k_BT)=600~k_BT$.
The black line is the potential energy time-series for the Rouse polymer with loops for 32-38~Mb of mouse chromosome 12,
corresponding to the simulation parameters given in Tables 1 and 2.
In this case, we see that the potential energy fluctuates about a mean value of 598 $k_BT$, also near the theoretical expectation.
This observation indicates that for the loop extrusion and Rouse parameters used in our simulations, Rouse-model polymers with loops remain in thermal equilibrium at temperature, $T$.
We also carried out simulations for which all of the rates in the loop extrusion simulations were increased by a factor of  either ten or one hundred.
The resulting potential energy versus time traces are shown in blue and green, respectively.
The corresponding mean potential energies are 600 and 599 $k_BT$, respectively.
Thus, we see that even for loop extrusion that is one hundred times faster than estimated, the potential energy does not noticeably exceed the value expected from equipartition.

\begin{figure}[htp]
      \includegraphics[width=0.45\textwidth]{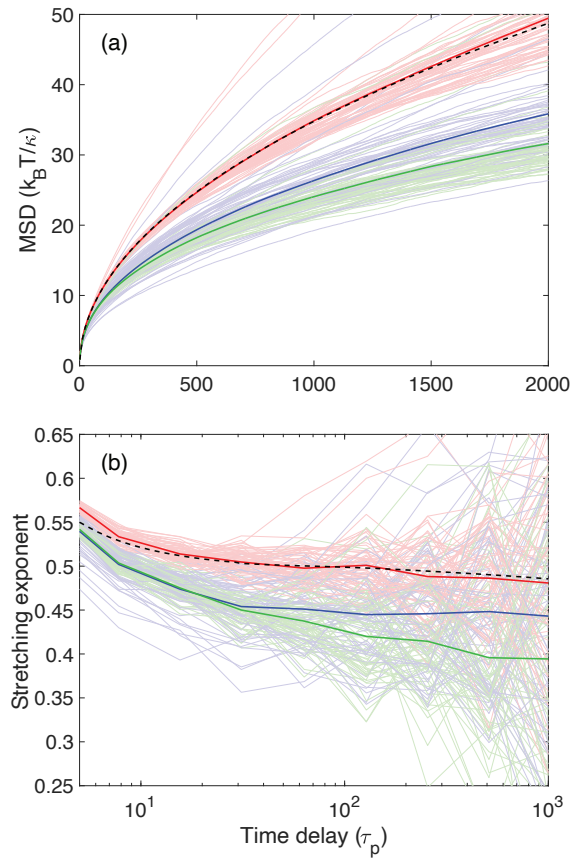}
    \caption{\label{fig1.1} 
    (a) Simulated MSDs of Rouse-model polymers with free boundary conditions for the classical Rouse model with no loops, shown in red, for the CTCF model with loops, shown in blue, and for the random loop model, shown in green.
    (b) The corresponding simulated MSD stretching exponent, \(\alpha(t)\), calculated according to Eq.~\ref{alphacalc} for no loops (red), for the CTCF model (blue), and for the random loop model (green).
    The dashed black line in (a) shows the theoretical mean MSD given by Eq.~\ref{MSD1}, and then the theoretical mean stretching exponent is calculated using Eq.~\ref{alphacalc} and shown in (b).
    In both (a) and (b), the thin lines correspond to the results for 60 individual beads,  uniformly distributed along the chromatin.
    The thick lines correspond to the average across all 60 beads.
    Each thin line is an average over 30 independent simulations with identical simulation parameters.
    The LEF and Rouse simulation parameters used are given in Table~\ref{LEFparam} and \ref{Rouseparam}.
    }
\end{figure}

\section{results and discussions}
\label{R&D}
\subsection{Chromatin mobility in the presence of loops}
\label{loopvsnoloop}
Fig.~\ref{fig1.1}(a) compares the MSDs versus time of a Rouse-model polymer without
loops to the MSDs of Rouse-model polymers with loops, generated by loop extrusion, all under free boundary condition.
Shown in red are simulated MSDs for the classical Rouse model without loops.
The thin red lines show the
MSDs, averaged over thirty independent simulations, for sixty individual beads, equally spaced along the polymer, while the thicker red line shows these MSDs averaged over all
sixty of these beads.
In comparison, the dashed black line in Fig.~\ref{fig1.1}(a) shows the theoretical mean MSD (Eq.~\ref{MSD1}). 
Evidently, the simulation closely matches the analytic theory.
Shown in blue and green are simulated MSDs for the CTCF model and the random loop model, respectively,
both with 48 LEFs, with free boundary conditions, and averaged over thirty independent simulations.
In these cases too,
the thin lines show the
MSDs for 60  individual beads, while the thicker lines show the MSDs averaged over these beads.
Importantly, in both cases, the MSD in the presence of loops is significantly smaller than the MSD for the classical
Rouse model without loops, indicating that the presence of loops significantly constrains chromatin mobility.
The difference between the MSDs of the two different loop extrusion models is small, although clearly the MSDs of the random loop model are more reduced than the
MSDs of the CTCF model.

MSDs versus time are often described using a ``stretching exponent''.
Fig.~\ref{fig1.1}(b) shows the time-dependent stretching exponent, $\alpha(t)$, corresponding to the MSDs shown
in  Fig.~\ref{fig1.1}(a),
defined as
\begin{equation}
    \alpha(t) = \frac{\log\left[\text{MSD}(t+\Delta t)\right] - \log\left[\text{MSD}(t)\right]}
    {\log(t+\Delta t) - \log(t)},
    \label{alphacalc}
\end{equation}
where $\Delta t$ is the time step of the Rouse simulations.
For ranges of time over which $\alpha (t)$ is constant, the MSD approximates power law behavior versus time.
For the classical Rouse model without loops (red curves), with increasing time, $\alpha(t)$ decreases from a value of 1 at the earliest times (data not shown) to achieve a plateau
value close to 0.5 for times greater than a few tens of $\tau_p$.
This is consistent with the theoretically predicted $\alpha(t)$, shown as the black, dashed line in Fig.~\ref{fig1.1}(b), which was calculated from Eq.~\ref{MSD1}.
For Rouse polymers with loops (blue and green curves), interestingly the stretching exponent falls significantly
below the value for the Rouse polymer without loops.
In the case of the CTCF model simulations, for times beyond a few tens of $\tau_p$, $\alpha$ achieves a plateau value close to
$\alpha=0.45$. For the random loop model simulations, $\alpha$
appears to decrease continuously with time from a value near 0.45 at 30~$\tau_p$ to a value near 0.4 for $10^3~\tau_p$.
\begin{figure}[t]
    \includegraphics[width=0.45\textwidth]{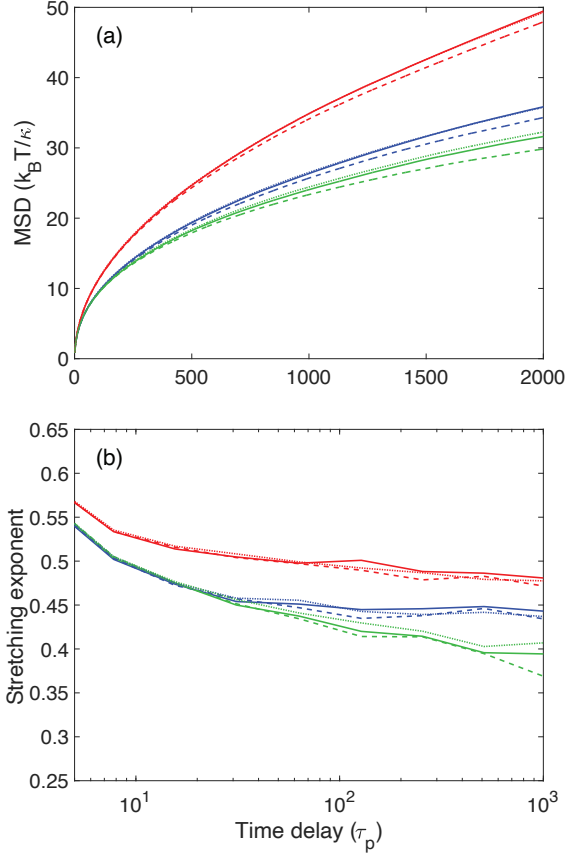}
    \caption{\label{fig1.4} 
    (a) Comparison between the simulated MSDs of Rouse-model polymers with free, periodic and fixed boundary conditions,
    shown as the solid, dashed, and dotted lines, respectively,
    for  the classical Rouse polymer with no loops, shown in red, for the CTCF model with 48 LEFs, shown in blue, and for the random loop model with 48 LEFs, shown in green.
    (b) Comparison between the corresponding simulated MSD stretching exponent, $\alpha(t)$, calculated according to Eq.~\ref{alphacalc} for no loops (red), for the CTCF model with
    48 LEFs (blue), and for the random loop model with 48 LEFs (green), each considered separately with the three different boundary conditions.
    In both (a) and (b), each line is averaged over 30 independent simulations and averaged over 60 individual beads, which are uniformly distributed along the chromatin.
    }
\end{figure}

The simulations of Fig.~\ref{fig1.1} correspond to polymers with free boundary conditions.
We also carried out comparable simulations for both closed polymer rings (periodic boundary conditions) and polymers with fixed ends (fixed boundary conditions).
A comparison of the mean MSDs for each possible boundary condition is shown in Fig.~\ref{fig1.4}(a) for different loop models.
It is clear from this figure that the MSDs for different boundary conditions are very similar to each other over the range of times studied.
Fig.~\ref{fig1.4}(b) shows the corresponding mean stretching exponents, which are also very similar to each other over the range of times studied.
Solutions to the classical Rouse model, subject to periodic and fixed boundary conditions, are given in Appendix~\ref{periodicbc} and Appendix~\ref{fixedbc}, respectively.

\begin{figure}[t]
      \includegraphics[width=0.45\textwidth]{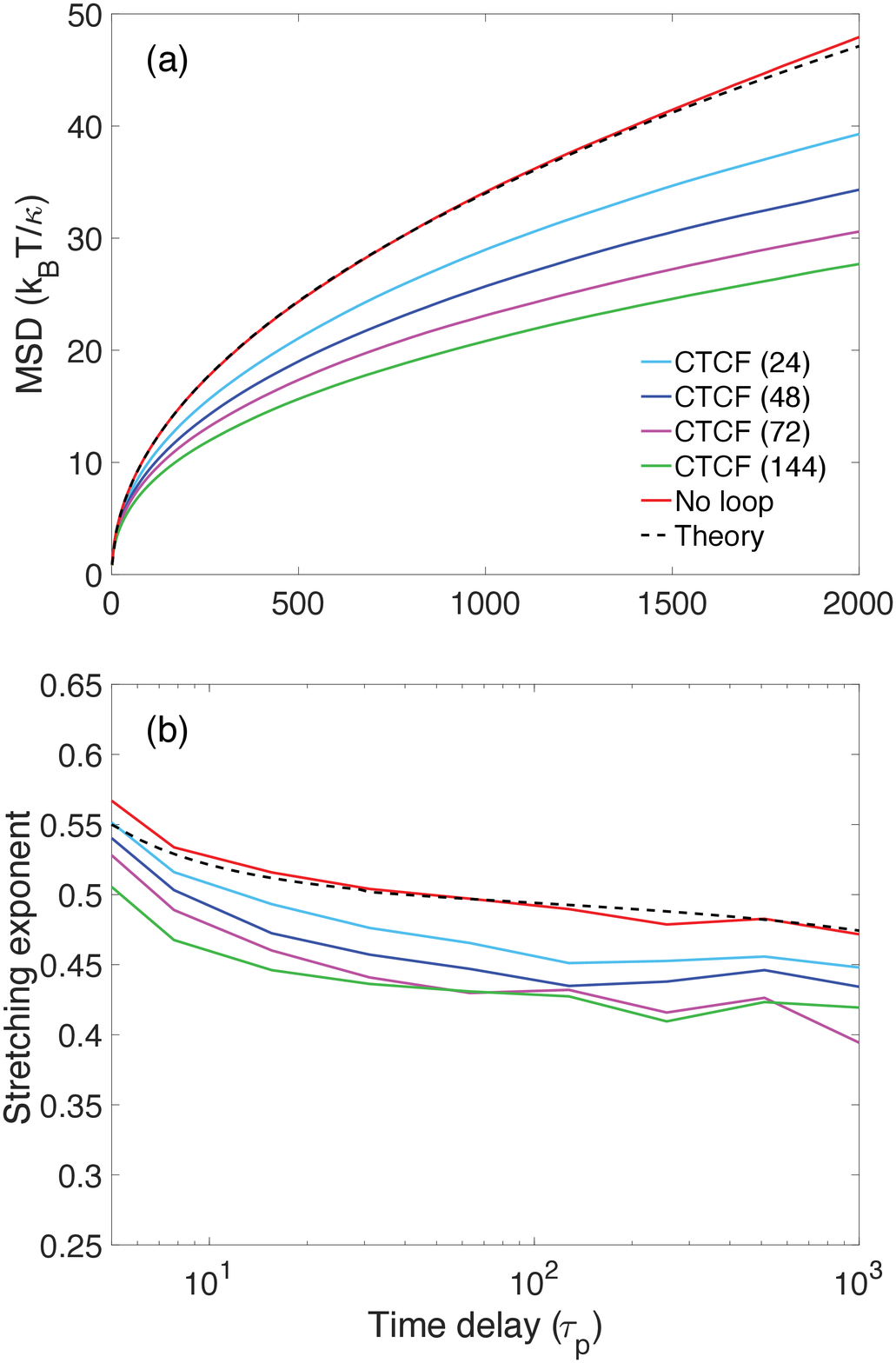}
    \caption{\label{fig1.2} 
    (a) Simulated MSDs of Rouse-model polymers with periodic boundary conditions for the classical Rouse model with no loops, shown in red,
    and for the CTCF model with different number of LEFs, shown in  cyan, blue, magenta, and green, corresponding to  24, 48, 72 and 144 LEFs, respectively.
    (b) The corresponding simulated MSD stretching exponent, \(\alpha(t)\), for no loops (red), and for the CTCF models with 24 (cyan), 48 (blue) 72 (magenta), and 144 (green) LEFs.
    The dashed black line in (a) shows the theoretical mean MSD given by Eq.~\ref{MSD1periodic}.
    The dashed line in (b) is the corresponding theoretical mean stretching exponent, calculated using Eq.~\ref{alphacalc}, applied to Eq.~\ref{MSD1periodic}.
    In both (a) and (b), each line is averaged over 30 independent simulations and averaged over 60 individual beads, which are uniformly distributed along the chromatin.
    }
\end{figure}

How MSD depends on the number of LEFs is illustrated in Fig.~\ref{fig1.2},
which depicts MSDs (Fig.~\ref{fig1.2}(a)) and stretching exponents (Fig.~\ref{fig1.2}(b))
for CTCF-model polymers with 0 LEFs, shown as the red line, 24 LEFs, shown as the cyan line, 48 LEFs, shown as the blue line, 72 LEFs, shown as the magenta line, and 144 LEFs, shown as the green line.
It is clear from this figure that the MSDs are progressively repressed as the number of LEFs increases.
Fig.~\ref{fig1.2}(b) displays the corresponding stretching exponents for different number of LEFs, revealing that the stretching exponent also initially decreases progressively as the number of LEFs increases as the number of LEFs increases.
Interestingly, the stretching exponent at intermediate times appears to show a limiting value of 0.43 for large number of LEFs.

\begin{figure}[htp]
      \includegraphics[width=0.45\textwidth]{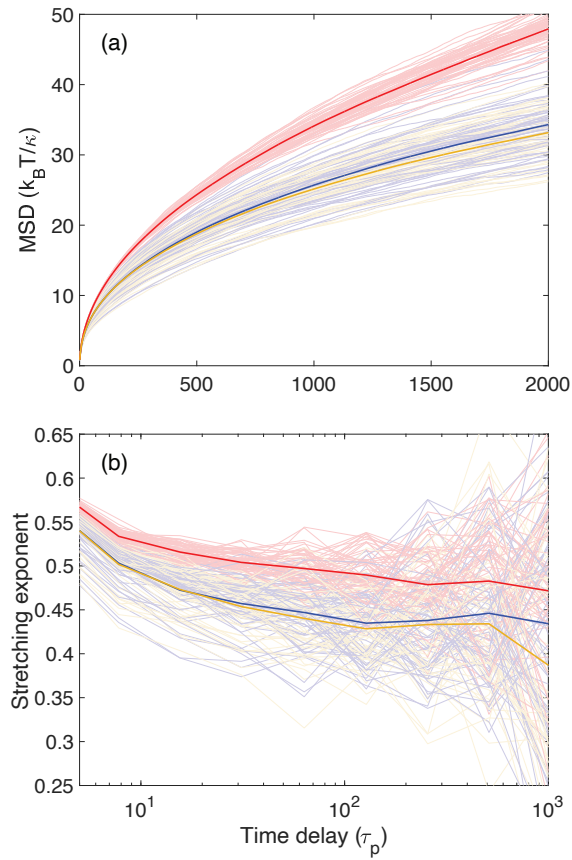}
    \caption{\label{fig:staticloop} 
    (a) Comparison between the MSDs versus time of polymers with 48 randomly-located static loops (orange) and 
    polymers with dynamic loops, evolving according to the CTCF model
    with 48 LEFs (blue). 
    (b) The corresponding time-dependent stretching exponents. 
    The LEF and Rouse simulation parameters used are given in Table~\ref{LEFparam} and \ref{Rouseparam}.
    In both (a) and (b), the thin lines correspond to results of 60 individual beads, uniformly distributed along the polymer; the thicker lines are averages over the 60 beads.
    The result for each individual bead is averaged over 30 independent simulations with identical simulation parameters.
    }
\end{figure}

LEFs actively extrude loops along the chromatin. 
Therefore, although chromatin mobility is reduced overall by the introduction of loops, it is interesting to investigate the extent to which the dynamics of loop extrusion might make a positive contribution to chromatin mobility.
To identify any possible contribution to chromatin dynamics from LEF dynamics, we investigated
Rouse polymers with randomly-located static loops, for which there are no LEF dynamics.
Each static loop configuration studied corresponds to one randomly-picked time point from the steady state of the CTCF model
with 48 LEFs on 32-38 Mb of mouse Chr 12.
For each such static loop configuration, a Rouse simulation was then carried out to determine its MSDs and stretching exponents.
Finally, we averaged the MSDs and stretching exponents from 30 different static loop configurations.
The resultant mean MSDs and stretching exponents for looped polymers with random, static loops (orange)
are compared in
Fig.~\ref{fig:staticloop} to the 
simulated MSDs and stretching exponents of looped polymers with dynamic loops (blue), that evolve in time according to the CTCF model with 48 LEFs in the same genomic region.
The MSDs of individual beads, equally separated along the polymer, shown as thin lines in Fig.~\ref{fig:staticloop}(a), show similar distributions and trends for both the static and dynamic loops.
The stretching exponents in the case of static and dynamic loops, shown in Fig.~\ref{fig:staticloop}(b), also behave similarly to each other.
It is apparent from this comparison that the MSDs and stretching exponents of polymers with dynamic, on one hand, and static loops,
on the other, show insignificant differences  over the range of times studied,
suggesting in turn that loop extrusion dynamics do not contribute significantly to chromatin dynamics,
at least for the chosen parameter values.

Our simulational results clearly demonstrate that the presence of loops significantly reduces chromatin mobility.
We believe this behavior
follows from the additional, positional constraints that loops impose, specifically that the two beads at the base of each loop must
necessarily
lie close to each other.
Notably, the simulational results, presented in this paper, are consistent with our recent  experimental measurements of the MSDs of gene loci in fission yeast,
described in Ref.~\cite{bailey2023}, where we found that fluorescently-labelled gene loci in yeast strains with absent or functionally-compromised versions of
the putative LEFs,
cohesin and condensin, exhibited larger MSDs than the same gene loci in wild-type strains with properly functioning cohesin and
condensin. Thus, Ref.~\cite{bailey2023}, in conjunction with the results of this paper, provides clear experimental support for the idea that loops reduce chromatin mobility, as characterized by the MSD of a gene locus.

A less expected result to emerge from our simulations is the observation that the stretching exponent in polymers with loops is noticeably smaller than that for the classical Rouse model without loops.
Strikingly, however, in our own measurements of the MSDs of fluorescently-labelled gene loci in fission yeast, we also observed stretching exponents with values  smaller than 0.5 (about 0.45), further connecting our simulations and experiments \cite{bailey2023}.
Our simulational results also recall the experimental observation in Ref.~\cite{weber2010a} that the MSD of fluorescently-labelled gene loci in
{\em Escherichia coli}
shows a stretching exponent of about $\alpha=0.4$.
In Ref.~\cite{weber2010a}, this value was explained  by envisioning the
bacterial genome to be an unlooped Rouse-model polymer within a viscoelastic medium. 
The calculations of Ref.~\cite{weber2010b} showed that
such a viscoelastic background could reduce the classical Rouse model
stretching exponent from 0.5 to the observed value.
However, a population of loops across the {\em E. coli} genome would also contribute to a smaller stretching exponent than 0.5 in this case too.
Polymer simulations of chromosome regions with high compaction show stretching exponents of around 0.3~\cite{dipierro2018,salari2022}, which may also partly originate from a population of loops.

\begin{figure}[htp]
    \includegraphics[width=0.45\textwidth]{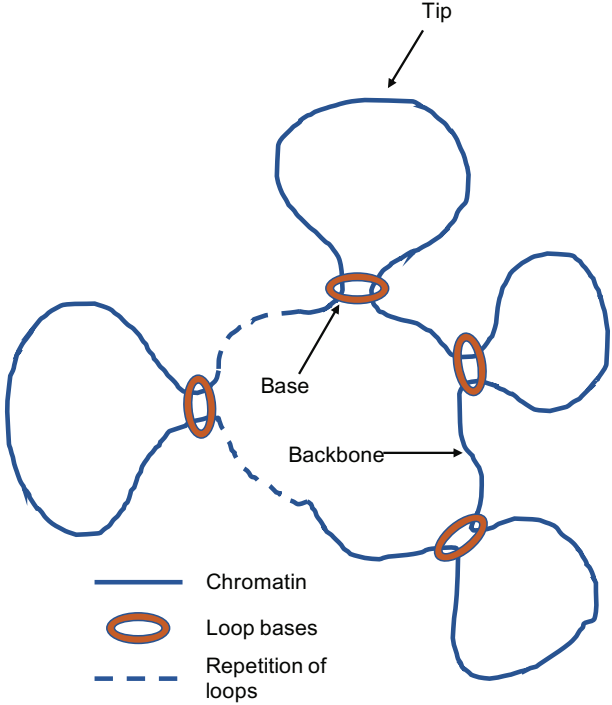}
    \caption{\label{fig:3.0} The rosette configuration investigated in Sec.~\ref{rosette}. 
    Shown in blue is the chromatin polymer itself; loop bases are highlighted as the brown circles. 
    Each loop is 17 lattice sites long, and each backbone segment between neighboring loops is 8 lattice sites long.
    Thus, the fraction of the chromatin polymer inside loops is 0.68 in this configuration.
    There are 24 repetitions of the loop-backbone structure within the periodic chromatin polymer, making a total of 600 lattice sites,  representing a genome of 6 Mb.
    The dashed blue line represents repetitions of the identical loop-backbone structures.
    Examples of each family of locus locations are labelled tip, base and backbone.
    }
\end{figure}
\begin{figure}[htp]
    \includegraphics[width=0.45\textwidth]{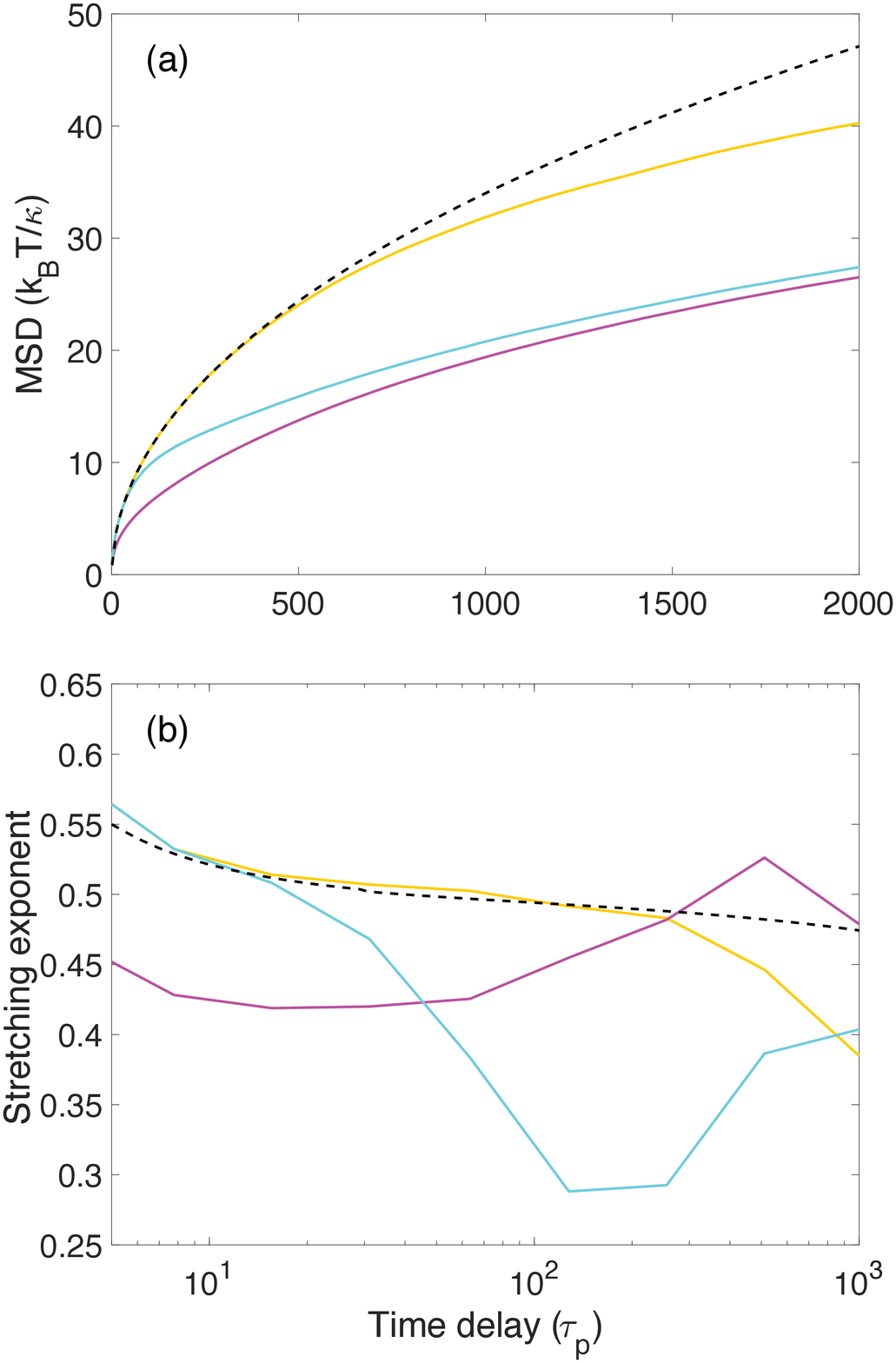}
    \caption{\label{fig:3.1} 
        (a) Simulated MSDs of the Rouse-model polymer with the rosette structure illustrated in Fig.~\ref{fig:3.0} for three different families of locations: tip (yellow), base (magenta), and backbone (cyan).
        (b) The corresponding simulated MSD stretching exponent for different families of locations on the rosette-structured polymer: tip (yellow), base (magenta), and backbone (cyan).
        In both (a) and (b), each line shows the averaged result of 720 independent simulations for beads in each family. 
        The black, dashed lines in (a) and (b) show the theoretical MSD and stretching exponent, respectively, of a Rouse-model polymer without loops, with periodic boundary condition, as given in Eq.~\ref{MSD1periodic} and \ref{alphacalc}.
    }
\end{figure}
\subsection{Locus dynamics around a chromatin rosette}
\label{rosette}
Our Rouse simulation method is applicable to any configuration of loops.
As noted in the introduction, genomic rosettes have been suggested to be an important motif for \textit{E. coli} nucleoids~\cite{hinnebusch1997,macvanin2012} and replication factories~\cite{newport1996,ma1998,jackson1998,frouin2003,saner2013,mangiameli2018,guillou2010}.
Accordingly, we have investigated the MSDs and stretching exponents for the static, rosette configuration,
illustrated in Fig.~\ref{fig:3.0}, in which
24 loops are periodically and equally spaced about a closed chromatin polymer.
For our simulations, each loop contains 17 beads,
and is separated from its neighboring loops by a segment of the backbone containing 8 beads.
By symmetry, we can expect the mean dynamics within different repeating units of the rosette to be identical.
Thus, because we can average the behavior of corresponding loci in different repeating units, the rosette facilitates 
 investigating the dynamics
at the midpoint of a loop, at the base of a loop, and at the midpoint of a backbone region, for example.
Each family of locations is illustrated and labelled as examples in Fig.~\ref{fig:3.0}.

In Fig.~\ref{fig:3.1}, shown in yellow, magenta, and cyan, respectively, are the MSDs and stretching exponents for beads at the tip of the loops (yellow), base of the loops (magenta), and midpoint of the backbone (cyan).
By symmetry of the rosette structure, beads belonging to the same family are statistically equivalent. 
Therefore, the yellow, magenta, and cyan lines shown in Fig.~\ref{fig:3.1} are averaged results of 720 independent simulations for the tip, base, and backbone families, respectively, indifferent about the loop index.
The theoretical MSD and stretching exponent, calculated by Eq.~\ref{MSD1periodic} and \ref{alphacalc}, are shown as the black, dashed lines in Fig.~\ref{fig:3.1}(a) and (b), respectively.

It is clear from Fig.~\ref{fig:3.1}(a) that loop bases (magenta) exhibit the smallest mean MSD at all times studied, taking on a value that is approximately one-half the theoretical MSD for periodic Rouse-polymer without loops, while loop tips show the largest MSDs.
At early times, the mean MSD for backbone beads show similar behavior to the mean MSD for loop tip beads.
However, with increasing time, the mean MSD for backbone beads deviate (at about 50~$\tau_p$) to smaller values to
eventually achieve (by about 1000~$\tau_P$) similar behavior to the mean MSD for the bases beads.
The mean MSD for loop tips closely follow the theoretical MSD for a polymer without loops until about 500~$\tau_p$, when it deviates below the theoretical MSD. 

The corresponding stretching exponents, shown in Fig.~\ref{fig:3.1}(b), also reveal elaborate behavior.
The loop tips in the rosette structure exhibit the same mean stretching exponent as the theoretical stretching exponent for polymers without loops at early times, until about 200~$\tau_p$. During this early time period, the mean stretching exponent for the loop tips decrease from value of 1 at the earliest time (not shown), to a level of roughly 0.5.
At times larger than 200~$\tau_p$, the mean stretching exponent for the loop tips starts to deviate to a lower level compared to the theory. 
The mean stretching exponent for the backbone beads has similar values to that for the loop tips and the theoretical value for polymers without loops;
however, it starts to deviate to a lower level much earlier, at about 10~$\tau_p$, and decreases to a much reduced value of about 0.3 at a few hundred $\tau_p$, at which it starts to recover and increase back to 0.4 at very late times.
The mean stretching exponent for the loop bases reside between 0.4 and 0.45 at earlier times, before $t = 100~\tau_p$, then increase beyond the 0.5 level at around 500~$\tau_p$, and finally drop back to the level around 0.5, similar to the theoretical stretching exponent for polymers without loops at the very later times.
It is clear from the results shown in this figure, that the dynamics of a locus are dictated
by the position of the locus in question relative to the positions of loops.
This behavior does not emerge in the case of dynamic loops, because commonly in a loop extrusion steady state
any particular locus randomly alternates between being in the backbone, being at a loop base, 
being at a loop tip, {\em etc.}

\begin{figure}
    \includegraphics[width=0.45\textwidth]{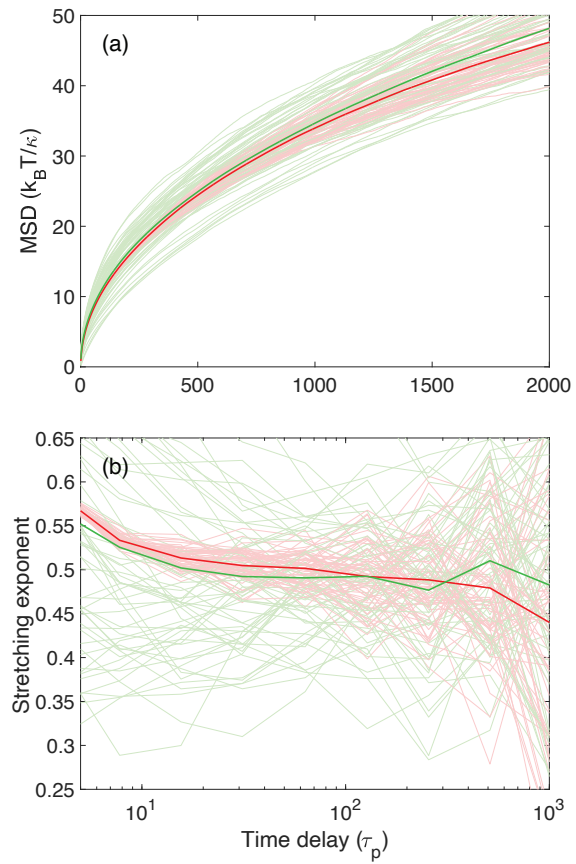}
    \caption{\label{fig:5.1}
    (a) Comparison between the MSDs of periodic polymers without loops, subject either to uniform friction, shown in red, or to log-normal-distributed (non-uniform) friction, shown in green. 
    For the polymer subject to log-normal-distributed non-uniform friction, each bead has a fixed friction coefficient, drawn from the scaled Lognormal(0, log(2)) with mean of $\zeta$, given in Table~\ref{Rouseparam}.
    (b) The corresponding MSD stretching exponents, calculated according to Eq.~\ref{alphacalc}, for uniform friction case (red) and log-normal-distributed (non-uniform) case (green).
    In both (a) and (b), thin lines represent 60 individual beads uniformly distributed along the polymer, and each thin line is an average of 30 independent simulations. 
    Each thicker line is an average of all corresponding thin lines.
    }
\end{figure}
\begin{figure}
    \includegraphics[width=0.4\textwidth]{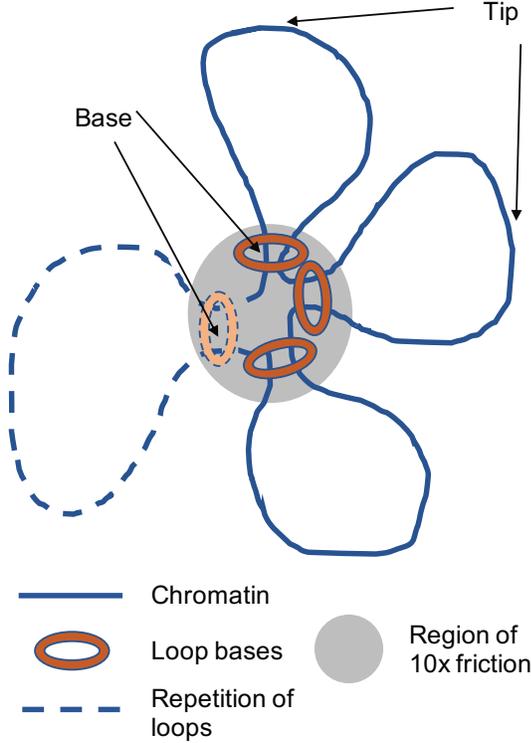}
    \caption{\label{fig:5.0} The  hypothetical replication factory  configuration discussed in Sec. \ref{NUFEnv}. 
    The shaded region is where the friction coefficient is 10 times higher than in the unshaded region.
    Loci experiencing higher friction coefficients are all located at the bases of the loops.
    The midpoints of each loop are labelled as ``tip'', and the bases of each loop are labeled as ``base''.
    Total number of LEFs is 24. Each loop has size of 24. The backbone length between neighboring loops is 1, which is the minimal length a backbone can form given that the LEFs will block each other at distance 1.
    The dashed line represents repetitions of the same loop structure.}
\end{figure}
\begin{figure}
    \includegraphics[width=0.45\textwidth]{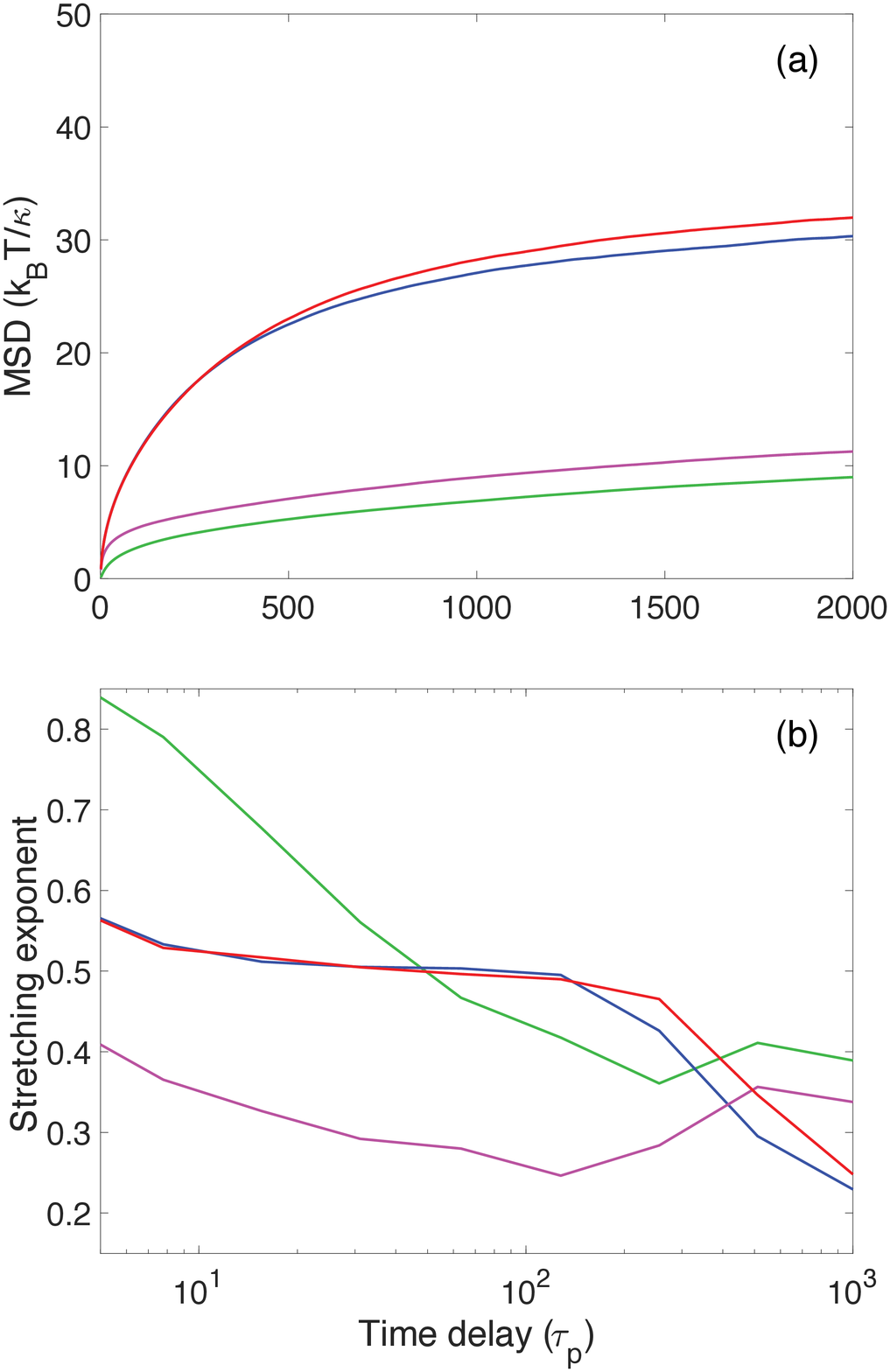}
    \caption{\label{fig:5.2} (a) Shown in red and magenta lines are the mean MSDs for the loop tips and bases, respectively, of a rosette polymer in the non-uniform-friction envionrment, as shown in Fig.~\ref{fig:5.0}. 
    Shown in blue and green lines are the mean MSDs for the loop tips and bases, respectively, of the same rosette polymer, but in a uniform-friction envioronment, i.e. without the region of ten-fold friction shown in Fig.~\ref{fig:5.0}.
    Each line is an average of 720 independent simulations for the tip beads or bases beads, indifferent about the loop repetitions to which each bead belongs, as a result of symmetry.
    (b) The corresponding averaged stretching exponents. 
    }
\end{figure}

\subsection{Rouse polymer with non-uniform friction}
\label{LogNormal}
The theory described in Sec.~\ref{nbead} also allows us to study polymers with friction coefficients that vary from bead to bead.
Accordingly, we first examined a linear Rouse polymer with each bead's friction coefficients drawn from a log-normal distribution. 
We use a log-normal distribution proportional to Lognormal(0,log(2)), so that the ratio of the standard deviation and the mean is 1,
and scale it to have a mean of $\zeta$, which is the friction coefficient in Table~\ref{Rouseparam}.
Each bead is assigned a fixed friction coefficient drawn from such a scaled log-normal distribution.
Fig.~\ref{fig:5.1}(a) compares the MSDs of a periodic Rouse polymer with friction coefficients distributed according to a log-normal
distribution to the MSDs of a periodic Rouse polymer with uniform friction coefficient. 
There are no loops in both cases.
The thin lines show the MSDs for 60 individual beads, equally spaced along the polymer, while the thicker lines show the averaged MSDs over all sixty of these beads.
Each thin line is also an average of 30 independent simulations.
The mean MSDs for the uniform friction coefficient and log-normal-distributed friction coefficient are very similar across all time-scales considered here, with a small discrepancy that the mean MSD for the non-uniform-friction case is slightly higher than that for the uniform-friction case at all time scales considered.
Fig.~\ref{fig:5.1}(b) shows the corresponding stretching exponents, and the mean stretching exponents for the two cases are, again, almost identical.
Both mean stretching exponents decreases from 1, at the earliest time (not shown here), to around 0.5, at 100~$\tau_p$, with the small discrepancy between the two cases such that the stretching exponent for the polymer with non-uniform friction is slightly lower than the stretching exponent for the uniform-friction polymer. 
After 100~$\tau_p$, the stretching exponent for the non-uniform-friction polymer fluctuates around the 0.5 level while the stretching exponent for the uniform-friction polymer continues to decrease to a level below 0.45 at 100~$\tau_p$. 
Even though the mean MSDs and stretching exponents are similar for polymers with uniform and non-uniform friction,
the MSDs and stretching exponents of individual beads in the non-uniform-friction polymer show a greater spread than those of uniform friction case. 
This is consistent with the simulation setup in which each bead experiences a different but fixed friction coefficient drawn from the log-normal distribution.
The above results indicate that fluctuations in the friction coefficient does not strongly affect the mean MSD and mean stretching exponent, at least up to the
spread of the beads' friction coefficients considered here.

\subsection{Dynamics of a rosette that models a replication factory}
\label{NUFEnv}
In a replication factory, chromatin is hypothesized to be folded into a rosette-like structure, at whose center  transcription factors and SMCs may form a phase-separated droplet and assist the simultaneous replication of the DNA within the loops ~\cite{newport1996,ma1998,jackson1998,frouin2003,saner2013,mangiameli2018,guillou2010}.
A simple version of such an organization is illustrated in Fig.~\ref{fig:5.0}.
The blue lines shown represent the chromatin polymer segment, and the brown circles represent the putative SMCs, cohesin or condensin molecules, that establish the replication-factory configuration.
In our simulations, we model a replication factory as consisting of 24 loops, each of which contains 24 beads.
Plausibly, the phase-separated droplet, shown as the grey region in this figure, which covers the loop bases,
may give rise to a high local viscosity.
In this case, we model the high viscosity region by using a ten-fold higher friction coefficient for the loop bases than for the beads in the loops.

Fig.~\ref{fig:5.2} compares the averaged (a) MSDs and (b) stretching exponents for the loop tips and bases in the replication-factory configuration (Fig.~\ref{fig:5.0}) with a uniform-friction or with a non-uniform-friction environment.
Because each loop repetition is statistically equivalent, the simulation is indifferent to the loop repetition each loop tip or base belongs to.
Therefore, for each line in Fig.~\ref{fig:5.2}, we take an average of 720 independent simulations, regardless of the loop index.
The time axis in the plot is measured in units of $\tau_p$, calculated using an unmodified friction coefficient  ($1.08\times 10^{-6}$~Ns/m).
In Fig.~\ref{fig:5.2}(a), the MSDs for the replication-factory polymer with uniform friction are shown in red and magenta, for the loop tips and bases, respectively.
In comparison, the MSDs for the replication-factory polymer with ten-fold higher friction at the loop bases are shown in blue and green, for loop tips and loop bases, respectively.
Unsurprisingly, the difference between the mean MSDs at loop tips for the two cases is small, with the mean MSDs nearly identical at early times ($ \lesssim 400~\tau_p$).
The loop tips in the replication-factory polymer admitting non-uniform friction have slightly lower mean MSD than those in a uniform-friction polymer, for times beyond about $400~\tau_p$.
Also unsurprisingly, the MSDs at loop bases are significantly smaller when the loop bases experience enhanced friction, than when they do not.

Fig.~\ref{fig:5.2}(b) shows
the corresponding stretching exponents. 
The mean stretching exponents for the loop tips in uniform (red) and non-uniform (blue) friction environments show similar behaviors across the entire range of times studied.
From the earliest time (not shown) to about 10~$\tau_p$, both stretching exponents decrease from 1 to the level around 0.5 and maintain that level of 0.5 until 100~$\tau_p$. 
From 100~$\tau_p$ to 1000~$\tau_p$, both start to decrease, in a similar manner, from the level around 0.5 to the level between 0.2 and 0.3.
By constrast, the stretching exponent for the loop bases appears very differently for the two cases:
in the uniform friction case, the loop bases display a stretching exponent that decreases from 0.4 to $\sim$0.25 as the time increases from a few $\tau_p$ to $\sim100~\tau_p$. It then appears to reverse this trend, recovering to 0.35 for times $\gtrsim 300 \tau_p$.
For loop bases with ten-fold higher friction, the stretching exponent for the loop bases decreases from 0.85  at early times (a few $\tau_p$) to reach 0.4 
(for times $\gtrsim 300~\tau_p$) where it appears to plateau.
We can understand the initial decrease of the stretching exponent from a large value in the enhanced  friction case by first recognizing that in actuality all stretching exponents start  from  1 at the earliest times. 
This behavior usually occurs at smaller values of the time than we plot. 
However, because the polymer time of the loop bases is effectively increased by a factor of 10, as a result of the ten-fold higher friction that they experience, it follows that the evolution of the stretching exponent from 1 to smaller values now occurs within the range of times covered by our simulations.
According to Eq.~\ref{RouseDiffusionCoefficient}, the diffusion coefficient, and thus the MSD amplitude, is proportional to $1/\sqrt{\zeta}$. 
Therefore, in theory, when the friction coefficient is increased by a factor of 10, the MSD amplitude should decrease by a factor of about 3. 
By examining the discrepancy between the mean MSD for loop bases with normal friction coefficient $\zeta$ (magenta) and the mean MSD for loop bases with friction coefficient of 10~$\zeta$ (green) in Fig.~\ref{fig:5.2}(a), we see that at very early times, from 0~$\tau_p$ to around 50~$\tau_p$, the mean MSD for loop bases with friction $\zeta$ is about 3 times greater than that with friction 10~$\zeta$, coinciding with the theoretical prediction.
However, the dependence of MSD amplitude on the friction coefficient becomes less significant for times greater than 50~$\tau_p$, and the two MSD amplitudes only differ by about 20\% at 2000~$\tau_p$.

\section{Conclusions}
\label{Concl}
We have incorporated loops and loop extrusion into the Rouse model of polymer dynamics in a fashion that permits exact simulations of the resultant looped polymer.
By carrying out simulations for polymers whose loop configuration evolves in time according to the loop extrusion factor (LEF) model, we have
demonstrated that 
chromatin loops in a dynamic steady state reduce chromatin mobility, as measured by the time- and ensemble-averaged MSDs of gene loci.
We have also shown that this reduction in mobility increases with increasing LEF density.
Also, in contrast to the classical Rouse model stretching exponent, $\alpha$, which admits the value of $1/2$ in the absence of loops, we have shown that loops reduce the stretching exponent at early times.
The reduction in the stretching exponent also increases with increasing LEF density, but achieves a value near 0.45 for best-estimate loop densities, estimated from an
analysis of Hi-C experiments.
Remarkably, the simulated stretching exponent of 0.45 is consistent with the value measured in our own recent experiments that study the
MSDs of several fluorescently-labelled gene loci in living fission yeast.
Active loop extrusion via the LEF model automatically ensures that  our simulations sample across an ensemble of  different loop configurations.
However, by finding indistinguishable results, when we explicitly average MSDs over an ensemble of different static loop configurations, 
we can infer that LEF dynamics themselves contribute negligibly to the polymer dynamics of chromatin, given that the polymer time is sufficiently smaller than the time scale of loop extrusion.

By studying a static rosette configuration (Fig.~\ref{fig:3.0}), we have shown that the MSD for loop bases is significantly suppressed compared to that of loop tips, which, at early times, behave akin to a polymer without loops.
The chromatin backbone resembles the loop tips at early times and resemble the loop bases at later times, in terms of both MSD and stretching exponent.

Our exact simulation method also allows us to examine Rouse models with non-uniform friction.
By assigning each bead of an unlooped polymer a fixed friction coefficient drawn from a scaled log-normal distribution, we have shown that a non-uniform friction environment does not affect the averaged dynamics predicted by the Rouse model.
We have also shown that chromatin mobility at the center of a replication factory is much reduced if a high-friction environment is present at the center.
More generally, our Rouse model simulation can provide predictions and evidence for chromatin organization and dynamics that are subject to loops or other genomic organizations and functions.

\begin{acknowledgments}
This research was supported by the NSF via EFRI CEE 1830904. M.L.P.B. was supported by NIH Grant No. T32EB019941 and the NSF GRFP.
\end{acknowledgments}
\appendix
\section{Mean squared radius of gyration}
\label{RoGformula}
The mean squared radius of gyration of a chromatin polymer segment, that extends from genomic location A to genomic location B, given the equivalent definition of radius of gyration from \cite{fixman1962}, can be represented as 
\begin{equation}
    \langle R_G^2\rangle = \frac{1}{2N(N+1)}\sum_{i = A}^B  \sum_{j = A}^B \langle (\bm{r}_i - \bm{r}_j)^2\rangle
    \label{RoGeqn}
\end{equation}
where $\bm{r}_i$ and $\bm{r}_j$ are the position vectors of genomic location $i$ and $j$, respectively.
At length scales longer than the Kuhn length, $l_k$, we treat chromatin as a Gaussian polymer, so that
the probability distribution of separation vector between $i$ and $j$ in $d$-dimension, namely $\bm{r}_{ij}$, is given by 
\begin{equation}
    \mathcal{P}(\bm{r}_{ij}) = (\frac{d}{2\pi N_{ij}l_k^2})^{d/2} e^{-d\bm{r}_{ij}^2/2N_{ij}l_k^2},
    \label{sepdistn}
\end{equation}
where $N_{ij}$ is the number of Kuhn segments between $i$ and $j$, and the mean squared separation between $i$ and $j$ is given by 
\begin{equation}
    \langle(\bm{r}_i - \bm{r}_j)^2\rangle = N_{ij}l_k^2.
\end{equation}
Due to the existence of loops, and thus a change of chromatin polymer topology, the separation between any two locations $i$ and $j$ could be reduced by intervening loops such that their separation vector follows a probability distribution given in Eq.~\ref{sepdistn} but with a smaller $N_{ij}$ value, or the effective genomic distance $(N_{eff})_{ij}$. 
The effective genomic distance can be calculated given only the loop configuration of the polymer and is expressed in the unit of number of Kuhn segments here.
Therefore, the mean squared separation between $i$ and $j$ with loop configuration is given by
\begin{equation}
    \langle(\bm{r}_i - \bm{r}_j)^2\rangle = (N_{eff})_{ij}l_k^2,
\end{equation}
and the mean squared radius of gyration in (\ref{RoGeqn}) reads as 
\begin{equation}
    \langle R_G^2\rangle = \frac{l_k^2}{2N(N+1)} \sum_{i = A}^B \sum_{j = A}^B (N_{eff})_{ij}.
    \label{rogFormula2}
\end{equation}
In the absence of loops, $(N_{eff})_{ij}$ reduces to $N_{ij} = |i-j|$.
Therefore, Eq.~\ref{rogFormula2} becomes
\begin{align}
    \langle R_G^2\rangle 
    &= \frac{l_k^2}{2N(N+1)} \sum_{i = 1}^N \sum_{j = 1}^N |i-j| \\
    &= \frac{l_k^2}{2N(N+1)} \frac{(N-1)N(N+1)}{3} \\
    &= \frac{(N-1)l_k^2}{6},
\end{align}
which exactly recovers the mean radius of gyration of a Gaussian polymer with $N$-beads ($N-1$ segments) and Kuhn length $l_k$.

\section{Rouse model for periodic boundary conditions}
\label{periodicbc}
In this case, the two ends of the polymer are connected together via a nearest-neighbor spring, and the matrix \(A\) (\(N\)-by-\(N\)) becomes 
\begin{equation}
    \label{matrixKclosed}
    A_{periodic}=
    \begin{bmatrix}
        2 & -1 & 0 & \cdots & \cdots & -1 \\
        -1 & 2 & -1 & \cdots & \cdots & 0 \\
        0 & -1 & 2 & \cdots & \cdots & 0 \\
        \vdots & \vdots & \vdots & \ddots & \ddots & \vdots \\
        \vdots & \vdots & \vdots & \ddots & 2 & -1 \\
        -1 & 0 & 0 & \cdots & -1 & 2
    \end{bmatrix}
    ,
\end{equation}
which is a circulant matrix. The eigenvectors of circulant matrices contain entries of solutions of \(N\)-th root of unity; more precisely, the complex eigenvector matrix of (\ref{matrixKclosed}) is given by
\begin{equation}
    U_{jk}=e^{\frac{2\pi i}{N}jk},
    \quad
    j,k\in \{0,1,...,N-1\}.
\end{equation}
Since (\ref{matrixKclosed}) is real, the real and imaginary parts of \(U\) each gives a set of eigenvectors, although neither of them is complete, due to the parities of sine and cosine. However, a linear combination of the real and imaginary parts breaks the parity and thus forms a set of linearly independent and complete eigenvectors. Here we choose a symmetric linear combination of cosine and sine, which results in the real eigenvector matrix: 
\begin{align}
    U_{jk}
    &=\frac{N_0}{2}\left(\sqrt{2}\cos(\frac{2\pi jk}{N})
    +\sqrt{2}\sin(\frac{2\pi jk}{N})\right) \\
    &=N_0\sin\left(\frac{2\pi jk}{N}+\frac{\pi}{4}\right) \\
    &=\sqrt{\frac{2}{N}}
    \sin\left(\frac{2\pi jk}{N}+\frac{\pi}{4}\right)
\end{align}
where the normalization constant \(N_0=\sqrt{\frac{2}{N}}\) ensures unitarity of \(U\). The eigenvalue corresponding to \(k\)-th column of \(U\) (the \(k\)-th eigenvector) is given by
\begin{align}
    \lambda_k&=\sum_{j=0}^{N-1}A_{0j}U_{jk}
    =\sum_{j=0}^{N-1}A_{0j}e^{\frac{2\pi i}{N}jk} \\
    &=A_{00}+A_{01}e^{\frac{2\pi i}{N}k}
    +A_{0,N-1}e^{\frac{2\pi i}{N}k(N-1)} \\
    &=2-e^{\frac{2\pi i}{N}k}-e^{-\frac{2\pi i}{N}k} \\
    &=2-2\cos\left(\frac{2\pi k}{N}\right).
\end{align}
Using the above results, Eq.~\ref{EQ18} and Eq.~\ref{EQ20} become 
\begin{align}
    \langle X_n^2(0) \rangle &= \langle X_n^2(t) \rangle \nonumber \\
    &= \frac{k_BT}{\kappa N} \sum_{k=1}^{N-1}
    \sin^2 \left(\frac{2\pi nk}{N}+\frac{\pi}{4}\right)
    \frac{1}{\left[1-\cos\left(\frac{2\pi k}{N}\right)\right]} \nonumber \\
    &= \frac{k_BT}{12\kappa N} (N^2-1),
    \label{MSD0periodic}
\end{align}
and
\begin{align}
    \langle &\left [ X_n(t)-X_n(0)\right ] ^2 \rangle = \nonumber \\
    &\frac{2k_BT}{\kappa N} 
    \sum_{k=1}^{N-1} \sin^2\left(\frac{2\pi nk}{N}+\frac{\pi}{4}\right)
    \frac{1-e^{-\frac{2\kappa}{\zeta} (1-\cos (\frac{2\pi k}{N})) \abs*{t}}}
    {\left[1-\cos \left(\frac{2\pi k}{N}\right)\right]},
    \label{MSD1periodic}
\end{align}
respectively. 
The first bead starts with index $n=0$. 
Note that the mean squared position $\langle X_n^2(0) \rangle$ (relative to Rouse polymer's center of mass) of the beads under periodic boundary condition does not depend on the bead index, 
which makes intuitive sense since the position of one bead relative to the others in the Rouse chain is identical for all beads.

\begin{figure}[h!]
    \includegraphics[width=0.45\textwidth]{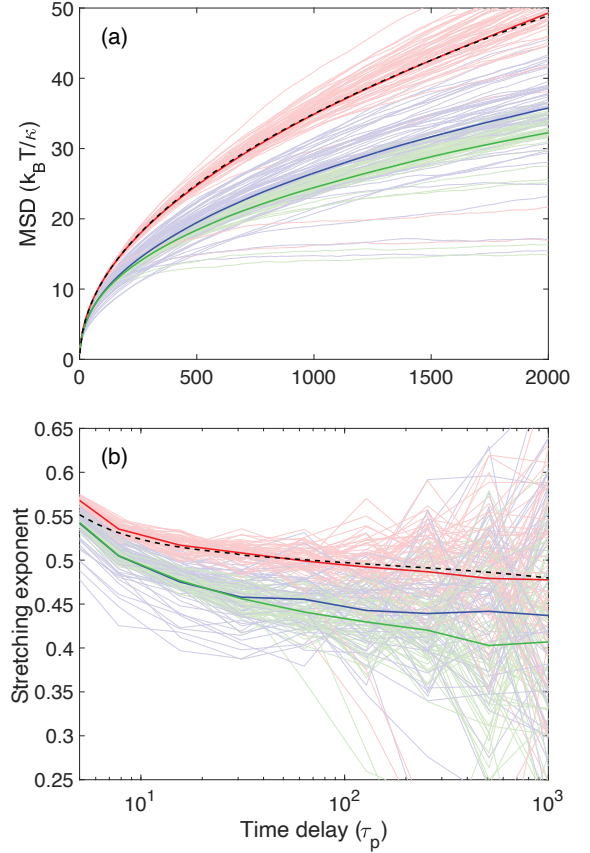}
    \caption{\label{fig1.3}
    (a) Simulated MSDs of Rouse-model polymers with fixed boundary conditions for the classical Rouse model with no loops, shown in red, for the CTCF model with loops, shown in blue, and for the random loop model, shown in green.
    (b) The corresponding simulated MSD stretching exponent, \(\alpha(t)\), calculated according to Eq.~\ref{alphacalc} for no loops (red), for the CTCF model (blue), and for the random loop model (green).
    The dashed black line in (a) shows the theoretical mean MSD given by Eq.~\ref{MSD1fixed}, and then the theoretical mean stretching exponent is calculated using Eq.~\ref{alphacalc} and shown in (b).
    In both (a) and (b), the thin lines correspond to the results for 60 individual beads,  uniformly distributed along the chromatin.
    The thick lines correspond to the average across all 60 beads.
    Each thin line is an average over 30 independent simulations with identical simulation parameters.
    The LEF and Rouse simulation parameters used are given in Table~\ref{LEFparam} and \ref{Rouseparam}.}
\end{figure}
\section{Rouse model for fixed boundary condition}
\label{fixedbc}
In this case, the two ends of the polymer are attached to two fixed location via additional springs
and
the  matrix \(A\) (\(N\)-by-\(N\)) becomes 
\begin{equation}
    \label{matrixKopen}
    A_{fixed}=
    \begin{bmatrix}
        2 & -1 & 0 & \cdots & \cdots & 0 \\
        -1 & 2 & -1 & \cdots & \cdots & 0 \\
        0 & -1 & 2 & \cdots & \cdots & 0 \\
        \vdots & \vdots & \vdots & \ddots & \ddots & \vdots \\
        \vdots & \vdots & \vdots & \ddots & 2 & -1 \\
        0 & 0 & 0 & \cdots & -1 & 2
    \end{bmatrix}
    .
\end{equation}
The eigenvector matrix is given by 
\begin{equation}
 S=\sqrt{\frac{2}{N+1}} 
    \begin{bmatrix}
        \sin\frac{\pi}{N+1} & \sin\frac{2\pi}{N+1} & \sin\frac{3\pi}{N+1} & \cdots \\
        \sin\frac{2\pi}{N+1} & \sin\frac{4\pi}{N+1} & \sin\frac{6\pi}{N+1} & \cdots \\
        \sin\frac{3\pi}{N+1} & \sin\frac{6\pi}{N+1} & \sin\frac{9\pi}{N+1} & \cdots \\
        \vdots & \vdots & \vdots & \vdots \\
        \vdots & \vdots & \vdots & \vdots \\
    \end{bmatrix}
    ,
\end{equation}
i.e. 
\begin{equation}
    \label{fixedU}
   S_{mn}=\sqrt{\frac{2}{N+1}}\sin\frac{mn\pi}{N+1}.
\end{equation}
$S$ diagonalizes \(A \), {\em i.e.} \(SAS=D_1\).
The resultant diagonal matrix \(D_1\) contains the eigenvalues of $A$ on its diagonal: 
\begin{equation}
\Lambda =    \begin{bmatrix}
        2-2\cos\frac{\pi}{N+1} & 0 & 0 & \cdots \\
        0 & 2-2\cos\frac{2\pi}{N+1} & 0 & \cdots \\
        0 & 0 & \ddots & \cdots \\
        \vdots & \vdots & \vdots & \ddots \\
    \end{bmatrix}
    ,
\end{equation}
i.e.
\begin{equation}
    \label{eigenVfixed}
    \Lambda_m=2-2\cos\frac{m\pi}{N+1}.
\end{equation}

Using these results, Eq.~\ref{EQ18} and Eq.~\ref{EQ20} become
\begin{align}
    \langle X_n^2(0) \rangle &= \langle X_n^2(t) \rangle \nonumber \\
   &= \frac{k_BT}{(N+1)\kappa} \sum_{k=1}^N
    \sin^2 \left(\frac{nk\pi}{N+1}\right)
    \frac{1}{\left[1-\cos\left(\frac{k\pi}{N+1}\right)\right]} \nonumber \\
    &= \frac{k_BT}{\kappa} n \left(1-\frac{n}{N+1}\right),
    \label{MSD0fixed}
\end{align}
and
\begin{align}
    \langle &\left [ X_n(t)-X_n(0)\right ] ^2 \rangle = \nonumber \\
    &\frac{2k_BT}{(N+1)\kappa} 
    \sum_{k=1}^N \sin^2 \left(\frac{nk\pi}{N+1}\right)
    \frac{1-e^{-\frac{2\kappa}{\zeta} (1-\cos (\frac{k\pi}{N+1})) \abs*{t}}}
    {\left[1-\cos \left(\frac{k\pi}{N+1}\right)\right]},
    \label{MSD1fixed}
\end{align}
respectively. Note that the bead indexing $n$ starts from 1 to N in this case.

Fig.~\ref{fig1.3} shows the simulated MSD and stretching exponent of a Rouse-model polymer with loops and without loops, under fixed boundary condition. 
The dashed black line gives the theoretical mean MSD and stretching exponent calculated using Eq.~\ref{MSD1fixed} and Eq.~\ref{alphacalc}, as a comparison to the simulation results for classical Rouse polymer under fixed boundary condition.

\bibliography{reference_rouse}

\begin{thebibliography}{87}%
\makeatletter
\providecommand \@ifxundefined [1]{%
 \@ifx{#1\undefined}
}%
\providecommand \@ifnum [1]{%
 \ifnum #1\expandafter \@firstoftwo
 \else \expandafter \@secondoftwo
 \fi
}%
\providecommand \@ifx [1]{%
 \ifx #1\expandafter \@firstoftwo
 \else \expandafter \@secondoftwo
 \fi
}%
\providecommand \natexlab [1]{#1}%
\providecommand \enquote  [1]{``#1''}%
\providecommand \bibnamefont  [1]{#1}%
\providecommand \bibfnamefont [1]{#1}%
\providecommand \citenamefont [1]{#1}%
\providecommand \href@noop [0]{\@secondoftwo}%
\providecommand \href [0]{\begingroup \@sanitize@url \@href}%
\providecommand \@href[1]{\@@startlink{#1}\@@href}%
\providecommand \@@href[1]{\endgroup#1\@@endlink}%
\providecommand \@sanitize@url [0]{\catcode `\\12\catcode `\$12\catcode
  `\&12\catcode `\#12\catcode `\^12\catcode `\_12\catcode `\%12\relax}%
\providecommand \@@startlink[1]{}%
\providecommand \@@endlink[0]{}%
\providecommand \url  [0]{\begingroup\@sanitize@url \@url }%
\providecommand \@url [1]{\endgroup\@href {#1}{\urlprefix }}%
\providecommand \urlprefix  [0]{URL }%
\providecommand \Eprint [0]{\href }%
\providecommand \doibase [0]{https://doi.org/}%
\providecommand \selectlanguage [0]{\@gobble}%
\providecommand \bibinfo  [0]{\@secondoftwo}%
\providecommand \bibfield  [0]{\@secondoftwo}%
\providecommand \translation [1]{[#1]}%
\providecommand \BibitemOpen [0]{}%
\providecommand \bibitemStop [0]{}%
\providecommand \bibitemNoStop [0]{.\EOS\space}%
\providecommand \EOS [0]{\spacefactor3000\relax}%
\providecommand \BibitemShut  [1]{\csname bibitem#1\endcsname}%
\let\auto@bib@innerbib\@empty
\bibitem [{\citenamefont {Rouse~Jr}(1953)}]{rouse1953}%
  \BibitemOpen
  \bibfield  {author} {\bibinfo {author} {\bibfnamefont {P.~E.}\ \bibnamefont
  {Rouse~Jr}},\ }\bibfield  {title} {\bibinfo {title} {A theory of the linear
  viscoelastic properties of dilute solutions of coiling polymers},\
  }\href@noop {} {\bibfield  {journal} {\bibinfo  {journal} {J. Chem. Phys.}\
  }\textbf {\bibinfo {volume} {21}},\ \bibinfo {pages} {1272} (\bibinfo {year}
  {1953})}\BibitemShut {NoStop}%
\bibitem [{\citenamefont {De~Gennes}\ and\ \citenamefont
  {Gennes}(1979)}]{deGennes1979}%
  \BibitemOpen
  \bibfield  {author} {\bibinfo {author} {\bibfnamefont {P.-G.}\ \bibnamefont
  {De~Gennes}}\ and\ \bibinfo {author} {\bibfnamefont {P.-G.}\ \bibnamefont
  {Gennes}},\ }\href@noop {} {\emph {\bibinfo {title} {Scaling concepts in
  polymer physics}}}\ (\bibinfo  {publisher} {Cornell University Press},\
  \bibinfo {year} {1979})\BibitemShut {NoStop}%
\bibitem [{\citenamefont {Ahlrichs}\ \emph {et~al.}(2001)\citenamefont
  {Ahlrichs}, \citenamefont {Everaers},\ and\ \citenamefont
  {D{\"u}nweg}}]{ahlrichs2001}%
  \BibitemOpen
  \bibfield  {author} {\bibinfo {author} {\bibfnamefont {P.}~\bibnamefont
  {Ahlrichs}}, \bibinfo {author} {\bibfnamefont {R.}~\bibnamefont {Everaers}},\
  and\ \bibinfo {author} {\bibfnamefont {B.}~\bibnamefont {D{\"u}nweg}},\
  }\bibfield  {title} {\bibinfo {title} {Screening of hydrodynamic interactions
  in semidilute polymer solutions: A computer simulation study},\ }\href@noop
  {} {\bibfield  {journal} {\bibinfo  {journal} {Phys. Rev. E}\ }\textbf
  {\bibinfo {volume} {64}},\ \bibinfo {pages} {040501} (\bibinfo {year}
  {2001})}\BibitemShut {NoStop}%
\bibitem [{\citenamefont {P{\"u}tz}\ \emph {et~al.}(2000)\citenamefont
  {P{\"u}tz}, \citenamefont {Kremer},\ and\ \citenamefont {Grest}}]{putz2000}%
  \BibitemOpen
  \bibfield  {author} {\bibinfo {author} {\bibfnamefont {M.}~\bibnamefont
  {P{\"u}tz}}, \bibinfo {author} {\bibfnamefont {K.}~\bibnamefont {Kremer}},\
  and\ \bibinfo {author} {\bibfnamefont {G.~S.}\ \bibnamefont {Grest}},\
  }\bibfield  {title} {\bibinfo {title} {What is the entanglement length in a
  polymer melt?},\ }\href@noop {} {\bibfield  {journal} {\bibinfo  {journal}
  {Europhys. Lett.}\ }\textbf {\bibinfo {volume} {49}},\ \bibinfo {pages} {735}
  (\bibinfo {year} {2000})}\BibitemShut {NoStop}%
\bibitem [{\citenamefont {Richter}\ \emph {et~al.}(1993)\citenamefont
  {Richter}, \citenamefont {Willner}, \citenamefont {Zirkel}, \citenamefont
  {Farago}, \citenamefont {Fetters},\ and\ \citenamefont
  {Huang}}]{richter1993}%
  \BibitemOpen
  \bibfield  {author} {\bibinfo {author} {\bibfnamefont {D.}~\bibnamefont
  {Richter}}, \bibinfo {author} {\bibfnamefont {L.}~\bibnamefont {Willner}},
  \bibinfo {author} {\bibfnamefont {A.}~\bibnamefont {Zirkel}}, \bibinfo
  {author} {\bibfnamefont {B.}~\bibnamefont {Farago}}, \bibinfo {author}
  {\bibfnamefont {L.~J.}\ \bibnamefont {Fetters}},\ and\ \bibinfo {author}
  {\bibfnamefont {J.~S.}\ \bibnamefont {Huang}},\ }\bibfield  {title} {\bibinfo
  {title} {Onset of topological constraints in polymer melts: A mode analysis
  by neutron spin echo spectroscopy},\ }\href
  {https://doi.org/10.1103/PhysRevLett.71.4158} {\bibfield  {journal} {\bibinfo
   {journal} {Phys. Rev. Lett.}\ }\textbf {\bibinfo {volume} {71}},\ \bibinfo
  {pages} {4158} (\bibinfo {year} {1993})}\BibitemShut {NoStop}%
\bibitem [{\citenamefont {Pearson}\ \emph {et~al.}(1994)\citenamefont
  {Pearson}, \citenamefont {Fetters}, \citenamefont {Graessley}, \citenamefont
  {Ver~Strate},\ and\ \citenamefont {von Meerwall}}]{pearson1994}%
  \BibitemOpen
  \bibfield  {author} {\bibinfo {author} {\bibfnamefont {D.~S.}\ \bibnamefont
  {Pearson}}, \bibinfo {author} {\bibfnamefont {L.~J.}\ \bibnamefont
  {Fetters}}, \bibinfo {author} {\bibfnamefont {W.~W.}\ \bibnamefont
  {Graessley}}, \bibinfo {author} {\bibfnamefont {G.}~\bibnamefont
  {Ver~Strate}},\ and\ \bibinfo {author} {\bibfnamefont {E.}~\bibnamefont {von
  Meerwall}},\ }\bibfield  {title} {\bibinfo {title} {Viscosity and
  self-diffusion coefficient of hydrogenated polybutadiene},\ }\href@noop {}
  {\bibfield  {journal} {\bibinfo  {journal} {Macromolecules}\ }\textbf
  {\bibinfo {volume} {27}},\ \bibinfo {pages} {711} (\bibinfo {year}
  {1994})}\BibitemShut {NoStop}%
\bibitem [{\citenamefont {Wischnewski}\ \emph {et~al.}(2003)\citenamefont
  {Wischnewski}, \citenamefont {Monkenbusch}, \citenamefont {Willner},
  \citenamefont {Richter},\ and\ \citenamefont {Kali}}]{wischnewski2003}%
  \BibitemOpen
  \bibfield  {author} {\bibinfo {author} {\bibfnamefont {A.}~\bibnamefont
  {Wischnewski}}, \bibinfo {author} {\bibfnamefont {M.}~\bibnamefont
  {Monkenbusch}}, \bibinfo {author} {\bibfnamefont {L.}~\bibnamefont
  {Willner}}, \bibinfo {author} {\bibfnamefont {D.}~\bibnamefont {Richter}},\
  and\ \bibinfo {author} {\bibfnamefont {G.}~\bibnamefont {Kali}},\ }\bibfield
  {title} {\bibinfo {title} {Direct observation of the transition from free to
  constrained single-segment motion in entangled polymer melts},\ }\href@noop
  {} {\bibfield  {journal} {\bibinfo  {journal} {Phys. Rev. Lett.}\ }\textbf
  {\bibinfo {volume} {90}},\ \bibinfo {pages} {058302} (\bibinfo {year}
  {2003})}\BibitemShut {NoStop}%
\bibitem [{\citenamefont {Weiss}\ \emph {et~al.}(2004)\citenamefont {Weiss},
  \citenamefont {Elsner}, \citenamefont {Kartberg},\ and\ \citenamefont
  {Nilsson}}]{weiss2004}%
  \BibitemOpen
  \bibfield  {author} {\bibinfo {author} {\bibfnamefont {M.}~\bibnamefont
  {Weiss}}, \bibinfo {author} {\bibfnamefont {M.}~\bibnamefont {Elsner}},
  \bibinfo {author} {\bibfnamefont {F.}~\bibnamefont {Kartberg}},\ and\
  \bibinfo {author} {\bibfnamefont {T.}~\bibnamefont {Nilsson}},\ }\bibfield
  {title} {\bibinfo {title} {Anomalous subdiffusion is a measure for
  cytoplasmic crowding in living cells},\ }\href@noop {} {\bibfield  {journal}
  {\bibinfo  {journal} {Biophys. J.}\ }\textbf {\bibinfo {volume} {87}},\
  \bibinfo {pages} {3518} (\bibinfo {year} {2004})}\BibitemShut {NoStop}%
\bibitem [{\citenamefont {Di~Pierro}\ \emph
  {et~al.}(2018{\natexlab{a}})\citenamefont {Di~Pierro}, \citenamefont
  {Potoyan}, \citenamefont {Wolynes},\ and\ \citenamefont {Onuchic}}]{di2018}%
  \BibitemOpen
  \bibfield  {author} {\bibinfo {author} {\bibfnamefont {M.}~\bibnamefont
  {Di~Pierro}}, \bibinfo {author} {\bibfnamefont {D.~A.}\ \bibnamefont
  {Potoyan}}, \bibinfo {author} {\bibfnamefont {P.~G.}\ \bibnamefont
  {Wolynes}},\ and\ \bibinfo {author} {\bibfnamefont {J.~N.}\ \bibnamefont
  {Onuchic}},\ }\bibfield  {title} {\bibinfo {title} {Anomalous diffusion,
  spatial coherence, and viscoelasticity from the energy landscape of human
  chromosomes},\ }\href@noop {} {\bibfield  {journal} {\bibinfo  {journal}
  {Proc. Natl. Acad. Sci. USA}\ }\textbf {\bibinfo {volume} {115}},\ \bibinfo
  {pages} {7753} (\bibinfo {year} {2018}{\natexlab{a}})}\BibitemShut {NoStop}%
\bibitem [{\citenamefont {Tamm}\ \emph {et~al.}(2015)\citenamefont {Tamm},
  \citenamefont {Nazarov}, \citenamefont {Gavrilov},\ and\ \citenamefont
  {Chertovich}}]{tamm2015}%
  \BibitemOpen
  \bibfield  {author} {\bibinfo {author} {\bibfnamefont {M.~V.}\ \bibnamefont
  {Tamm}}, \bibinfo {author} {\bibfnamefont {L.~I.}\ \bibnamefont {Nazarov}},
  \bibinfo {author} {\bibfnamefont {A.~A.}\ \bibnamefont {Gavrilov}},\ and\
  \bibinfo {author} {\bibfnamefont {A.~V.}\ \bibnamefont {Chertovich}},\
  }\bibfield  {title} {\bibinfo {title} {Anomalous diffusion in fractal
  globules},\ }\href {https://doi.org/10.1103/PhysRevLett.114.178102}
  {\bibfield  {journal} {\bibinfo  {journal} {Phys. Rev. Lett.}\ }\textbf
  {\bibinfo {volume} {114}},\ \bibinfo {pages} {178102} (\bibinfo {year}
  {2015})}\BibitemShut {NoStop}%
\bibitem [{\citenamefont {Hediger}\ \emph {et~al.}(2003)\citenamefont
  {Hediger}, \citenamefont {Taddei}, \citenamefont {Neumann},\ and\
  \citenamefont {Gasser}}]{hediger2003}%
  \BibitemOpen
  \bibfield  {author} {\bibinfo {author} {\bibfnamefont {F.}~\bibnamefont
  {Hediger}}, \bibinfo {author} {\bibfnamefont {A.}~\bibnamefont {Taddei}},
  \bibinfo {author} {\bibfnamefont {F.~R.}\ \bibnamefont {Neumann}},\ and\
  \bibinfo {author} {\bibfnamefont {S.~M.}\ \bibnamefont {Gasser}},\ }\bibfield
   {title} {\bibinfo {title} {Methods for visualizing chromatin dynamics in
  living yeast},\ }in\ \href@noop {} {\emph {\bibinfo {booktitle} {Methods in
  enzymology}}},\ Vol.\ \bibinfo {volume} {375}\ (\bibinfo  {publisher}
  {Elsevier},\ \bibinfo {year} {2003})\ pp.\ \bibinfo {pages}
  {345--365}\BibitemShut {NoStop}%
\bibitem [{\citenamefont {Cabal}\ \emph {et~al.}(2006)\citenamefont {Cabal},
  \citenamefont {Genovesio}, \citenamefont {Rodriguez-Navarro}, \citenamefont
  {Zimmer}, \citenamefont {Gadal}, \citenamefont {Lesne}, \citenamefont {Buc},
  \citenamefont {Feuerbach-Fournier}, \citenamefont {Olivo-Marin},
  \citenamefont {Hurt} \emph {et~al.}}]{cabal2006}%
  \BibitemOpen
  \bibfield  {author} {\bibinfo {author} {\bibfnamefont {G.~G.}\ \bibnamefont
  {Cabal}}, \bibinfo {author} {\bibfnamefont {A.}~\bibnamefont {Genovesio}},
  \bibinfo {author} {\bibfnamefont {S.}~\bibnamefont {Rodriguez-Navarro}},
  \bibinfo {author} {\bibfnamefont {C.}~\bibnamefont {Zimmer}}, \bibinfo
  {author} {\bibfnamefont {O.}~\bibnamefont {Gadal}}, \bibinfo {author}
  {\bibfnamefont {A.}~\bibnamefont {Lesne}}, \bibinfo {author} {\bibfnamefont
  {H.}~\bibnamefont {Buc}}, \bibinfo {author} {\bibfnamefont {F.}~\bibnamefont
  {Feuerbach-Fournier}}, \bibinfo {author} {\bibfnamefont {J.-C.}\ \bibnamefont
  {Olivo-Marin}}, \bibinfo {author} {\bibfnamefont {E.~C.}\ \bibnamefont
  {Hurt}}, \emph {et~al.},\ }\bibfield  {title} {\bibinfo {title} {Saga
  interacting factors confine sub-diffusion of transcribed genes to the nuclear
  envelope},\ }\href@noop {} {\bibfield  {journal} {\bibinfo  {journal}
  {Nature}\ }\textbf {\bibinfo {volume} {441}},\ \bibinfo {pages} {770}
  (\bibinfo {year} {2006})}\BibitemShut {NoStop}%
\bibitem [{\citenamefont {Weber}\ \emph
  {et~al.}(2010{\natexlab{a}})\citenamefont {Weber}, \citenamefont
  {Spakowitz},\ and\ \citenamefont {Theriot}}]{weber2010a}%
  \BibitemOpen
  \bibfield  {author} {\bibinfo {author} {\bibfnamefont {S.~C.}\ \bibnamefont
  {Weber}}, \bibinfo {author} {\bibfnamefont {A.~J.}\ \bibnamefont
  {Spakowitz}},\ and\ \bibinfo {author} {\bibfnamefont {J.~A.}\ \bibnamefont
  {Theriot}},\ }\bibfield  {title} {\bibinfo {title} {Bacterial chromosomal
  loci move subdiffusively through a viscoelastic cytoplasm},\ }\href
  {https://doi.org/10.1103/PhysRevLett.104.238102} {\bibfield  {journal}
  {\bibinfo  {journal} {Phys. Rev. Lett.}\ }\textbf {\bibinfo {volume} {104}},\
  \bibinfo {pages} {238102} (\bibinfo {year} {2010}{\natexlab{a}})}\BibitemShut
  {NoStop}%
\bibitem [{\citenamefont {Weber}\ \emph
  {et~al.}(2012{\natexlab{a}})\citenamefont {Weber}, \citenamefont {Thompson},
  \citenamefont {Moerner}, \citenamefont {Spakowitz},\ and\ \citenamefont
  {Theriot}}]{weber2012a}%
  \BibitemOpen
  \bibfield  {author} {\bibinfo {author} {\bibfnamefont {S.~C.}\ \bibnamefont
  {Weber}}, \bibinfo {author} {\bibfnamefont {M.~A.}\ \bibnamefont {Thompson}},
  \bibinfo {author} {\bibfnamefont {W.~E.}\ \bibnamefont {Moerner}}, \bibinfo
  {author} {\bibfnamefont {A.~J.}\ \bibnamefont {Spakowitz}},\ and\ \bibinfo
  {author} {\bibfnamefont {J.~A.}\ \bibnamefont {Theriot}},\ }\bibfield
  {title} {\bibinfo {title} {Analytical tools to distinguish the effects of
  localization error, confinement, and medium elasticity on the velocity
  autocorrelation function},\ }\href@noop {} {\bibfield  {journal} {\bibinfo
  {journal} {Biophys. J.}\ }\textbf {\bibinfo {volume} {102}},\ \bibinfo
  {pages} {2443} (\bibinfo {year} {2012}{\natexlab{a}})}\BibitemShut {NoStop}%
\bibitem [{\citenamefont {Weber}\ \emph
  {et~al.}(2012{\natexlab{b}})\citenamefont {Weber}, \citenamefont
  {Spakowitz},\ and\ \citenamefont {Theriot}}]{weber2012b}%
  \BibitemOpen
  \bibfield  {author} {\bibinfo {author} {\bibfnamefont {S.~C.}\ \bibnamefont
  {Weber}}, \bibinfo {author} {\bibfnamefont {A.~J.}\ \bibnamefont
  {Spakowitz}},\ and\ \bibinfo {author} {\bibfnamefont {J.~A.}\ \bibnamefont
  {Theriot}},\ }\bibfield  {title} {\bibinfo {title} {Nonthermal
  {ATP}-dependent fluctuations contribute to the in vivo motion of chromosomal
  loci},\ }\href@noop {} {\bibfield  {journal} {\bibinfo  {journal} {Proc.
  Natl. Acad. Sci. USA}\ }\textbf {\bibinfo {volume} {109}},\ \bibinfo {pages}
  {7338} (\bibinfo {year} {2012}{\natexlab{b}})}\BibitemShut {NoStop}%
\bibitem [{\citenamefont {Hajjoul}\ \emph {et~al.}(2013)\citenamefont
  {Hajjoul}, \citenamefont {Mathon}, \citenamefont {Ranchon}, \citenamefont
  {Goiffon}, \citenamefont {Mozziconacci}, \citenamefont {Albert},
  \citenamefont {Carrivain}, \citenamefont {Victor}, \citenamefont {Gadal},
  \citenamefont {Bystricky} \emph {et~al.}}]{hajjoul2013}%
  \BibitemOpen
  \bibfield  {author} {\bibinfo {author} {\bibfnamefont {H.}~\bibnamefont
  {Hajjoul}}, \bibinfo {author} {\bibfnamefont {J.}~\bibnamefont {Mathon}},
  \bibinfo {author} {\bibfnamefont {H.}~\bibnamefont {Ranchon}}, \bibinfo
  {author} {\bibfnamefont {I.}~\bibnamefont {Goiffon}}, \bibinfo {author}
  {\bibfnamefont {J.}~\bibnamefont {Mozziconacci}}, \bibinfo {author}
  {\bibfnamefont {B.}~\bibnamefont {Albert}}, \bibinfo {author} {\bibfnamefont
  {P.}~\bibnamefont {Carrivain}}, \bibinfo {author} {\bibfnamefont {J.-M.}\
  \bibnamefont {Victor}}, \bibinfo {author} {\bibfnamefont {O.}~\bibnamefont
  {Gadal}}, \bibinfo {author} {\bibfnamefont {K.}~\bibnamefont {Bystricky}},
  \emph {et~al.},\ }\bibfield  {title} {\bibinfo {title} {High-throughput
  chromatin motion tracking in living yeast reveals the flexibility of the
  fiber throughout the genome},\ }\href@noop {} {\bibfield  {journal} {\bibinfo
   {journal} {Genome Res.}\ }\textbf {\bibinfo {volume} {23}},\ \bibinfo
  {pages} {1829} (\bibinfo {year} {2013})}\BibitemShut {NoStop}%
\bibitem [{\citenamefont {Verdaasdonk}\ \emph {et~al.}(2013)\citenamefont
  {Verdaasdonk}, \citenamefont {Vasquez}, \citenamefont {Barry}, \citenamefont
  {Barry}, \citenamefont {Goodwin}, \citenamefont {Forest},\ and\ \citenamefont
  {Bloom}}]{verdaasdonk2013}%
  \BibitemOpen
  \bibfield  {author} {\bibinfo {author} {\bibfnamefont {J.~S.}\ \bibnamefont
  {Verdaasdonk}}, \bibinfo {author} {\bibfnamefont {P.~A.}\ \bibnamefont
  {Vasquez}}, \bibinfo {author} {\bibfnamefont {R.~M.}\ \bibnamefont {Barry}},
  \bibinfo {author} {\bibfnamefont {T.}~\bibnamefont {Barry}}, \bibinfo
  {author} {\bibfnamefont {S.}~\bibnamefont {Goodwin}}, \bibinfo {author}
  {\bibfnamefont {M.~G.}\ \bibnamefont {Forest}},\ and\ \bibinfo {author}
  {\bibfnamefont {K.}~\bibnamefont {Bloom}},\ }\bibfield  {title} {\bibinfo
  {title} {Centromere tethering confines chromosome domains},\ }\href@noop {}
  {\bibfield  {journal} {\bibinfo  {journal} {Mol. Cell}\ }\textbf {\bibinfo
  {volume} {52}},\ \bibinfo {pages} {819} (\bibinfo {year} {2013})}\BibitemShut
  {NoStop}%
\bibitem [{\citenamefont {Backlund}\ \emph {et~al.}(2014)\citenamefont
  {Backlund}, \citenamefont {Joyner}, \citenamefont {Weis},\ and\ \citenamefont
  {Moerner}}]{backlund2014}%
  \BibitemOpen
  \bibfield  {author} {\bibinfo {author} {\bibfnamefont {M.~P.}\ \bibnamefont
  {Backlund}}, \bibinfo {author} {\bibfnamefont {R.}~\bibnamefont {Joyner}},
  \bibinfo {author} {\bibfnamefont {K.}~\bibnamefont {Weis}},\ and\ \bibinfo
  {author} {\bibfnamefont {W.}~\bibnamefont {Moerner}},\ }\bibfield  {title}
  {\bibinfo {title} {Correlations of three-dimensional motion of chromosomal
  loci in yeast revealed by the double-helix point spread function
  microscope},\ }\href@noop {} {\bibfield  {journal} {\bibinfo  {journal} {Mol.
  Biol. Cell}\ }\textbf {\bibinfo {volume} {25}},\ \bibinfo {pages} {3619}
  (\bibinfo {year} {2014})}\BibitemShut {NoStop}%
\bibitem [{\citenamefont {Backlund}\ \emph {et~al.}(2015)\citenamefont
  {Backlund}, \citenamefont {Joyner},\ and\ \citenamefont
  {Moerner}}]{backlund2015}%
  \BibitemOpen
  \bibfield  {author} {\bibinfo {author} {\bibfnamefont {M.~P.}\ \bibnamefont
  {Backlund}}, \bibinfo {author} {\bibfnamefont {R.}~\bibnamefont {Joyner}},\
  and\ \bibinfo {author} {\bibfnamefont {W.}~\bibnamefont {Moerner}},\
  }\bibfield  {title} {\bibinfo {title} {Chromosomal locus tracking with proper
  accounting of static and dynamic errors},\ }\href@noop {} {\bibfield
  {journal} {\bibinfo  {journal} {Phys. Rev. E}\ }\textbf {\bibinfo {volume}
  {91}},\ \bibinfo {pages} {062716} (\bibinfo {year} {2015})}\BibitemShut
  {NoStop}%
\bibitem [{\citenamefont {Wang}\ \emph {et~al.}(2015)\citenamefont {Wang},
  \citenamefont {Mozziconacci}, \citenamefont {Bancaud},\ and\ \citenamefont
  {Gadal}}]{wang2015}%
  \BibitemOpen
  \bibfield  {author} {\bibinfo {author} {\bibfnamefont {R.}~\bibnamefont
  {Wang}}, \bibinfo {author} {\bibfnamefont {J.}~\bibnamefont {Mozziconacci}},
  \bibinfo {author} {\bibfnamefont {A.}~\bibnamefont {Bancaud}},\ and\ \bibinfo
  {author} {\bibfnamefont {O.}~\bibnamefont {Gadal}},\ }\bibfield  {title}
  {\bibinfo {title} {Principles of chromatin organization in yeast: relevance
  of polymer models to describe nuclear organization and dynamics},\
  }\href@noop {} {\bibfield  {journal} {\bibinfo  {journal} {Curr. Opin. Cell
  Biol.}\ }\textbf {\bibinfo {volume} {34}},\ \bibinfo {pages} {54} (\bibinfo
  {year} {2015})}\BibitemShut {NoStop}%
\bibitem [{\citenamefont {Rolls}\ \emph {et~al.}(2017)\citenamefont {Rolls},
  \citenamefont {Togashi},\ and\ \citenamefont {Erban}}]{rolls2017}%
  \BibitemOpen
  \bibfield  {author} {\bibinfo {author} {\bibfnamefont {E.}~\bibnamefont
  {Rolls}}, \bibinfo {author} {\bibfnamefont {Y.}~\bibnamefont {Togashi}},\
  and\ \bibinfo {author} {\bibfnamefont {R.}~\bibnamefont {Erban}},\ }\bibfield
   {title} {\bibinfo {title} {Varying the resolution of the {R}ouse model on
  temporal and spatial scales: application to multiscale modeling of {DNA}
  dynamics},\ }\href@noop {} {\bibfield  {journal} {\bibinfo  {journal}
  {Multiscale Model. Simul.}\ }\textbf {\bibinfo {volume} {15}},\ \bibinfo
  {pages} {1672} (\bibinfo {year} {2017})}\BibitemShut {NoStop}%
\bibitem [{\citenamefont {Shukron}\ and\ \citenamefont
  {Holcman}(2017)}]{shukron2017}%
  \BibitemOpen
  \bibfield  {author} {\bibinfo {author} {\bibfnamefont {O.}~\bibnamefont
  {Shukron}}\ and\ \bibinfo {author} {\bibfnamefont {D.}~\bibnamefont
  {Holcman}},\ }\bibfield  {title} {\bibinfo {title} {Transient chromatin
  properties revealed by polymer models and stochastic simulations constructed
  from chromosomal capture data},\ }\href@noop {} {\bibfield  {journal}
  {\bibinfo  {journal} {PLoS Comput. Biol.}\ }\textbf {\bibinfo {volume}
  {13}},\ \bibinfo {pages} {e1005469} (\bibinfo {year} {2017})}\BibitemShut
  {NoStop}%
\bibitem [{\citenamefont {Osmanovi{\'c}}\ and\ \citenamefont
  {Rabin}(2017)}]{osmanovic2017}%
  \BibitemOpen
  \bibfield  {author} {\bibinfo {author} {\bibfnamefont {D.}~\bibnamefont
  {Osmanovi{\'c}}}\ and\ \bibinfo {author} {\bibfnamefont {Y.}~\bibnamefont
  {Rabin}},\ }\bibfield  {title} {\bibinfo {title} {Dynamics of active rouse
  chains},\ }\href@noop {} {\bibfield  {journal} {\bibinfo  {journal} {Soft
  Matter}\ }\textbf {\bibinfo {volume} {13}},\ \bibinfo {pages} {963} (\bibinfo
  {year} {2017})}\BibitemShut {NoStop}%
\bibitem [{\citenamefont {Socol}\ \emph {et~al.}(2019)\citenamefont {Socol},
  \citenamefont {Wang}, \citenamefont {Jost}, \citenamefont {Carrivain},
  \citenamefont {Vaillant}, \citenamefont {Le~Cam}, \citenamefont {Dahirel},
  \citenamefont {Normand}, \citenamefont {Bystricky}, \citenamefont {Victor}
  \emph {et~al.}}]{socol2019}%
  \BibitemOpen
  \bibfield  {author} {\bibinfo {author} {\bibfnamefont {M.}~\bibnamefont
  {Socol}}, \bibinfo {author} {\bibfnamefont {R.}~\bibnamefont {Wang}},
  \bibinfo {author} {\bibfnamefont {D.}~\bibnamefont {Jost}}, \bibinfo {author}
  {\bibfnamefont {P.}~\bibnamefont {Carrivain}}, \bibinfo {author}
  {\bibfnamefont {C.}~\bibnamefont {Vaillant}}, \bibinfo {author}
  {\bibfnamefont {E.}~\bibnamefont {Le~Cam}}, \bibinfo {author} {\bibfnamefont
  {V.}~\bibnamefont {Dahirel}}, \bibinfo {author} {\bibfnamefont
  {C.}~\bibnamefont {Normand}}, \bibinfo {author} {\bibfnamefont
  {K.}~\bibnamefont {Bystricky}}, \bibinfo {author} {\bibfnamefont {J.-M.}\
  \bibnamefont {Victor}}, \emph {et~al.},\ }\bibfield  {title} {\bibinfo
  {title} {Rouse model with transient intramolecular contacts on a timescale of
  seconds recapitulates folding and fluctuation of yeast chromosomes},\
  }\href@noop {} {\bibfield  {journal} {\bibinfo  {journal} {Nucleic Acids
  Res.}\ }\textbf {\bibinfo {volume} {47}},\ \bibinfo {pages} {6195} (\bibinfo
  {year} {2019})}\BibitemShut {NoStop}%
\bibitem [{\citenamefont {Kaplan}\ \emph {et~al.}(2009)\citenamefont {Kaplan},
  \citenamefont {Moore}, \citenamefont {Fondufe-Mittendorf}, \citenamefont
  {Gossett}, \citenamefont {Tillo}, \citenamefont {Field}, \citenamefont
  {LeProust}, \citenamefont {Hughes}, \citenamefont {Lieb}, \citenamefont
  {Widom} \emph {et~al.}}]{kaplan2009}%
  \BibitemOpen
  \bibfield  {author} {\bibinfo {author} {\bibfnamefont {N.}~\bibnamefont
  {Kaplan}}, \bibinfo {author} {\bibfnamefont {I.~K.}\ \bibnamefont {Moore}},
  \bibinfo {author} {\bibfnamefont {Y.}~\bibnamefont {Fondufe-Mittendorf}},
  \bibinfo {author} {\bibfnamefont {A.~J.}\ \bibnamefont {Gossett}}, \bibinfo
  {author} {\bibfnamefont {D.}~\bibnamefont {Tillo}}, \bibinfo {author}
  {\bibfnamefont {Y.}~\bibnamefont {Field}}, \bibinfo {author} {\bibfnamefont
  {E.~M.}\ \bibnamefont {LeProust}}, \bibinfo {author} {\bibfnamefont {T.~R.}\
  \bibnamefont {Hughes}}, \bibinfo {author} {\bibfnamefont {J.~D.}\
  \bibnamefont {Lieb}}, \bibinfo {author} {\bibfnamefont {J.}~\bibnamefont
  {Widom}}, \emph {et~al.},\ }\bibfield  {title} {\bibinfo {title} {The
  {DNA}-encoded nucleosome organization of a eukaryotic genome},\ }\href@noop
  {} {\bibfield  {journal} {\bibinfo  {journal} {Nature}\ }\textbf {\bibinfo
  {volume} {458}},\ \bibinfo {pages} {362} (\bibinfo {year}
  {2009})}\BibitemShut {NoStop}%
\bibitem [{\citenamefont {Zhang}\ \emph {et~al.}(2011)\citenamefont {Zhang},
  \citenamefont {Wippo}, \citenamefont {Wal}, \citenamefont {Ward},
  \citenamefont {Korber},\ and\ \citenamefont {Pugh}}]{zhang2011}%
  \BibitemOpen
  \bibfield  {author} {\bibinfo {author} {\bibfnamefont {Z.}~\bibnamefont
  {Zhang}}, \bibinfo {author} {\bibfnamefont {C.~J.}\ \bibnamefont {Wippo}},
  \bibinfo {author} {\bibfnamefont {M.}~\bibnamefont {Wal}}, \bibinfo {author}
  {\bibfnamefont {E.}~\bibnamefont {Ward}}, \bibinfo {author} {\bibfnamefont
  {P.}~\bibnamefont {Korber}},\ and\ \bibinfo {author} {\bibfnamefont {B.~F.}\
  \bibnamefont {Pugh}},\ }\bibfield  {title} {\bibinfo {title} {A packing
  mechanism for nucleosome organization reconstituted across a eukaryotic
  genome},\ }\href@noop {} {\bibfield  {journal} {\bibinfo  {journal}
  {Science}\ }\textbf {\bibinfo {volume} {332}},\ \bibinfo {pages} {977}
  (\bibinfo {year} {2011})}\BibitemShut {NoStop}%
\bibitem [{\citenamefont {Garc{\'\i}a}\ \emph {et~al.}(2017)\citenamefont
  {Garc{\'\i}a}, \citenamefont {Gonz{\'a}lez},\ and\ \citenamefont
  {Antequera}}]{garcia2017}%
  \BibitemOpen
  \bibfield  {author} {\bibinfo {author} {\bibfnamefont {A.}~\bibnamefont
  {Garc{\'\i}a}}, \bibinfo {author} {\bibfnamefont {S.}~\bibnamefont
  {Gonz{\'a}lez}},\ and\ \bibinfo {author} {\bibfnamefont {F.}~\bibnamefont
  {Antequera}},\ }\bibfield  {title} {\bibinfo {title} {Nucleosomal
  organization and {DNA} base composition patterns},\ }\href@noop {} {\bibfield
   {journal} {\bibinfo  {journal} {Nucleus}\ }\textbf {\bibinfo {volume} {8}},\
  \bibinfo {pages} {469} (\bibinfo {year} {2017})}\BibitemShut {NoStop}%
\bibitem [{\citenamefont {Zink}\ \emph {et~al.}(1998)\citenamefont {Zink},
  \citenamefont {Cremer}, \citenamefont {Saffrich}, \citenamefont {Fischer},
  \citenamefont {Trendelenburg}, \citenamefont {Ansorge},\ and\ \citenamefont
  {Stelzer}}]{zink1998}%
  \BibitemOpen
  \bibfield  {author} {\bibinfo {author} {\bibfnamefont {D.}~\bibnamefont
  {Zink}}, \bibinfo {author} {\bibfnamefont {T.}~\bibnamefont {Cremer}},
  \bibinfo {author} {\bibfnamefont {R.}~\bibnamefont {Saffrich}}, \bibinfo
  {author} {\bibfnamefont {R.}~\bibnamefont {Fischer}}, \bibinfo {author}
  {\bibfnamefont {M.~F.}\ \bibnamefont {Trendelenburg}}, \bibinfo {author}
  {\bibfnamefont {W.}~\bibnamefont {Ansorge}},\ and\ \bibinfo {author}
  {\bibfnamefont {E.~H.}\ \bibnamefont {Stelzer}},\ }\bibfield  {title}
  {\bibinfo {title} {Structure and dynamics of human interphase chromosome
  territories in vivo},\ }\href@noop {} {\bibfield  {journal} {\bibinfo
  {journal} {Hum. Genet.}\ }\textbf {\bibinfo {volume} {102}},\ \bibinfo
  {pages} {241} (\bibinfo {year} {1998})}\BibitemShut {NoStop}%
\bibitem [{\citenamefont {Fritz}\ \emph {et~al.}(2019)\citenamefont {Fritz},
  \citenamefont {Sehgal}, \citenamefont {Pliss}, \citenamefont {Xu},\ and\
  \citenamefont {Berezney}}]{fritz2019}%
  \BibitemOpen
  \bibfield  {author} {\bibinfo {author} {\bibfnamefont {A.~J.}\ \bibnamefont
  {Fritz}}, \bibinfo {author} {\bibfnamefont {N.}~\bibnamefont {Sehgal}},
  \bibinfo {author} {\bibfnamefont {A.}~\bibnamefont {Pliss}}, \bibinfo
  {author} {\bibfnamefont {J.}~\bibnamefont {Xu}},\ and\ \bibinfo {author}
  {\bibfnamefont {R.}~\bibnamefont {Berezney}},\ }\bibfield  {title} {\bibinfo
  {title} {Chromosome territories and the global regulation of the genome},\
  }\href@noop {} {\bibfield  {journal} {\bibinfo  {journal} {Genes Chromosomes
  Cancer}\ }\textbf {\bibinfo {volume} {58}},\ \bibinfo {pages} {407} (\bibinfo
  {year} {2019})}\BibitemShut {NoStop}%
\bibitem [{\citenamefont {Ghosh}\ and\ \citenamefont
  {Meyer}(2021)}]{ghosh2021}%
  \BibitemOpen
  \bibfield  {author} {\bibinfo {author} {\bibfnamefont {R.~P.}\ \bibnamefont
  {Ghosh}}\ and\ \bibinfo {author} {\bibfnamefont {B.~J.}\ \bibnamefont
  {Meyer}},\ }\bibfield  {title} {\bibinfo {title} {Spatial organization of
  chromatin: Emergence of chromatin structure during development},\ }\href@noop
  {} {\bibfield  {journal} {\bibinfo  {journal} {Annu. Rev. Cell Dev. Biol.}\
  }\textbf {\bibinfo {volume} {37}},\ \bibinfo {pages} {199} (\bibinfo {year}
  {2021})}\BibitemShut {NoStop}%
\bibitem [{\citenamefont {Lieberman-Aiden}\ \emph {et~al.}(2009)\citenamefont
  {Lieberman-Aiden}, \citenamefont {Van~Berkum}, \citenamefont {Williams},
  \citenamefont {Imakaev}, \citenamefont {Ragoczy}, \citenamefont {Telling},
  \citenamefont {Amit}, \citenamefont {Lajoie}, \citenamefont {Sabo},
  \citenamefont {Dorschner} \emph {et~al.}}]{aiden2009}%
  \BibitemOpen
  \bibfield  {author} {\bibinfo {author} {\bibfnamefont {E.}~\bibnamefont
  {Lieberman-Aiden}}, \bibinfo {author} {\bibfnamefont {N.~L.}\ \bibnamefont
  {Van~Berkum}}, \bibinfo {author} {\bibfnamefont {L.}~\bibnamefont
  {Williams}}, \bibinfo {author} {\bibfnamefont {M.}~\bibnamefont {Imakaev}},
  \bibinfo {author} {\bibfnamefont {T.}~\bibnamefont {Ragoczy}}, \bibinfo
  {author} {\bibfnamefont {A.}~\bibnamefont {Telling}}, \bibinfo {author}
  {\bibfnamefont {I.}~\bibnamefont {Amit}}, \bibinfo {author} {\bibfnamefont
  {B.~R.}\ \bibnamefont {Lajoie}}, \bibinfo {author} {\bibfnamefont {P.~J.}\
  \bibnamefont {Sabo}}, \bibinfo {author} {\bibfnamefont {M.~O.}\ \bibnamefont
  {Dorschner}}, \emph {et~al.},\ }\bibfield  {title} {\bibinfo {title}
  {Comprehensive mapping of long-range interactions reveals folding principles
  of the human genome},\ }\href@noop {} {\bibfield  {journal} {\bibinfo
  {journal} {Science}\ }\textbf {\bibinfo {volume} {326}},\ \bibinfo {pages}
  {289} (\bibinfo {year} {2009})}\BibitemShut {NoStop}%
\bibitem [{\citenamefont {van Berkum}\ \emph {et~al.}(2010)\citenamefont {van
  Berkum}, \citenamefont {Lieberman-Aiden}, \citenamefont {Williams},
  \citenamefont {Imakaev}, \citenamefont {Gnirke}, \citenamefont {Mirny},
  \citenamefont {Dekker},\ and\ \citenamefont {Lander}}]{berkum2010}%
  \BibitemOpen
  \bibfield  {author} {\bibinfo {author} {\bibfnamefont {N.~L.}\ \bibnamefont
  {van Berkum}}, \bibinfo {author} {\bibfnamefont {E.}~\bibnamefont
  {Lieberman-Aiden}}, \bibinfo {author} {\bibfnamefont {L.}~\bibnamefont
  {Williams}}, \bibinfo {author} {\bibfnamefont {M.}~\bibnamefont {Imakaev}},
  \bibinfo {author} {\bibfnamefont {A.}~\bibnamefont {Gnirke}}, \bibinfo
  {author} {\bibfnamefont {L.~A.}\ \bibnamefont {Mirny}}, \bibinfo {author}
  {\bibfnamefont {J.}~\bibnamefont {Dekker}},\ and\ \bibinfo {author}
  {\bibfnamefont {E.~S.}\ \bibnamefont {Lander}},\ }\bibfield  {title}
  {\bibinfo {title} {Hi-c: A method to study the three-dimensional architecture
  of genomes.},\ }\href@noop {} {\bibfield  {journal} {\bibinfo  {journal} {J.
  Vis. Exp.}\ } (\bibinfo {year} {2010})}\BibitemShut {NoStop}%
\bibitem [{\citenamefont {Dixon}\ \emph {et~al.}(2012)\citenamefont {Dixon},
  \citenamefont {Selvaraj}, \citenamefont {Yue}, \citenamefont {Kim},
  \citenamefont {Li}, \citenamefont {Shen}, \citenamefont {Hu}, \citenamefont
  {Liu},\ and\ \citenamefont {Ren}}]{dixon2012}%
  \BibitemOpen
  \bibfield  {author} {\bibinfo {author} {\bibfnamefont {J.~R.}\ \bibnamefont
  {Dixon}}, \bibinfo {author} {\bibfnamefont {S.}~\bibnamefont {Selvaraj}},
  \bibinfo {author} {\bibfnamefont {F.}~\bibnamefont {Yue}}, \bibinfo {author}
  {\bibfnamefont {A.}~\bibnamefont {Kim}}, \bibinfo {author} {\bibfnamefont
  {Y.}~\bibnamefont {Li}}, \bibinfo {author} {\bibfnamefont {Y.}~\bibnamefont
  {Shen}}, \bibinfo {author} {\bibfnamefont {M.}~\bibnamefont {Hu}}, \bibinfo
  {author} {\bibfnamefont {J.~S.}\ \bibnamefont {Liu}},\ and\ \bibinfo {author}
  {\bibfnamefont {B.}~\bibnamefont {Ren}},\ }\bibfield  {title} {\bibinfo
  {title} {Topological domains in mammalian genomes identified by analysis of
  chromatin interactions},\ }\href@noop {} {\bibfield  {journal} {\bibinfo
  {journal} {Nature}\ }\textbf {\bibinfo {volume} {485}},\ \bibinfo {pages}
  {376} (\bibinfo {year} {2012})}\BibitemShut {NoStop}%
\bibitem [{\citenamefont {Dixon}\ \emph {et~al.}(2016)\citenamefont {Dixon},
  \citenamefont {Gorkin},\ and\ \citenamefont {Ren}}]{dixon2016}%
  \BibitemOpen
  \bibfield  {author} {\bibinfo {author} {\bibfnamefont {J.~R.}\ \bibnamefont
  {Dixon}}, \bibinfo {author} {\bibfnamefont {D.~U.}\ \bibnamefont {Gorkin}},\
  and\ \bibinfo {author} {\bibfnamefont {B.}~\bibnamefont {Ren}},\ }\bibfield
  {title} {\bibinfo {title} {Chromatin domains: the unit of chromosome
  organization},\ }\href@noop {} {\bibfield  {journal} {\bibinfo  {journal}
  {Mol. Cell}\ }\textbf {\bibinfo {volume} {62}},\ \bibinfo {pages} {668}
  (\bibinfo {year} {2016})}\BibitemShut {NoStop}%
\bibitem [{\citenamefont {Sexton}\ \emph {et~al.}(2012)\citenamefont {Sexton},
  \citenamefont {Yaffe}, \citenamefont {Kenigsberg}, \citenamefont
  {Bantignies}, \citenamefont {Leblanc}, \citenamefont {Hoichman},
  \citenamefont {Parrinello}, \citenamefont {Tanay},\ and\ \citenamefont
  {Cavalli}}]{sexton2012}%
  \BibitemOpen
  \bibfield  {author} {\bibinfo {author} {\bibfnamefont {T.}~\bibnamefont
  {Sexton}}, \bibinfo {author} {\bibfnamefont {E.}~\bibnamefont {Yaffe}},
  \bibinfo {author} {\bibfnamefont {E.}~\bibnamefont {Kenigsberg}}, \bibinfo
  {author} {\bibfnamefont {F.}~\bibnamefont {Bantignies}}, \bibinfo {author}
  {\bibfnamefont {B.}~\bibnamefont {Leblanc}}, \bibinfo {author} {\bibfnamefont
  {M.}~\bibnamefont {Hoichman}}, \bibinfo {author} {\bibfnamefont
  {H.}~\bibnamefont {Parrinello}}, \bibinfo {author} {\bibfnamefont
  {A.}~\bibnamefont {Tanay}},\ and\ \bibinfo {author} {\bibfnamefont
  {G.}~\bibnamefont {Cavalli}},\ }\bibfield  {title} {\bibinfo {title}
  {Three-dimensional folding and functional organization principles of the
  {D}rosophila genome},\ }\href@noop {} {\bibfield  {journal} {\bibinfo
  {journal} {Cell}\ }\textbf {\bibinfo {volume} {148}},\ \bibinfo {pages} {458}
  (\bibinfo {year} {2012})}\BibitemShut {NoStop}%
\bibitem [{\citenamefont {Mizuguchi}\ \emph {et~al.}(2014)\citenamefont
  {Mizuguchi}, \citenamefont {Fudenberg}, \citenamefont {Mehta}, \citenamefont
  {Belton}, \citenamefont {Taneja}, \citenamefont {Folco}, \citenamefont
  {FitzGerald}, \citenamefont {Dekker}, \citenamefont {Mirny}, \citenamefont
  {Barrowman} \emph {et~al.}}]{mizuguchi2014}%
  \BibitemOpen
  \bibfield  {author} {\bibinfo {author} {\bibfnamefont {T.}~\bibnamefont
  {Mizuguchi}}, \bibinfo {author} {\bibfnamefont {G.}~\bibnamefont
  {Fudenberg}}, \bibinfo {author} {\bibfnamefont {S.}~\bibnamefont {Mehta}},
  \bibinfo {author} {\bibfnamefont {J.-M.}\ \bibnamefont {Belton}}, \bibinfo
  {author} {\bibfnamefont {N.}~\bibnamefont {Taneja}}, \bibinfo {author}
  {\bibfnamefont {H.~D.}\ \bibnamefont {Folco}}, \bibinfo {author}
  {\bibfnamefont {P.}~\bibnamefont {FitzGerald}}, \bibinfo {author}
  {\bibfnamefont {J.}~\bibnamefont {Dekker}}, \bibinfo {author} {\bibfnamefont
  {L.}~\bibnamefont {Mirny}}, \bibinfo {author} {\bibfnamefont
  {J.}~\bibnamefont {Barrowman}}, \emph {et~al.},\ }\bibfield  {title}
  {\bibinfo {title} {Cohesin-dependent globules and heterochromatin shape 3d
  genome architecture in s. pombe},\ }\href@noop {} {\bibfield  {journal}
  {\bibinfo  {journal} {Nature}\ }\textbf {\bibinfo {volume} {516}},\ \bibinfo
  {pages} {432} (\bibinfo {year} {2014})}\BibitemShut {NoStop}%
\bibitem [{\citenamefont {Dekker}(2014)}]{dekker2014}%
  \BibitemOpen
  \bibfield  {author} {\bibinfo {author} {\bibfnamefont {J.}~\bibnamefont
  {Dekker}},\ }\bibfield  {title} {\bibinfo {title} {Two ways to fold the
  genome during the cell cycle: insights obtained with chromosome conformation
  capture},\ }\href@noop {} {\bibfield  {journal} {\bibinfo  {journal}
  {Epigenetics Chromatin}\ }\textbf {\bibinfo {volume} {7}},\ \bibinfo {pages}
  {1} (\bibinfo {year} {2014})}\BibitemShut {NoStop}%
\bibitem [{\citenamefont {Dekker}\ and\ \citenamefont
  {Heard}(2015)}]{dekker2015}%
  \BibitemOpen
  \bibfield  {author} {\bibinfo {author} {\bibfnamefont {J.}~\bibnamefont
  {Dekker}}\ and\ \bibinfo {author} {\bibfnamefont {E.}~\bibnamefont {Heard}},\
  }\bibfield  {title} {\bibinfo {title} {Structural and functional diversity of
  topologically associating domains},\ }\href@noop {} {\bibfield  {journal}
  {\bibinfo  {journal} {FEBS Lett.}\ }\textbf {\bibinfo {volume} {589}},\
  \bibinfo {pages} {2877} (\bibinfo {year} {2015})}\BibitemShut {NoStop}%
\bibitem [{\citenamefont {Pollard}\ \emph {et~al.}(2016)\citenamefont
  {Pollard}, \citenamefont {Earnshaw}, \citenamefont {Lippincott-Schwartz},\
  and\ \citenamefont {Johnson}}]{pollard2016}%
  \BibitemOpen
  \bibfield  {author} {\bibinfo {author} {\bibfnamefont {T.~D.}\ \bibnamefont
  {Pollard}}, \bibinfo {author} {\bibfnamefont {W.~C.}\ \bibnamefont
  {Earnshaw}}, \bibinfo {author} {\bibfnamefont {J.}~\bibnamefont
  {Lippincott-Schwartz}},\ and\ \bibinfo {author} {\bibfnamefont
  {G.}~\bibnamefont {Johnson}},\ }\href@noop {} {\emph {\bibinfo {title} {Cell
  biology E-book}}}\ (\bibinfo  {publisher} {Elsevier Health Sciences},\
  \bibinfo {year} {2016})\BibitemShut {NoStop}%
\bibitem [{\citenamefont {Jerkovi{\'c}}\ and\ \citenamefont
  {Cavalli}(2021)}]{jerkovic2021}%
  \BibitemOpen
  \bibfield  {author} {\bibinfo {author} {\bibfnamefont {I.}~\bibnamefont
  {Jerkovi{\'c}}}\ and\ \bibinfo {author} {\bibfnamefont {G.}~\bibnamefont
  {Cavalli}},\ }\bibfield  {title} {\bibinfo {title} {Understanding 3d genome
  organization by multidisciplinary methods},\ }\href@noop {} {\bibfield
  {journal} {\bibinfo  {journal} {Nat. Rev. Mol. Cell Biol.}\ }\textbf
  {\bibinfo {volume} {22}},\ \bibinfo {pages} {511} (\bibinfo {year}
  {2021})}\BibitemShut {NoStop}%
\bibitem [{\citenamefont {Dekker}\ and\ \citenamefont
  {Mirny}(2016)}]{dekker2016}%
  \BibitemOpen
  \bibfield  {author} {\bibinfo {author} {\bibfnamefont {J.}~\bibnamefont
  {Dekker}}\ and\ \bibinfo {author} {\bibfnamefont {L.}~\bibnamefont {Mirny}},\
  }\bibfield  {title} {\bibinfo {title} {{The 3D genome as moderator of
  chromosomal communication}},\ }\href@noop {} {\bibfield  {journal} {\bibinfo
  {journal} {Cell}\ }\textbf {\bibinfo {volume} {164}},\ \bibinfo {pages}
  {1110} (\bibinfo {year} {2016})}\BibitemShut {NoStop}%
\bibitem [{\citenamefont {Schleif}(1992)}]{schleif1992}%
  \BibitemOpen
  \bibfield  {author} {\bibinfo {author} {\bibfnamefont {R.}~\bibnamefont
  {Schleif}},\ }\bibfield  {title} {\bibinfo {title} {{DNA} looping},\
  }\href@noop {} {\bibfield  {journal} {\bibinfo  {journal} {Annu. Rev.
  Biochem.}\ }\textbf {\bibinfo {volume} {61}},\ \bibinfo {pages} {199}
  (\bibinfo {year} {1992})}\BibitemShut {NoStop}%
\bibitem [{\citenamefont {Yokota}\ \emph {et~al.}(1995)\citenamefont {Yokota},
  \citenamefont {Van Den~Engh}, \citenamefont {Hearst}, \citenamefont {Sachs},\
  and\ \citenamefont {Trask}}]{yokota1995}%
  \BibitemOpen
  \bibfield  {author} {\bibinfo {author} {\bibfnamefont {H.}~\bibnamefont
  {Yokota}}, \bibinfo {author} {\bibfnamefont {G.}~\bibnamefont {Van
  Den~Engh}}, \bibinfo {author} {\bibfnamefont {J.~E.}\ \bibnamefont {Hearst}},
  \bibinfo {author} {\bibfnamefont {R.~K.}\ \bibnamefont {Sachs}},\ and\
  \bibinfo {author} {\bibfnamefont {B.~J.}\ \bibnamefont {Trask}},\ }\bibfield
  {title} {\bibinfo {title} {{Evidence for the organization of chromatin in
  megabase pair-sized loops arranged along a random walk path in the human
  G0/G1 interphase nucleus}},\ }\href@noop {} {\bibfield  {journal} {\bibinfo
  {journal} {J. Cell Biol.}\ }\textbf {\bibinfo {volume} {130}},\ \bibinfo
  {pages} {1239} (\bibinfo {year} {1995})}\BibitemShut {NoStop}%
\bibitem [{\citenamefont {Dekker}(2008)}]{dekker2008}%
  \BibitemOpen
  \bibfield  {author} {\bibinfo {author} {\bibfnamefont {J.}~\bibnamefont
  {Dekker}},\ }\bibfield  {title} {\bibinfo {title} {Mapping in vivo chromatin
  interactions in yeast suggests an extended chromatin fiber with regional
  variation in compaction},\ }\href@noop {} {\bibfield  {journal} {\bibinfo
  {journal} {J. Biol. Chem.}\ }\textbf {\bibinfo {volume} {283}},\ \bibinfo
  {pages} {34532} (\bibinfo {year} {2008})}\BibitemShut {NoStop}%
\bibitem [{\citenamefont {Alipour}\ and\ \citenamefont
  {Marko}(2012)}]{alipour2012}%
  \BibitemOpen
  \bibfield  {author} {\bibinfo {author} {\bibfnamefont {E.}~\bibnamefont
  {Alipour}}\ and\ \bibinfo {author} {\bibfnamefont {J.~F.}\ \bibnamefont
  {Marko}},\ }\bibfield  {title} {\bibinfo {title} {Self-organization of domain
  structures by {DNA}-loop-extruding enzymes},\ }\href@noop {} {\bibfield
  {journal} {\bibinfo  {journal} {Nucleic Acids Res.}\ }\textbf {\bibinfo
  {volume} {40}},\ \bibinfo {pages} {11202} (\bibinfo {year}
  {2012})}\BibitemShut {NoStop}%
\bibitem [{\citenamefont {Sanborn}\ \emph {et~al.}(2015)\citenamefont
  {Sanborn}, \citenamefont {Rao}, \citenamefont {Huang}, \citenamefont
  {Durand}, \citenamefont {Huntley}, \citenamefont {Jewett}, \citenamefont
  {Bochkov}, \citenamefont {Chinnappan}, \citenamefont {Cutkosky},
  \citenamefont {Li} \emph {et~al.}}]{sanborn2015}%
  \BibitemOpen
  \bibfield  {author} {\bibinfo {author} {\bibfnamefont {A.~L.}\ \bibnamefont
  {Sanborn}}, \bibinfo {author} {\bibfnamefont {S.~S.}\ \bibnamefont {Rao}},
  \bibinfo {author} {\bibfnamefont {S.-C.}\ \bibnamefont {Huang}}, \bibinfo
  {author} {\bibfnamefont {N.~C.}\ \bibnamefont {Durand}}, \bibinfo {author}
  {\bibfnamefont {M.~H.}\ \bibnamefont {Huntley}}, \bibinfo {author}
  {\bibfnamefont {A.~I.}\ \bibnamefont {Jewett}}, \bibinfo {author}
  {\bibfnamefont {I.~D.}\ \bibnamefont {Bochkov}}, \bibinfo {author}
  {\bibfnamefont {D.}~\bibnamefont {Chinnappan}}, \bibinfo {author}
  {\bibfnamefont {A.}~\bibnamefont {Cutkosky}}, \bibinfo {author}
  {\bibfnamefont {J.}~\bibnamefont {Li}}, \emph {et~al.},\ }\bibfield  {title}
  {\bibinfo {title} {Chromatin extrusion explains key features of loop and
  domain formation in wild-type and engineered genomes},\ }\href@noop {}
  {\bibfield  {journal} {\bibinfo  {journal} {Proc. Natl. Acad. Sci. USA}\
  }\textbf {\bibinfo {volume} {112}},\ \bibinfo {pages} {E6456} (\bibinfo
  {year} {2015})}\BibitemShut {NoStop}%
\bibitem [{\citenamefont {Fudenberg}\ \emph {et~al.}(2016)\citenamefont
  {Fudenberg}, \citenamefont {Imakaev}, \citenamefont {Lu}, \citenamefont
  {Goloborodko}, \citenamefont {Abdennur},\ and\ \citenamefont
  {Mirny}}]{fudenberg2016}%
  \BibitemOpen
  \bibfield  {author} {\bibinfo {author} {\bibfnamefont {G.}~\bibnamefont
  {Fudenberg}}, \bibinfo {author} {\bibfnamefont {M.}~\bibnamefont {Imakaev}},
  \bibinfo {author} {\bibfnamefont {C.}~\bibnamefont {Lu}}, \bibinfo {author}
  {\bibfnamefont {A.}~\bibnamefont {Goloborodko}}, \bibinfo {author}
  {\bibfnamefont {N.}~\bibnamefont {Abdennur}},\ and\ \bibinfo {author}
  {\bibfnamefont {L.~A.}\ \bibnamefont {Mirny}},\ }\bibfield  {title} {\bibinfo
  {title} {Formation of chromosomal domains by loop extrusion},\ }\href@noop {}
  {\bibfield  {journal} {\bibinfo  {journal} {Cell Rep.}\ }\textbf {\bibinfo
  {volume} {15}},\ \bibinfo {pages} {2038} (\bibinfo {year}
  {2016})}\BibitemShut {NoStop}%
\bibitem [{\citenamefont {Goloborodko}\ \emph
  {et~al.}(2016{\natexlab{a}})\citenamefont {Goloborodko}, \citenamefont
  {Marko},\ and\ \citenamefont {Mirny}}]{goloborodko2016.1}%
  \BibitemOpen
  \bibfield  {author} {\bibinfo {author} {\bibfnamefont {A.}~\bibnamefont
  {Goloborodko}}, \bibinfo {author} {\bibfnamefont {J.~F.}\ \bibnamefont
  {Marko}},\ and\ \bibinfo {author} {\bibfnamefont {L.~A.}\ \bibnamefont
  {Mirny}},\ }\bibfield  {title} {\bibinfo {title} {Chromosome compaction by
  active loop extrusion},\ }\href@noop {} {\bibfield  {journal} {\bibinfo
  {journal} {Biophys. J.}\ }\textbf {\bibinfo {volume} {110}},\ \bibinfo
  {pages} {2162} (\bibinfo {year} {2016}{\natexlab{a}})}\BibitemShut {NoStop}%
\bibitem [{\citenamefont {Goloborodko}\ \emph
  {et~al.}(2016{\natexlab{b}})\citenamefont {Goloborodko}, \citenamefont
  {Imakaev}, \citenamefont {Marko},\ and\ \citenamefont
  {Mirny}}]{goloborodko2016.2}%
  \BibitemOpen
  \bibfield  {author} {\bibinfo {author} {\bibfnamefont {A.}~\bibnamefont
  {Goloborodko}}, \bibinfo {author} {\bibfnamefont {M.~V.}\ \bibnamefont
  {Imakaev}}, \bibinfo {author} {\bibfnamefont {J.~F.}\ \bibnamefont {Marko}},\
  and\ \bibinfo {author} {\bibfnamefont {L.}~\bibnamefont {Mirny}},\ }\bibfield
   {title} {\bibinfo {title} {Compaction and segregation of sister chromatids
  via active loop extrusion},\ }\href@noop {} {\bibfield  {journal} {\bibinfo
  {journal} {Elife}\ }\textbf {\bibinfo {volume} {5}},\ \bibinfo {pages}
  {e14864} (\bibinfo {year} {2016}{\natexlab{b}})}\BibitemShut {NoStop}%
\bibitem [{\citenamefont {Nuebler}\ \emph {et~al.}(2018)\citenamefont
  {Nuebler}, \citenamefont {Fudenberg}, \citenamefont {Imakaev}, \citenamefont
  {Abdennur},\ and\ \citenamefont {Mirny}}]{nuebler2018}%
  \BibitemOpen
  \bibfield  {author} {\bibinfo {author} {\bibfnamefont {J.}~\bibnamefont
  {Nuebler}}, \bibinfo {author} {\bibfnamefont {G.}~\bibnamefont {Fudenberg}},
  \bibinfo {author} {\bibfnamefont {M.}~\bibnamefont {Imakaev}}, \bibinfo
  {author} {\bibfnamefont {N.}~\bibnamefont {Abdennur}},\ and\ \bibinfo
  {author} {\bibfnamefont {L.~A.}\ \bibnamefont {Mirny}},\ }\bibfield  {title}
  {\bibinfo {title} {Chromatin organization by an interplay of loop extrusion
  and compartmental segregation},\ }\href@noop {} {\bibfield  {journal}
  {\bibinfo  {journal} {Proc. Natl. Acad. Sci. USA}\ }\textbf {\bibinfo
  {volume} {115}},\ \bibinfo {pages} {E6697} (\bibinfo {year}
  {2018})}\BibitemShut {NoStop}%
\bibitem [{\citenamefont {Banigan}\ and\ \citenamefont
  {Mirny}(2020)}]{banigan2020}%
  \BibitemOpen
  \bibfield  {author} {\bibinfo {author} {\bibfnamefont {E.~J.}\ \bibnamefont
  {Banigan}}\ and\ \bibinfo {author} {\bibfnamefont {L.~A.}\ \bibnamefont
  {Mirny}},\ }\bibfield  {title} {\bibinfo {title} {Loop extrusion: theory
  meets single-molecule experiments},\ }\href@noop {} {\bibfield  {journal}
  {\bibinfo  {journal} {Curr. Opin. Cell Biol.}\ }\textbf {\bibinfo {volume}
  {64}},\ \bibinfo {pages} {124} (\bibinfo {year} {2020})}\BibitemShut
  {NoStop}%
\bibitem [{\citenamefont {Davidson}\ and\ \citenamefont
  {Peters}(2021)}]{davidson2021}%
  \BibitemOpen
  \bibfield  {author} {\bibinfo {author} {\bibfnamefont {I.~F.}\ \bibnamefont
  {Davidson}}\ and\ \bibinfo {author} {\bibfnamefont {J.-M.}\ \bibnamefont
  {Peters}},\ }\bibfield  {title} {\bibinfo {title} {Genome folding through
  loop extrusion by {SMC} complexes},\ }\href@noop {} {\bibfield  {journal}
  {\bibinfo  {journal} {Nat. Rev. Mol. Cell Biol.}\ }\textbf {\bibinfo {volume}
  {22}},\ \bibinfo {pages} {445} (\bibinfo {year} {2021})}\BibitemShut
  {NoStop}%
\bibitem [{\citenamefont {Oudelaar}\ and\ \citenamefont
  {Higgs}(2021)}]{oudelaar2021}%
  \BibitemOpen
  \bibfield  {author} {\bibinfo {author} {\bibfnamefont {A.~M.}\ \bibnamefont
  {Oudelaar}}\ and\ \bibinfo {author} {\bibfnamefont {D.~R.}\ \bibnamefont
  {Higgs}},\ }\bibfield  {title} {\bibinfo {title} {The relationship between
  genome structure and function},\ }\href@noop {} {\bibfield  {journal}
  {\bibinfo  {journal} {Nat. Rev. Genet.}\ }\textbf {\bibinfo {volume} {22}},\
  \bibinfo {pages} {154} (\bibinfo {year} {2021})}\BibitemShut {NoStop}%
\bibitem [{\citenamefont {Bonev}\ \emph {et~al.}(2017)\citenamefont {Bonev},
  \citenamefont {Cohen}, \citenamefont {Szabo}, \citenamefont {Fritsch},
  \citenamefont {Papadopoulos}, \citenamefont {Lubling}, \citenamefont {Xu},
  \citenamefont {Lv}, \citenamefont {Hugnot}, \citenamefont {Tanay} \emph
  {et~al.}}]{bonev2017}%
  \BibitemOpen
  \bibfield  {author} {\bibinfo {author} {\bibfnamefont {B.}~\bibnamefont
  {Bonev}}, \bibinfo {author} {\bibfnamefont {N.~M.}\ \bibnamefont {Cohen}},
  \bibinfo {author} {\bibfnamefont {Q.}~\bibnamefont {Szabo}}, \bibinfo
  {author} {\bibfnamefont {L.}~\bibnamefont {Fritsch}}, \bibinfo {author}
  {\bibfnamefont {G.~L.}\ \bibnamefont {Papadopoulos}}, \bibinfo {author}
  {\bibfnamefont {Y.}~\bibnamefont {Lubling}}, \bibinfo {author} {\bibfnamefont
  {X.}~\bibnamefont {Xu}}, \bibinfo {author} {\bibfnamefont {X.}~\bibnamefont
  {Lv}}, \bibinfo {author} {\bibfnamefont {J.-P.}\ \bibnamefont {Hugnot}},
  \bibinfo {author} {\bibfnamefont {A.}~\bibnamefont {Tanay}}, \emph {et~al.},\
  }\bibfield  {title} {\bibinfo {title} {Multiscale 3{D} genome rewiring during
  mouse neural development},\ }\href@noop {} {\bibfield  {journal} {\bibinfo
  {journal} {Cell}\ }\textbf {\bibinfo {volume} {171}},\ \bibinfo {pages} {557}
  (\bibinfo {year} {2017})}\BibitemShut {NoStop}%
\bibitem [{\citenamefont {Rao}\ \emph {et~al.}(2017)\citenamefont {Rao},
  \citenamefont {Huang}, \citenamefont {St~Hilaire}, \citenamefont {Engreitz},
  \citenamefont {Perez}, \citenamefont {Kieffer-Kwon}, \citenamefont {Sanborn},
  \citenamefont {Johnstone}, \citenamefont {Bascom}, \citenamefont {Bochkov}
  \emph {et~al.}}]{rao2017}%
  \BibitemOpen
  \bibfield  {author} {\bibinfo {author} {\bibfnamefont {S.~S.}\ \bibnamefont
  {Rao}}, \bibinfo {author} {\bibfnamefont {S.-C.}\ \bibnamefont {Huang}},
  \bibinfo {author} {\bibfnamefont {B.~G.}\ \bibnamefont {St~Hilaire}},
  \bibinfo {author} {\bibfnamefont {J.~M.}\ \bibnamefont {Engreitz}}, \bibinfo
  {author} {\bibfnamefont {E.~M.}\ \bibnamefont {Perez}}, \bibinfo {author}
  {\bibfnamefont {K.-R.}\ \bibnamefont {Kieffer-Kwon}}, \bibinfo {author}
  {\bibfnamefont {A.~L.}\ \bibnamefont {Sanborn}}, \bibinfo {author}
  {\bibfnamefont {S.~E.}\ \bibnamefont {Johnstone}}, \bibinfo {author}
  {\bibfnamefont {G.~D.}\ \bibnamefont {Bascom}}, \bibinfo {author}
  {\bibfnamefont {I.~D.}\ \bibnamefont {Bochkov}}, \emph {et~al.},\ }\bibfield
  {title} {\bibinfo {title} {Cohesin loss eliminates all loop domains},\
  }\href@noop {} {\bibfield  {journal} {\bibinfo  {journal} {Cell}\ }\textbf
  {\bibinfo {volume} {171}},\ \bibinfo {pages} {305} (\bibinfo {year}
  {2017})}\BibitemShut {NoStop}%
\bibitem [{\citenamefont {Wutz}\ \emph {et~al.}(2017)\citenamefont {Wutz},
  \citenamefont {V{\'a}rnai}, \citenamefont {Nagasaka}, \citenamefont
  {Cisneros}, \citenamefont {Stocsits}, \citenamefont {Tang}, \citenamefont
  {Schoenfelder}, \citenamefont {Jessberger}, \citenamefont {Muhar},
  \citenamefont {Hossain} \emph {et~al.}}]{wutz2017}%
  \BibitemOpen
  \bibfield  {author} {\bibinfo {author} {\bibfnamefont {G.}~\bibnamefont
  {Wutz}}, \bibinfo {author} {\bibfnamefont {C.}~\bibnamefont {V{\'a}rnai}},
  \bibinfo {author} {\bibfnamefont {K.}~\bibnamefont {Nagasaka}}, \bibinfo
  {author} {\bibfnamefont {D.~A.}\ \bibnamefont {Cisneros}}, \bibinfo {author}
  {\bibfnamefont {R.~R.}\ \bibnamefont {Stocsits}}, \bibinfo {author}
  {\bibfnamefont {W.}~\bibnamefont {Tang}}, \bibinfo {author} {\bibfnamefont
  {S.}~\bibnamefont {Schoenfelder}}, \bibinfo {author} {\bibfnamefont
  {G.}~\bibnamefont {Jessberger}}, \bibinfo {author} {\bibfnamefont
  {M.}~\bibnamefont {Muhar}}, \bibinfo {author} {\bibfnamefont {M.~J.}\
  \bibnamefont {Hossain}}, \emph {et~al.},\ }\bibfield  {title} {\bibinfo
  {title} {{Topologically associating domains and chromatin loops depend on
  cohesin and are regulated by CTCF, WAPL, and PDS5 proteins}},\ }\href@noop {}
  {\bibfield  {journal} {\bibinfo  {journal} {EMBO J.}\ }\textbf {\bibinfo
  {volume} {36}},\ \bibinfo {pages} {3573} (\bibinfo {year}
  {2017})}\BibitemShut {NoStop}%
\bibitem [{\citenamefont {Zhang}\ and\ \citenamefont
  {Kutateladze}(2019)}]{zhang2019}%
  \BibitemOpen
  \bibfield  {author} {\bibinfo {author} {\bibfnamefont {Y.}~\bibnamefont
  {Zhang}}\ and\ \bibinfo {author} {\bibfnamefont {T.~G.}\ \bibnamefont
  {Kutateladze}},\ }\bibfield  {title} {\bibinfo {title} {Liquid--liquid phase
  separation is an intrinsic physicochemical property of chromatin},\
  }\href@noop {} {\bibfield  {journal} {\bibinfo  {journal} {Nat. Struct. Mol.
  Biol.}\ }\textbf {\bibinfo {volume} {26}},\ \bibinfo {pages} {1085} (\bibinfo
  {year} {2019})}\BibitemShut {NoStop}%
\bibitem [{\citenamefont {Misteli}(2020)}]{misteli2020}%
  \BibitemOpen
  \bibfield  {author} {\bibinfo {author} {\bibfnamefont {T.}~\bibnamefont
  {Misteli}},\ }\bibfield  {title} {\bibinfo {title} {The self-organizing
  genome: principles of genome architecture and function},\ }\href@noop {}
  {\bibfield  {journal} {\bibinfo  {journal} {Cell}\ }\textbf {\bibinfo
  {volume} {183}},\ \bibinfo {pages} {28} (\bibinfo {year} {2020})}\BibitemShut
  {NoStop}%
\bibitem [{\citenamefont {Ahn}\ \emph {et~al.}(2021)\citenamefont {Ahn},
  \citenamefont {Davis}, \citenamefont {Daugird}, \citenamefont {Zhao},
  \citenamefont {Quiroga}, \citenamefont {Uryu}, \citenamefont {Li},
  \citenamefont {Storey}, \citenamefont {Tsai}, \citenamefont {Keeley} \emph
  {et~al.}}]{ahn2021}%
  \BibitemOpen
  \bibfield  {author} {\bibinfo {author} {\bibfnamefont {J.~H.}\ \bibnamefont
  {Ahn}}, \bibinfo {author} {\bibfnamefont {E.~S.}\ \bibnamefont {Davis}},
  \bibinfo {author} {\bibfnamefont {T.~A.}\ \bibnamefont {Daugird}}, \bibinfo
  {author} {\bibfnamefont {S.}~\bibnamefont {Zhao}}, \bibinfo {author}
  {\bibfnamefont {I.~Y.}\ \bibnamefont {Quiroga}}, \bibinfo {author}
  {\bibfnamefont {H.}~\bibnamefont {Uryu}}, \bibinfo {author} {\bibfnamefont
  {J.}~\bibnamefont {Li}}, \bibinfo {author} {\bibfnamefont {A.~J.}\
  \bibnamefont {Storey}}, \bibinfo {author} {\bibfnamefont {Y.-H.}\
  \bibnamefont {Tsai}}, \bibinfo {author} {\bibfnamefont {D.~P.}\ \bibnamefont
  {Keeley}}, \emph {et~al.},\ }\bibfield  {title} {\bibinfo {title} {Phase
  separation drives aberrant chromatin looping and cancer development},\
  }\href@noop {} {\bibfield  {journal} {\bibinfo  {journal} {Nature}\ }\textbf
  {\bibinfo {volume} {595}},\ \bibinfo {pages} {591} (\bibinfo {year}
  {2021})}\BibitemShut {NoStop}%
\bibitem [{\citenamefont {Erdel}\ and\ \citenamefont
  {Rippe}(2018)}]{erdel2018}%
  \BibitemOpen
  \bibfield  {author} {\bibinfo {author} {\bibfnamefont {F.}~\bibnamefont
  {Erdel}}\ and\ \bibinfo {author} {\bibfnamefont {K.}~\bibnamefont {Rippe}},\
  }\bibfield  {title} {\bibinfo {title} {Formation of chromatin subcompartments
  by phase separation},\ }\href@noop {} {\bibfield  {journal} {\bibinfo
  {journal} {Biophys. J.}\ }\textbf {\bibinfo {volume} {114}},\ \bibinfo
  {pages} {2262} (\bibinfo {year} {2018})}\BibitemShut {NoStop}%
\bibitem [{\citenamefont {Weber}\ \emph
  {et~al.}(2010{\natexlab{b}})\citenamefont {Weber}, \citenamefont {Theriot},\
  and\ \citenamefont {Spakowitz}}]{weber2010b}%
  \BibitemOpen
  \bibfield  {author} {\bibinfo {author} {\bibfnamefont {S.~C.}\ \bibnamefont
  {Weber}}, \bibinfo {author} {\bibfnamefont {J.~A.}\ \bibnamefont {Theriot}},\
  and\ \bibinfo {author} {\bibfnamefont {A.~J.}\ \bibnamefont {Spakowitz}},\
  }\bibfield  {title} {\bibinfo {title} {Subdiffusive motion of a polymer
  composed of subdiffusive monomers},\ }\href@noop {} {\bibfield  {journal}
  {\bibinfo  {journal} {Phys. Rev. E}\ }\textbf {\bibinfo {volume} {82}},\
  \bibinfo {pages} {011913} (\bibinfo {year} {2010}{\natexlab{b}})}\BibitemShut
  {NoStop}%
\bibitem [{\citenamefont {Gillespie}(1996)}]{gillespie1996}%
  \BibitemOpen
  \bibfield  {author} {\bibinfo {author} {\bibfnamefont {D.~T.}\ \bibnamefont
  {Gillespie}},\ }\bibfield  {title} {\bibinfo {title} {The mathematics of
  {Brownian} motion and {Johnson} noise},\ }\href@noop {} {\bibfield  {journal}
  {\bibinfo  {journal} {Am. J. Phys.}\ }\textbf {\bibinfo {volume} {64}},\
  \bibinfo {pages} {225} (\bibinfo {year} {1996})}\BibitemShut {NoStop}%
\bibitem [{\citenamefont {Bailey}\ \emph {et~al.}(ress)\citenamefont {Bailey},
  \citenamefont {Surovtsev}, \citenamefont {Williams}, \citenamefont {Yan},
  \citenamefont {Yuan}, \citenamefont {Mochrie},\ and\ \citenamefont
  {King}}]{bailey2023}%
  \BibitemOpen
  \bibfield  {author} {\bibinfo {author} {\bibfnamefont {M.~L.~P.}\
  \bibnamefont {Bailey}}, \bibinfo {author} {\bibfnamefont {I.}~\bibnamefont
  {Surovtsev}}, \bibinfo {author} {\bibfnamefont {J.~F.}\ \bibnamefont
  {Williams}}, \bibinfo {author} {\bibfnamefont {H.}~\bibnamefont {Yan}},
  \bibinfo {author} {\bibfnamefont {T.}~\bibnamefont {Yuan}}, \bibinfo {author}
  {\bibfnamefont {S.~G.}\ \bibnamefont {Mochrie}},\ and\ \bibinfo {author}
  {\bibfnamefont {M.~C.}\ \bibnamefont {King}},\ }\bibfield  {title} {\bibinfo
  {title} {Chromatin dynamics are constrained by loops and driven by the
  {INO80} nucleosome remodeler},\ }\href@noop {} {\bibfield  {journal}
  {\bibinfo  {journal} {Mol. Biol. Cell}\ } (\bibinfo {year} {in
  press})}\BibitemShut {NoStop}%
\bibitem [{\citenamefont {Hinnebusch}\ and\ \citenamefont
  {Bendich}(1997)}]{hinnebusch1997}%
  \BibitemOpen
  \bibfield  {author} {\bibinfo {author} {\bibfnamefont {B.~J.}\ \bibnamefont
  {Hinnebusch}}\ and\ \bibinfo {author} {\bibfnamefont {A.~J.}\ \bibnamefont
  {Bendich}},\ }\bibfield  {title} {\bibinfo {title} {The bacterial nucleoid
  visualized by fluorescence microscopy of cells lysed within agarose:
  comparison of {Escherichia coli} and spirochetes of the genus {Borrelia}},\
  }\href@noop {} {\bibfield  {journal} {\bibinfo  {journal} {J. Bacteriol.}\
  }\textbf {\bibinfo {volume} {179}},\ \bibinfo {pages} {2228} (\bibinfo {year}
  {1997})}\BibitemShut {NoStop}%
\bibitem [{\citenamefont {Macvanin}\ and\ \citenamefont
  {Adhya}(2012)}]{macvanin2012}%
  \BibitemOpen
  \bibfield  {author} {\bibinfo {author} {\bibfnamefont {M.}~\bibnamefont
  {Macvanin}}\ and\ \bibinfo {author} {\bibfnamefont {S.}~\bibnamefont
  {Adhya}},\ }\bibfield  {title} {\bibinfo {title} {Architectural organization
  in {E. coli} nucleoid},\ }\href@noop {} {\bibfield  {journal} {\bibinfo
  {journal} {Biochim. Biophys. Acta. Gene Regul. Mech.}\ }\textbf {\bibinfo
  {volume} {1819}},\ \bibinfo {pages} {830} (\bibinfo {year}
  {2012})}\BibitemShut {NoStop}%
\bibitem [{\citenamefont {Newport}\ and\ \citenamefont
  {Yan}(1996)}]{newport1996}%
  \BibitemOpen
  \bibfield  {author} {\bibinfo {author} {\bibfnamefont {J.}~\bibnamefont
  {Newport}}\ and\ \bibinfo {author} {\bibfnamefont {H.}~\bibnamefont {Yan}},\
  }\bibfield  {title} {\bibinfo {title} {Organization of {DNA} into foci during
  replication},\ }\href@noop {} {\bibfield  {journal} {\bibinfo  {journal}
  {Curr. Opin. Cell Biol.}\ }\textbf {\bibinfo {volume} {8}},\ \bibinfo {pages}
  {365} (\bibinfo {year} {1996})}\BibitemShut {NoStop}%
\bibitem [{\citenamefont {Ma}\ \emph {et~al.}(1998)\citenamefont {Ma},
  \citenamefont {Samarabandu}, \citenamefont {Devdhar}, \citenamefont
  {Acharya}, \citenamefont {Cheng}, \citenamefont {Meng},\ and\ \citenamefont
  {Berezney}}]{ma1998}%
  \BibitemOpen
  \bibfield  {author} {\bibinfo {author} {\bibfnamefont {H.}~\bibnamefont
  {Ma}}, \bibinfo {author} {\bibfnamefont {J.}~\bibnamefont {Samarabandu}},
  \bibinfo {author} {\bibfnamefont {R.~S.}\ \bibnamefont {Devdhar}}, \bibinfo
  {author} {\bibfnamefont {R.}~\bibnamefont {Acharya}}, \bibinfo {author}
  {\bibfnamefont {P.-c.}\ \bibnamefont {Cheng}}, \bibinfo {author}
  {\bibfnamefont {C.}~\bibnamefont {Meng}},\ and\ \bibinfo {author}
  {\bibfnamefont {R.}~\bibnamefont {Berezney}},\ }\bibfield  {title} {\bibinfo
  {title} {Spatial and temporal dynamics of {DNA} replication sites in
  mammalian cells},\ }\href@noop {} {\bibfield  {journal} {\bibinfo  {journal}
  {J. Cell Biol.}\ }\textbf {\bibinfo {volume} {143}},\ \bibinfo {pages} {1415}
  (\bibinfo {year} {1998})}\BibitemShut {NoStop}%
\bibitem [{\citenamefont {Jackson}\ and\ \citenamefont
  {Pombo}(1998)}]{jackson1998}%
  \BibitemOpen
  \bibfield  {author} {\bibinfo {author} {\bibfnamefont {D.~A.}\ \bibnamefont
  {Jackson}}\ and\ \bibinfo {author} {\bibfnamefont {A.}~\bibnamefont
  {Pombo}},\ }\bibfield  {title} {\bibinfo {title} {Replicon clusters are
  stable units of chromosome structure: evidence that nuclear organization
  contributes to the efficient activation and propagation of {S} phase in human
  cells},\ }\href@noop {} {\bibfield  {journal} {\bibinfo  {journal} {J. Cell
  Biol.}\ }\textbf {\bibinfo {volume} {140}},\ \bibinfo {pages} {1285}
  (\bibinfo {year} {1998})}\BibitemShut {NoStop}%
\bibitem [{\citenamefont {Frouin}\ \emph {et~al.}(2003)\citenamefont {Frouin},
  \citenamefont {Montecucco}, \citenamefont {Spadari},\ and\ \citenamefont
  {Maga}}]{frouin2003}%
  \BibitemOpen
  \bibfield  {author} {\bibinfo {author} {\bibfnamefont {I.}~\bibnamefont
  {Frouin}}, \bibinfo {author} {\bibfnamefont {A.}~\bibnamefont {Montecucco}},
  \bibinfo {author} {\bibfnamefont {S.}~\bibnamefont {Spadari}},\ and\ \bibinfo
  {author} {\bibfnamefont {G.}~\bibnamefont {Maga}},\ }\bibfield  {title}
  {\bibinfo {title} {{DNA} replication: a complex matter},\ }\href@noop {}
  {\bibfield  {journal} {\bibinfo  {journal} {EMBO Rep.}\ }\textbf {\bibinfo
  {volume} {4}},\ \bibinfo {pages} {666} (\bibinfo {year} {2003})}\BibitemShut
  {NoStop}%
\bibitem [{\citenamefont {Guillou}\ \emph {et~al.}(2010)\citenamefont
  {Guillou}, \citenamefont {Ibarra}, \citenamefont {Coulon}, \citenamefont
  {Casado-Vela}, \citenamefont {Rico}, \citenamefont {Casal}, \citenamefont
  {Schwob}, \citenamefont {Losada},\ and\ \citenamefont
  {M{\'e}ndez}}]{guillou2010}%
  \BibitemOpen
  \bibfield  {author} {\bibinfo {author} {\bibfnamefont {E.}~\bibnamefont
  {Guillou}}, \bibinfo {author} {\bibfnamefont {A.}~\bibnamefont {Ibarra}},
  \bibinfo {author} {\bibfnamefont {V.}~\bibnamefont {Coulon}}, \bibinfo
  {author} {\bibfnamefont {J.}~\bibnamefont {Casado-Vela}}, \bibinfo {author}
  {\bibfnamefont {D.}~\bibnamefont {Rico}}, \bibinfo {author} {\bibfnamefont
  {I.}~\bibnamefont {Casal}}, \bibinfo {author} {\bibfnamefont
  {E.}~\bibnamefont {Schwob}}, \bibinfo {author} {\bibfnamefont
  {A.}~\bibnamefont {Losada}},\ and\ \bibinfo {author} {\bibfnamefont
  {J.}~\bibnamefont {M{\'e}ndez}},\ }\bibfield  {title} {\bibinfo {title}
  {Cohesin organizes chromatin loops at {DNA} replication factories},\
  }\href@noop {} {\bibfield  {journal} {\bibinfo  {journal} {Genes Dev.}\
  }\textbf {\bibinfo {volume} {24}},\ \bibinfo {pages} {2812} (\bibinfo {year}
  {2010})}\BibitemShut {NoStop}%
\bibitem [{\citenamefont {Saner}\ \emph {et~al.}(2013)\citenamefont {Saner},
  \citenamefont {Karschau}, \citenamefont {Natsume}, \citenamefont
  {Gierli{\'n}ski}, \citenamefont {Retkute}, \citenamefont {Hawkins},
  \citenamefont {Nieduszynski}, \citenamefont {Blow}, \citenamefont
  {de~Moura},\ and\ \citenamefont {Tanaka}}]{saner2013}%
  \BibitemOpen
  \bibfield  {author} {\bibinfo {author} {\bibfnamefont {N.}~\bibnamefont
  {Saner}}, \bibinfo {author} {\bibfnamefont {J.}~\bibnamefont {Karschau}},
  \bibinfo {author} {\bibfnamefont {T.}~\bibnamefont {Natsume}}, \bibinfo
  {author} {\bibfnamefont {M.}~\bibnamefont {Gierli{\'n}ski}}, \bibinfo
  {author} {\bibfnamefont {R.}~\bibnamefont {Retkute}}, \bibinfo {author}
  {\bibfnamefont {M.}~\bibnamefont {Hawkins}}, \bibinfo {author} {\bibfnamefont
  {C.~A.}\ \bibnamefont {Nieduszynski}}, \bibinfo {author} {\bibfnamefont
  {J.~J.}\ \bibnamefont {Blow}}, \bibinfo {author} {\bibfnamefont {A.~P.}\
  \bibnamefont {de~Moura}},\ and\ \bibinfo {author} {\bibfnamefont {T.~U.}\
  \bibnamefont {Tanaka}},\ }\bibfield  {title} {\bibinfo {title} {Stochastic
  association of neighboring replicons creates replication factories in budding
  yeast},\ }\href@noop {} {\bibfield  {journal} {\bibinfo  {journal} {J. Cell
  Biol.}\ }\textbf {\bibinfo {volume} {202}},\ \bibinfo {pages} {1001}
  (\bibinfo {year} {2013})}\BibitemShut {NoStop}%
\bibitem [{\citenamefont {Mangiameli}\ \emph {et~al.}(2018)\citenamefont
  {Mangiameli}, \citenamefont {Cass}, \citenamefont {Merrikh},\ and\
  \citenamefont {Wiggins}}]{mangiameli2018}%
  \BibitemOpen
  \bibfield  {author} {\bibinfo {author} {\bibfnamefont {S.~M.}\ \bibnamefont
  {Mangiameli}}, \bibinfo {author} {\bibfnamefont {J.~A.}\ \bibnamefont
  {Cass}}, \bibinfo {author} {\bibfnamefont {H.}~\bibnamefont {Merrikh}},\ and\
  \bibinfo {author} {\bibfnamefont {P.~A.}\ \bibnamefont {Wiggins}},\
  }\bibfield  {title} {\bibinfo {title} {The bacterial replisome has
  factory-like localization},\ }\href@noop {} {\bibfield  {journal} {\bibinfo
  {journal} {Curr. Genet.}\ }\textbf {\bibinfo {volume} {64}},\ \bibinfo
  {pages} {1029} (\bibinfo {year} {2018})}\BibitemShut {NoStop}%
\bibitem [{\citenamefont {Geiger}\ \emph {et~al.}(2021)\citenamefont {Geiger},
  \citenamefont {Acker}, \citenamefont {Papa}, \citenamefont {Wang},
  \citenamefont {Arter}, \citenamefont {Saar}, \citenamefont {Erkamp},
  \citenamefont {Qi}, \citenamefont {Bravo}, \citenamefont {Strauss} \emph
  {et~al.}}]{geiger2021}%
  \BibitemOpen
  \bibfield  {author} {\bibinfo {author} {\bibfnamefont {F.}~\bibnamefont
  {Geiger}}, \bibinfo {author} {\bibfnamefont {J.}~\bibnamefont {Acker}},
  \bibinfo {author} {\bibfnamefont {G.}~\bibnamefont {Papa}}, \bibinfo {author}
  {\bibfnamefont {X.}~\bibnamefont {Wang}}, \bibinfo {author} {\bibfnamefont
  {W.~E.}\ \bibnamefont {Arter}}, \bibinfo {author} {\bibfnamefont {K.~L.}\
  \bibnamefont {Saar}}, \bibinfo {author} {\bibfnamefont {N.~A.}\ \bibnamefont
  {Erkamp}}, \bibinfo {author} {\bibfnamefont {R.}~\bibnamefont {Qi}}, \bibinfo
  {author} {\bibfnamefont {J.~P.}\ \bibnamefont {Bravo}}, \bibinfo {author}
  {\bibfnamefont {S.}~\bibnamefont {Strauss}}, \emph {et~al.},\ }\bibfield
  {title} {\bibinfo {title} {Liquid--liquid phase separation underpins the
  formation of replication factories in rotaviruses},\ }\href@noop {}
  {\bibfield  {journal} {\bibinfo  {journal} {EMBO J.}\ }\textbf {\bibinfo
  {volume} {40}},\ \bibinfo {pages} {e107711} (\bibinfo {year}
  {2021})}\BibitemShut {NoStop}%
\bibitem [{\citenamefont {Weber}(2019)}]{weber2019}%
  \BibitemOpen
  \bibfield  {author} {\bibinfo {author} {\bibfnamefont {S.~C.}\ \bibnamefont
  {Weber}},\ }\bibfield  {title} {\bibinfo {title} {Evidence for and against
  liquid-liquid phase separation in the nucleus},\ }\href@noop {} {\bibfield
  {journal} {\bibinfo  {journal} {Noncoding RNA}\ }\textbf {\bibinfo {volume}
  {5}},\ \bibinfo {pages} {50} (\bibinfo {year} {2019})}\BibitemShut {NoStop}%
\bibitem [{\citenamefont {Strom}\ and\ \citenamefont
  {Brangwynne}(2019)}]{strom2019}%
  \BibitemOpen
  \bibfield  {author} {\bibinfo {author} {\bibfnamefont {A.~R.}\ \bibnamefont
  {Strom}}\ and\ \bibinfo {author} {\bibfnamefont {C.~P.}\ \bibnamefont
  {Brangwynne}},\ }\bibfield  {title} {\bibinfo {title} {The liquid
  nucleome--phase transitions in the nucleus at a glance},\ }\href@noop {}
  {\bibfield  {journal} {\bibinfo  {journal} {J. Cell Sci.}\ }\textbf {\bibinfo
  {volume} {132}},\ \bibinfo {pages} {jcs235093} (\bibinfo {year}
  {2019})}\BibitemShut {NoStop}%
\bibitem [{\citenamefont {Razin}\ and\ \citenamefont
  {Gavrilov}(2020)}]{razin2020}%
  \BibitemOpen
  \bibfield  {author} {\bibinfo {author} {\bibfnamefont {S.}~\bibnamefont
  {Razin}}\ and\ \bibinfo {author} {\bibfnamefont {A.}~\bibnamefont
  {Gavrilov}},\ }\bibfield  {title} {\bibinfo {title} {The role of
  liquid--liquid phase separation in the compartmentalization of cell nucleus
  and spatial genome organization},\ }\href@noop {} {\bibfield  {journal}
  {\bibinfo  {journal} {Biochem. (Mosc.)}\ }\textbf {\bibinfo {volume} {85}},\
  \bibinfo {pages} {643} (\bibinfo {year} {2020})}\BibitemShut {NoStop}%
\bibitem [{\citenamefont {Strobl}\ and\ \citenamefont
  {Strobl}(1997)}]{strobl1997}%
  \BibitemOpen
  \bibfield  {author} {\bibinfo {author} {\bibfnamefont {G.~R.}\ \bibnamefont
  {Strobl}}\ and\ \bibinfo {author} {\bibfnamefont {G.~R.}\ \bibnamefont
  {Strobl}},\ }\href@noop {} {\emph {\bibinfo {title} {The physics of
  polymers}}},\ Vol.~\bibinfo {volume} {2}\ (\bibinfo  {publisher} {Springer},\
  \bibinfo {year} {1997})\BibitemShut {NoStop}%
\bibitem [{\citenamefont {Keesman}\ \emph {et~al.}(2013)\citenamefont
  {Keesman}, \citenamefont {Barkema},\ and\ \citenamefont
  {Panja}}]{keesman2013}%
  \BibitemOpen
  \bibfield  {author} {\bibinfo {author} {\bibfnamefont {R.}~\bibnamefont
  {Keesman}}, \bibinfo {author} {\bibfnamefont {G.~T.}\ \bibnamefont
  {Barkema}},\ and\ \bibinfo {author} {\bibfnamefont {D.}~\bibnamefont
  {Panja}},\ }\bibfield  {title} {\bibinfo {title} {Dynamical eigenmodes of
  star and tadpole polymers},\ }\href@noop {} {\bibfield  {journal} {\bibinfo
  {journal} {J. Stat. Mech.}\ }\textbf {\bibinfo {volume} {2013}},\ \bibinfo
  {pages} {P02021} (\bibinfo {year} {2013})}\BibitemShut {NoStop}%
\bibitem [{\citenamefont {Sato}\ \emph {et~al.}(2021)\citenamefont {Sato},
  \citenamefont {Kwon}, \citenamefont {Matsumiya},\ and\ \citenamefont
  {Watanabe}}]{sato2021}%
  \BibitemOpen
  \bibfield  {author} {\bibinfo {author} {\bibfnamefont {T.}~\bibnamefont
  {Sato}}, \bibinfo {author} {\bibfnamefont {Y.}~\bibnamefont {Kwon}}, \bibinfo
  {author} {\bibfnamefont {Y.}~\bibnamefont {Matsumiya}},\ and\ \bibinfo
  {author} {\bibfnamefont {H.}~\bibnamefont {Watanabe}},\ }\bibfield  {title}
  {\bibinfo {title} {A constitutive equation for {R}ouse model modified for
  variations of spring stiffness, bead friction, and brownian force intensity
  under flow},\ }\href@noop {} {\bibfield  {journal} {\bibinfo  {journal}
  {Phys. Fluids}\ }\textbf {\bibinfo {volume} {33}},\ \bibinfo {pages} {063106}
  (\bibinfo {year} {2021})}\BibitemShut {NoStop}%
\bibitem [{\citenamefont {Doi}\ \emph {et~al.}(1988)\citenamefont {Doi},
  \citenamefont {Edwards},\ and\ \citenamefont {Edwards}}]{doi1988}%
  \BibitemOpen
  \bibfield  {author} {\bibinfo {author} {\bibfnamefont {M.}~\bibnamefont
  {Doi}}, \bibinfo {author} {\bibfnamefont {S.~F.}\ \bibnamefont {Edwards}},\
  and\ \bibinfo {author} {\bibfnamefont {S.~F.}\ \bibnamefont {Edwards}},\
  }\href@noop {} {\emph {\bibinfo {title} {The theory of polymer dynamics}}},\
  Vol.~\bibinfo {volume} {73}\ (\bibinfo  {publisher} {Oxford University
  Press},\ \bibinfo {year} {1988})\BibitemShut {NoStop}%
\bibitem [{\citenamefont {Arbona}\ \emph {et~al.}(2017)\citenamefont {Arbona},
  \citenamefont {Herbert}, \citenamefont {Fabre},\ and\ \citenamefont
  {Zimmer}}]{arbona2017}%
  \BibitemOpen
  \bibfield  {author} {\bibinfo {author} {\bibfnamefont {J.-M.}\ \bibnamefont
  {Arbona}}, \bibinfo {author} {\bibfnamefont {S.}~\bibnamefont {Herbert}},
  \bibinfo {author} {\bibfnamefont {E.}~\bibnamefont {Fabre}},\ and\ \bibinfo
  {author} {\bibfnamefont {C.}~\bibnamefont {Zimmer}},\ }\bibfield  {title}
  {\bibinfo {title} {Inferring the physical properties of yeast chromatin
  through {Bayesian} analysis of whole nucleus simulations},\ }\href@noop {}
  {\bibfield  {journal} {\bibinfo  {journal} {Genome Biol.}\ }\textbf {\bibinfo
  {volume} {18}},\ \bibinfo {pages} {1} (\bibinfo {year} {2017})}\BibitemShut
  {NoStop}%
\bibitem [{\citenamefont {Gillespie}(1977)}]{gillespie1977}%
  \BibitemOpen
  \bibfield  {author} {\bibinfo {author} {\bibfnamefont {D.~T.}\ \bibnamefont
  {Gillespie}},\ }\bibfield  {title} {\bibinfo {title} {Exact stochastic
  simulation of coupled chemical reactions},\ }\href@noop {} {\bibfield
  {journal} {\bibinfo  {journal} {J. Phys. Chem.}\ }\textbf {\bibinfo {volume}
  {81}},\ \bibinfo {pages} {2340} (\bibinfo {year} {1977})}\BibitemShut
  {NoStop}%
\bibitem [{\citenamefont {Gabriele}\ \emph {et~al.}(2022)\citenamefont
  {Gabriele}, \citenamefont {Brand{\~a}o}, \citenamefont {Grosse-Holz},
  \citenamefont {Jha}, \citenamefont {Dailey}, \citenamefont {Cattoglio},
  \citenamefont {Hsieh}, \citenamefont {Mirny}, \citenamefont {Zechner},\ and\
  \citenamefont {Hansen}}]{gabriele2022}%
  \BibitemOpen
  \bibfield  {author} {\bibinfo {author} {\bibfnamefont {M.}~\bibnamefont
  {Gabriele}}, \bibinfo {author} {\bibfnamefont {H.~B.}\ \bibnamefont
  {Brand{\~a}o}}, \bibinfo {author} {\bibfnamefont {S.}~\bibnamefont
  {Grosse-Holz}}, \bibinfo {author} {\bibfnamefont {A.}~\bibnamefont {Jha}},
  \bibinfo {author} {\bibfnamefont {G.~M.}\ \bibnamefont {Dailey}}, \bibinfo
  {author} {\bibfnamefont {C.}~\bibnamefont {Cattoglio}}, \bibinfo {author}
  {\bibfnamefont {T.-H.~S.}\ \bibnamefont {Hsieh}}, \bibinfo {author}
  {\bibfnamefont {L.}~\bibnamefont {Mirny}}, \bibinfo {author} {\bibfnamefont
  {C.}~\bibnamefont {Zechner}},\ and\ \bibinfo {author} {\bibfnamefont {A.~S.}\
  \bibnamefont {Hansen}},\ }\bibfield  {title} {\bibinfo {title} {Dynamics of
  {CTCF}-and cohesin-mediated chromatin looping revealed by live-cell
  imaging},\ }\href@noop {} {\bibfield  {journal} {\bibinfo  {journal}
  {Science}\ }\textbf {\bibinfo {volume} {376}},\ \bibinfo {pages} {496}
  (\bibinfo {year} {2022})}\BibitemShut {NoStop}%
\bibitem [{\citenamefont {Bundschuh}\ and\ \citenamefont
  {Hwa}(2002)}]{bundschuh2002}%
  \BibitemOpen
  \bibfield  {author} {\bibinfo {author} {\bibfnamefont {R.}~\bibnamefont
  {Bundschuh}}\ and\ \bibinfo {author} {\bibfnamefont {T.}~\bibnamefont
  {Hwa}},\ }\bibfield  {title} {\bibinfo {title} {Statistical mechanics of
  secondary structures formed by random rna sequences},\ }\href@noop {}
  {\bibfield  {journal} {\bibinfo  {journal} {Phys. Rev. E}\ }\textbf {\bibinfo
  {volume} {65}},\ \bibinfo {pages} {031903} (\bibinfo {year}
  {2002})}\BibitemShut {NoStop}%
\bibitem [{\citenamefont {Di~Pierro}\ \emph
  {et~al.}(2018{\natexlab{b}})\citenamefont {Di~Pierro}, \citenamefont
  {Potoyan}, \citenamefont {Wolynes},\ and\ \citenamefont
  {Onuchic}}]{dipierro2018}%
  \BibitemOpen
  \bibfield  {author} {\bibinfo {author} {\bibfnamefont {M.}~\bibnamefont
  {Di~Pierro}}, \bibinfo {author} {\bibfnamefont {D.~A.}\ \bibnamefont
  {Potoyan}}, \bibinfo {author} {\bibfnamefont {P.~G.}\ \bibnamefont
  {Wolynes}},\ and\ \bibinfo {author} {\bibfnamefont {J.~N.}\ \bibnamefont
  {Onuchic}},\ }\bibfield  {title} {\bibinfo {title} {Anomalous diffusion,
  spatial coherence, and viscoelasticity from the energy landscape of human
  chromosomes},\ }\href@noop {} {\bibfield  {journal} {\bibinfo  {journal}
  {Proc. Natl. Acad. Sci. USA}\ }\textbf {\bibinfo {volume} {115}},\ \bibinfo
  {pages} {7753} (\bibinfo {year} {2018}{\natexlab{b}})}\BibitemShut {NoStop}%
\bibitem [{\citenamefont {Salari}\ \emph {et~al.}(2022)\citenamefont {Salari},
  \citenamefont {Di~Stefano},\ and\ \citenamefont {Jost}}]{salari2022}%
  \BibitemOpen
  \bibfield  {author} {\bibinfo {author} {\bibfnamefont {H.}~\bibnamefont
  {Salari}}, \bibinfo {author} {\bibfnamefont {M.}~\bibnamefont {Di~Stefano}},\
  and\ \bibinfo {author} {\bibfnamefont {D.}~\bibnamefont {Jost}},\ }\bibfield
  {title} {\bibinfo {title} {Spatial organization of chromosomes leads to
  heterogeneous chromatin motion and drives the liquid-or gel-like dynamical
  behavior of chromatin},\ }\href@noop {} {\bibfield  {journal} {\bibinfo
  {journal} {Genome Res.}\ }\textbf {\bibinfo {volume} {32}},\ \bibinfo {pages}
  {28} (\bibinfo {year} {2022})}\BibitemShut {NoStop}%
\bibitem [{\citenamefont {Fixman}(1962)}]{fixman1962}%
  \BibitemOpen
  \bibfield  {author} {\bibinfo {author} {\bibfnamefont {M.}~\bibnamefont
  {Fixman}},\ }\bibfield  {title} {\bibinfo {title} {Radius of gyration of
  polymer chains},\ }\href@noop {} {\bibfield  {journal} {\bibinfo  {journal}
  {J. Chem. Phys.}\ }\textbf {\bibinfo {volume} {36}},\ \bibinfo {pages} {306}
  (\bibinfo {year} {1962})}\BibitemShut {NoStop}%
\end{thebibliography}%
\end{document}